\documentclass[ALICE,manyauthors]{cernphprep}

\newcommand{\pp}{p\kern-0.05em p}
\newcommand{\ppbar}{\mathrm{p}\kern-0.05em \overline{\mathrm{p}}}

\newcommand{\PbPb}{\ensuremath{\mbox{Pb--Pb}}}
\newcommand{\GeVc}{\ensuremath{\mathrm{GeV}\kern-0.05em/\kern-0.02em c}}

\newcommand{\pT}{\ensuremath{p_{\mathrm{T}}}}

\newcommand{\pTtrack}{\ensuremath{p_{\mathrm{T,track}}}}

\newcommand{\etajet}{\ensuremath{\eta_{\mathrm{jet}}}}
\newcommand{\pTjet}{\ensuremath{p_{\mathrm{T}}^{\mathrm{jet}}}}
\newcommand{\pTchjet}{\ensuremath{p_{\mathrm{T}}^{\mathrm{ch\; jet}}}}

\newcommand{\pTdet}{\ensuremath{p_{\mathrm{T,det}}^{\mathrm{ch\; jet}}}}

\newcommand{\Ninc}{\ensuremath{N_{\mathrm{jets}}}}
\newcommand{\Sinc}{\ensuremath{\sigma_{\mathrm{jets}}}}

\newcommand{\zcut}{\ensuremath{z_{\mathrm{cut}}}}

\newcommand{\DeltaR}{\ensuremath{\Delta R_{\mathrm{axis}}}}

\usepackage[comma,square,numbers,sort&compress]{natbib}
\usepackage{hyperref}
\usepackage{lineno}
\usepackage{amsmath}
\usepackage{bigints}  
\usepackage{afterpage}  
\usepackage{multirow}
\usepackage{xcolor}
\usepackage[Symbol]{upgreek}
\usepackage[T1]{fontenc}
\usepackage{orcidlink}

\begin{document}%

\begin{titlepage}
\PHyear{2022}
\PHnumber{242}      
\PHdate{08 November}  
%

\title{Measurement of the angle between jet axes in \pp{} collisions at $\mathbf{\sqrt{\textit s}=5.02}  \ \mathbf{TeV}$}

\ShortTitle{Measurement of the angle between jet axes in \pp{} collisions at $\sqrt{s} = 5.02$ TeV}   

\Collaboration{ALICE Collaboration\thanks{See Appendix~\ref{app:collab} for the list of collaboration members}}
\ShortAuthor{ALICE Collaboration} 

\begin{abstract}

This article reports measurements of the angle between differently defined jet axes
 in \pp{} collisions at $\sqrt s = 5.02 \ \mathrm{TeV}$ carried out by the ALICE Collaboration. Charged particles at midrapidity are clustered into jets with resolution parameters $R=0.2$ and 0.4. 
 The jet axis, before and after Soft Drop grooming, is
 compared to the jet axis from
 the Winner-Takes-All (WTA)  recombination scheme.  The angle between these axes, $\DeltaR$, probes a wide phase space of the jet formation and evolution, ranging from the initial high-momentum-transfer scattering to the hadronization process. The $\DeltaR$ observable is presented for $20 < \pTchjet < 100 \ \GeVc$, and  compared to predictions from the PYTHIA~8 and Herwig~7 event generators. 
 The distributions can also be calculated analytically with a leading hadronization correction  related to the non-perturbative component of the Collins\textendash{}Soper\textendash{}Sterman (CSS) evolution kernel. Comparisons to analytical predictions at next-to-leading-logarithmic accuracy with leading hadronization correction implemented from experimental extractions of the CSS kernel in Drell\textendash{}Yan measurements are presented. The analytical predictions describe the measured data within 20\% in the perturbative regime, with surprising agreement in the non-perturbative regime as well. These results are compatible with the universality of the CSS kernel in the context of jet substructure. 

\end{abstract}
\end{titlepage}
\setcounter{page}{2}

\section{Introduction}

Jets play a fundamental role in the study of quantum chromodynamics (QCD). Jets form when partons (quarks and gluons) scattered in high-momentum-transfer (hard) processes fragment into lower-energy (softer) partons. The fragmentation creates a shower of partons, until the average energy per particle falls below the scale at which color-neutral hadrons 
emerge.
Jet substructure, which studies the radiation patterns inside jets,
is a prolific field in both  experiment~\cite{Kogler:2018hem} and theory~\cite{Larkoski:2017jix}.
The large difference between the energy scale of the hard-scattered parton and the measured hadrons leaves a significant phase space for jet formation and evolution~\cite{Marzani:2019hun}. Therefore, a multitude of jet-substructure measurements probing different regions of this phase space is needed to characterize the internal substructure of jets and advance our understanding of QCD.
Numerous analyses have been carried out by the ALICE~\cite{ALICE:2014dla,ALICE:2018gni,ALICE:2018ype,ALICE:2019cbr,ALICE:2019ykw,ALICE:2021njq,ALICE:2021obz,ALICE:2021vrw,ALICE:2021aqk},
ATLAS~\cite{ATLAS:2012am,ATLAS:2012nnf,ATLAS:2013uet,ATLAS:2017zda,ATLAS:2019kwg,ATLAS:2019rqw,ATLAS:2019mgf,ATLAS:2020bbn,Aad:2019wdr},
CMS~\cite{CMS:2012oyn,CMS:2012nro,CMS:2014jjt,CMS:2017yer,CMS:2017qlm,CMS:2018vzn,CMS:2018jco,CMS:2021iwu}, and
LHCb~\cite{LHCb:2017llq,LHCb:2019qoc} Collaborations at the LHC,
as well as at RHIC~\cite{STAR:2020ejj,STAR:2021lvw,STAR:2021kjt}.

\begin{figure}[b]
    \centering
    \includegraphics[width=\textwidth]{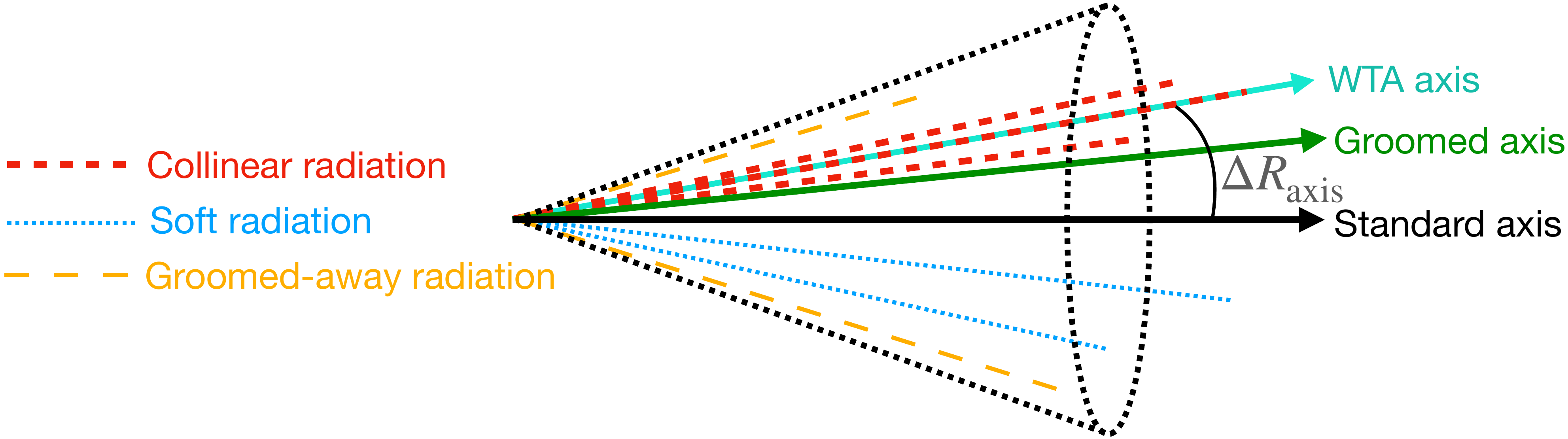}
    \caption{Representation of different jet axes. The colored dashed lines represent particles comprising the jet in the initial sample constructed with the anti-$k_{\rm T}$ algorithm and $E$ recombination scheme. These define the ``Standard'' jet axis. The grooming procedure removes soft wide-angle radiation (represented by orange long-dashed lines), and the resulting axis (``Groomed (SD)'') is defined by the remaining particles (red dashed and blue dotted lines). Finally, the ``WTA'' axis, which is determined with all the particles in the initial (ungroomed) jet, tends to be aligned with the most-energetic particle in the jet. The \DeltaR{} observable is determined from the angle between any pair of these axes.
    }
    \label{fig:cartoon}
\end{figure}

In this article, a novel jet-substructure observable proposed in Ref.~\citenum{Cal:2019gxa} corresponding to the angle between two definitions of the axis of a jet:
\begin{equation}
    \Delta R_{\rm axis} \equiv \sqrt{(y_{\rm axis \, 1}-y_{\rm axis \, 2})^2+(\varphi_{\rm axis \, 1}-\varphi_{\rm axis \, 2})^2}
    \label{eq:ang_dist}
\end{equation}
is studied. Here, a given axis corresponds to a set of coordinates in the rapidity ($y$) and azimuth ($\varphi$) plane. This is illustrated in Fig.~\ref{fig:cartoon}. The ``Standard'' axis is determined by clustering the jet constituents with the anti-$k_{\rm T}$ algorithm~\cite{Cacciari:2008gp} and the $E$ recombination scheme. Alternatively, the ``Groomed'' axis can be determined by first using a systematic procedure to remove the soft wide-angle radiation in the jet and then determining the axis of the anti-$k_{\rm T}$ jet (with $E$ recombination scheme), clustering only the constituents that remain after grooming.
Given the removal of soft wide-angle radiation, this quantity is less sensitive to non-perturbative effects. The grooming procedure used in this analysis is reviewed below.

A third way to define the jet axis corresponds to reclustering the jet (initially clustered with the anti-$k_{\rm T}$ algorithm and $E$ recombination scheme) with the Cambridge\textendash{}Aachen (C/A) algorithm~\cite{Dokshitzer:1997in,Wobisch:1998wt}
ensuring that the resulting jet includes all constituents from the original jet.
The C/A algorithm clusters particles exclusively based on their spatial separation (in the $y$\textendash{}$\varphi$ plane) without taking into account their energies/momenta.
Thus, particles closest in distance are clustered first, which results in an angular-ordered clustering sequence.
Subsequently, the constituents are recombined with the Winner-Takes-All (WTA) transverse-momentum recombination scheme~\cite{Bertolini:2013iqa}.
This consists of going through the clustering history and combining the pair of prongs in each $2 \rightarrow 1$ merging by assigning to the merged branch the direction of the
harder of the two prongs and transverse momentum (\pT) corresponding to the sum of the two transverse momenta. The WTA scheme is infrared and collinear (IRC) safe and the resulting axis is insensitive to soft radiation at leading power of \DeltaR, which makes it amenable to perturbative calculations~\cite{Cal:2019gxa}.

The \DeltaR{} observable is IRC safe and can be analytically calculated~\cite{Cal:2019gxa}. The sensitivity of \DeltaR{} to non-perturbative effects can be controlled by changing the pair of axes that are used to determine the observable and
also by varying the grooming and the jet resolution parameter. For instance, the difference between the Standard and Groomed axes specifically probes the influence of the groomed soft wide-angle radiation on the jet direction.
Similarly, the WTA axis constitutes a robust reference against soft-radiation effects. The angle between the WTA and Standard/SD axes offers a proxy for studying how the fragmenting parton that initiated the jet gets distributed inside the jet cone.
Consequently, this observable can be used both to test the performance of analytic predictions and to constrain non-perturbative models used in event generators.

The \DeltaR{} is sensitive to Transverse-Momentum-Dependent (TMD) physics~\cite{Cal:2019gxa}.
The leading hadronization correction for \DeltaR{} is related to the non-perturbative component of the Collins\textendash{}Soper\textendash{}Sterman (CSS) evolution kernel or rapidity anomalous dimension.
Here, leading means logarithmically enhanced in the calculation of \DeltaR, in contrast to other nonperturbative components that are not logarithmically enhanced and thus expected to be small.
The CSS kernel is a process-independent non-perturbative function constrained from fits to measured data that governs the non-perturbative evolution in rapidity and encodes information about soft-gluon exchanges between partons in vacuum~\cite{Vladimirov:2020umg}.
As such, extractions from diverse physics processes such as Drell\textendash{}Yan and semi-inclusive deep-inelastic scattering~\cite{Landry:2002ix,Sun:2014dqm,Bacchetta:2017gcc,Bertone:2019nxa}
should adequately describe the hadronization corrections for the jet substructure observable presented in this study.
This measurement can be included in future global fits to further constrain the CSS kernel.
Furthermore, it has been proposed that the CSS kernel can be determined from lattice QCD~\cite{Ebert:2018gzl,Shanahan:2020zxr}, and such results can be benchmarked with our measurements.

The measurements presented here can also serve as a reference for measurements in heavy-ion collisions. The internal structure of jets fragmenting in the strongly interacting, deconfined state of matter formed in heavy-ion collisions 
is modified relative to jet fragmentation in vacuum~\cite{Jacak:2012dx, Busza:2018rrf}. Comparing to pp collisions allows us to study the medium properties. 

In this analysis, the Soft Drop grooming procedure~\cite{Larkoski:2014wba} is used to determine the Groomed axis (referred to as ``SD'' axis from here on). The jet is first reclustered with the C/A algorithm, with the same resolution parameter used in the original clustering of the anti-$k_{\rm T}$-based sample.
This tree is then recursively declustered starting from the largest-distance splitting and at each $1\rightarrow2$ splitting, the condition

\begin{equation}
    \frac{\mathrm{min}(p_{\rm T,1},p_{\rm T,2})}{p_{\rm T,1}+p_{\rm T,2}}>\zcut\Big(\frac{\Delta R_{1,2}}{R}\Big)^{\beta}
\end{equation}
is checked. Here, $p_{\rm T,1}$ and $p_{\rm T,2}$ are the transverse momenta of each prong, and $\Delta R_{1,2}$ is the angular distance between them, calculated as in Eq.~\ref{eq:ang_dist}. The jet resolution parameter is denoted by $R$. The free parameters \zcut{} and $\beta$ determine how asymmetric and how wide the splitting can be, respectively. If the condition is not satisfied at a given splitting, the softer branch is removed from the tree and the procedure continues through the harder prong. When a splitting that satisfies the condition is found, the Soft Drop procedure stops and the branch parent to the splitting defines the groomed jet.
If there is not a single splitting that satisfies the Soft Drop condition, then the grooming procedure fails.
These jets are labeled ``untagged'' and their abundance is used in the normalization convention used in this analysis. A scan in Soft Drop parameters is carried out, both fixing $\zcut = 0.1$ and varying $\beta$ from 0 to 3 in increments of 1 and fixing $\beta=1$ and varying $\zcut$ from 0.1 to 0.3 in increments of 0.1.

This article presents the first measurement of the \DeltaR{} observable, which is carried out in \pp{} collisions at a center-of-mass energy $\sqrt{s}=5.02$ TeV with the ALICE detectors using charged-particle jets.
The document is structured as follows.
Section~\ref{sec:exp_data} describes the experimental setup used for the measurement and details of the datasets.
Section~\ref{sec:analysis} outlines the steps taken in the data analysis.
Section~\ref{sec:systematics} describes variations implemented in the analysis to determine systematic uncertainties.
Section~\ref{sec:results} presents our results and a discussion of the findings.
Section~\ref{sec:conclusions} summarizes our findings and introduces the conclusions of this work.
 
\section{Experimental setup and data sets}
\label{sec:exp_data}

This analysis uses proton\textendash{}proton (pp) collisions at $\sqrt{s} = 5.02$ TeV, recorded in 2017 with the ALICE detector at the CERN LHC~\cite{Evans:2008zzb}. 
A detailed description of the ALICE detector and its performance can be found in Refs.~\citenum{ALICE:2014sbx,ALICE:2008ngc}. The event sample consists of minimum-bias events triggered by a coincidence of hits in the two V0 scintillator detectors~\cite{ALICE:2013axi}, which cover an azimuthal acceptance of $0 < \varphi < 2\uppi$ and pseudorapidity 
$2.8 < \eta < 5.1$ (V0A) and $-3.7 < \eta < -1.7$ (V0C).
Pileup effects are removed by discarding events that contain multiple reconstructed vertices~\cite{ALICE:2019qyj}.
The primary event vertex is required to be within $\pm 10$ cm of the nominal interaction point.
The data sample contains 870 million events, corresponding to an integrated luminosity of 18.0(4) nb$^{-1}$~\cite{ppXsec}.

Tracks were reconstructed using the Inner Tracking System (ITS)~\cite{ALICE:2010tia} and Time Projection Chamber (TPC)~\cite{Alme:2010ke}. The track sample consists of two track categories. In the first category, tracks are required to include at least one hit in the silicon pixel detector (SPD) of the ITS, cross at least 70 (out of 159) TPC readout pad rows, and have at least 80$\%$ of the geometrically findable space points in the TPC, among other quality criteria~\cite{ALICE:2019wqv}. In the second category, tracks that do not contain any SPD hits but otherwise satisfy the track-selection criteria are refit with a constraint to the primary vertex. The inclusion of this second category produces a track sample approximately uniform in azimuth, while preserving a similar \pT{} resolution to tracks with SPD hits. Tracks with $\pTtrack > 0.15 \ \GeVc$, $|\eta|<0.9$, and $0 < \varphi < 2\uppi$ are assigned the charged-pion mass and included in the analysis.

The ALICE detector tracking efficiency in pp collisions grows from $\approx67\%$ at $\pTtrack=0.15$ \GeVc{} to $\approx84\%$ at $\pTtrack=1$ \GeVc{}, and remains above $\approx75\%$ for higher \pTtrack{}~\cite{ALICE:2014sbx}. The track momentum resolution increases from $\approx 1\%$ at $\pTtrack=1$ \GeVc{} to $\approx 4\%$ at $\pTtrack=4$ \GeVc.

\section{Analysis method}
\label{sec:analysis}

\subsection{Jet reconstruction}
Charged-particle jets were reconstructed from tracks with $0.15 < \pTtrack < 100$ \GeVc{},
using the anti-$k_{\rm T}$ algorithm~\cite{Cacciari:2008gp}, and the $E$ recombination scheme, for resolution parameters $R=0.2$ and $0.4$ using the FastJet package~\cite{Cacciari:2011ma}.
Only jets within the fiducial acceptance of the TPC ($|\etajet|<0.9-R$) were analyzed,
and the resulting distributions are reported in four \pTchjet{} intervals with edges at $[20,40,60,80,100]$ \GeVc.
The different axes used to construct the \DeltaR{} observable are determined from the jets in this sample. 
The underlying-event was not subtracted.

The jet reconstruction performance was estimated using
PYTHIA~8~\cite{Sjostrand:2014zea} with Monash 2013 tune~\cite{Skands:2014pea} events (``truth" level) propagated through a GEANT~3~\cite{Brun:1119728} model of the ALICE detector (``detector" level). 
In this sample, final-state particles are defined as those with a mean proper lifetime $c\tau>1 \ {\rm cm}$~\cite{ALICE:2017hcy}.
Particles (tracks) at truth (detector) level were individually clustered into jets, and the jets were matched between the two populations by requiring that the distance between their Standard axes in the $y$\textendash{}$\varphi$ plane satisfied $\Delta R < 0.6 R$ and that the match was unique~\cite{ALICE:2019qyj}. The matched jets were used to construct the jet energy scale (${\rm JES} \equiv (p^{\rm ch \, jet}_{\rm T, \, det}-p^{\rm ch \, jet}_{\rm T, \, truth})/p^{\rm ch \, jet}_{\rm T, \, truth}$), jet energy resolution (${\rm JER}\equiv\sigma(p^{\rm ch \, jet}_{\rm T, \, det})/p^{\rm ch \, jet}_{\rm T, \, truth}$), 
and jet reconstruction efficiency ($\epsilon_{\rm reco}$), where $p^{\rm ch \, jet}_{\rm T, \, truth}$ and $p^{\rm ch \, jet}_{\rm T, \, det}$ refer to the transverse momentum of the truth- and detector-level jets, respectively, and $\sigma(p^{\rm ch \, jet}_{\rm T, \, det})$ corresponds to the standard deviation of the $p^{\rm ch \, jet}_{\rm T, \, det}$ spectrum for a given value of $p^{\rm ch \, jet}_{\rm T, \, truth}$.
Table~\ref{tab:jetPerf} shows values characterizing the jet-reconstruction performance. In the case of the JES, this distribution is peaked at 0, but has an asymmetric tail that shifts the mean value, $\varDelta_{\rm JES}$, due to tracking inefficiencies. The unfolding procedure described in the next section corrects for these JES and JER effects.

\begin{table}[!ht]
\centering
\caption{Approximate values for the figures of merit characterizing the jet-reconstruction performance in this analysis. $\varDelta_{\rm JES}$ is the mean value of the JES spectrum. See text for details.}
\begin{tabular}{ccccc}
\hline \hline
$R$                    & \pTchjet \ (\GeVc) & $\varDelta_{\rm JES}$ (\%) & JER (\%) & $\epsilon_{\rm reco}$ (\%) \\
\hline
\multirow{5}{*}{0.4} & 20  & $-13$     & 21 & 97  \\
                     & 40  & $-16$     & 21 & 100 \\
                     & 60  & $-19$     & 22 & 100 \\
                     & 80  & $-21$     & 23 & 100 \\
                     & 100 & $-25$     & 24 & 100 \\
\hline
\multirow{5}{*}{0.2} & 20  & $-12$     & 21 & 94  \\
                     & 40  & $-16$     & 23 & 100 \\
                     & 60  & $-20$     & 24 & 100 \\
                     & 80  & $-23$     & 25 & 100 \\
                     & 100 & $-27$     & 26 & 100 \\
\hline
\end{tabular}
\label{tab:jetPerf}
\end{table}

\subsection{Corrections}

To obtain distributions free of detector effects (truth level), the measured spectra (detector level) were unfolded using an iterative procedure based on Bayes' theorem~\cite{DAgostini,DAgostini:2010hil}, implemented in the RooUnfold package~\cite{roounfold}. This unfolding procedure accounts for effects such as track \pT{} resolution, tracking inefficiencies, and particle interactions in the detector volume. The input to this procedure consisted of a 4-dimensional response matrix (RM) that maps the correspondence between detector and truth levels for \pTchjet{} and \DeltaR{}.
The RM was created using the simulated jets described in the previous section.
The \pTdet{} axes were constructed in the range $[10,130]$ \GeVc{} to capture bin-migration effects.
In the \DeltaR{} cases in which one of the two axes is the SD axis, the untagged jets were included in the unfolding procedure.
The unfolding iterations were fixed at the number that minimizes the quadratic sum of the statistical and systematic uncertainties. In some cases, a few more iterations were carried out until a convergence within 5\% between subsequent iterations was achieved.

The unfolding procedure was validated by performing a series of closure tests. A refolding test was performed to check that the detector-level spectrum can be recovered by reversing the unfolding procedure. In this test, the unfolded result was multiplied by the response matrix and the resulting spectrum was compared to the detector-level spectrum. Additionally, a statistical-closure test was performed to confirm that the unfolding procedure is robust against statistical fluctuations in the data. In this test, the simulated detector-level spectrum was smeared by an amount equal to the statistical uncertainty of the measured data. Subsequently, the smeared spectrum was unfolded, and the agreement between the resulting distribution and the truth-level simulated distribution was assessed. Finally, a shape-closure test was performed to account for the fact that the true detector-level distribution may be different than that from the generator, and to evaluate the robustness against this shape. In this test, the shapes of the simulated detector- and truth-level spectra were scaled by the prior-scaling function described in the systematic-uncertainty section. Subsequently, the scaled detector-level spectrum was unfolded, and the resulting spectrum was compared to the scaled truth-level distribution. In all these tests, closure within the statistical uncertainties was achieved.
 
\section{Systematic uncertainties}
\label{sec:systematics}

Three sources of systematic uncertainties are considered in this analysis:
those arising from the unfolding procedure, uncertainties 
in the tracking efficiency, and 
the choice of event generator used to produce the response matrix.

The unfolding uncertainty is estimated by performing four variations of the unfolding procedure:

\begin{enumerate}
    \item The analysis is repeated with a number of iterations varied by $\pm 2$ around the nominal value and the average difference with respect to the nominal spectrum is taken as an uncertainty.
    \item The analysis is repeated with a prior distribution multiplied by \\ $(\pTchjet \, [{\rm GeV}/c])^{\pm0.5}\times (\pm A\DeltaR \mp B)$, where the parameters $A$ and $B$ are selected for each \DeltaR{} variation such that the endpoints of the spectrum are scaled by $\approx \pm 20\%$. The maximum difference between these two variations and the nominal result is taken as an uncertainty.
    \item The analysis is repeated with the transverse-momentum range $\pTdet\in[5,135]$ \GeVc{} and the difference with respect to the nominal analysis ($\pTdet\in[10,130]$ \GeVc) is taken as an uncertainty.
    \item The analysis is repeated with a different \DeltaR{} binning at detector level (slightly finer or coarser granularity) and the difference with respect to the main result is taken as an uncertainty.
\end{enumerate}

Given that these four variations probe the same underlying source of uncertainty, the total unfolding uncertainty is defined as the standard deviation $\sigma_{\rm unfolding} \equiv \sqrt{\sum_{i=1}^{4} \sigma_i^2/4}$, where $\sigma_i$ corresponds to the uncertainty from the individual variations.

The uncertainty on the ALICE tracking efficiency is 3\textendash{}4$\%$, determined by varying the track selection parameters and possible imperfections in the description of the ITS\textendash{}TPC matching efficiency in the simulation. Consequently, the analysis was repeated with a RM populated with jets clustered over a track sample with $4\%$ of all tracks randomly discarded, taking the more conservative value.
Differences with respect to the nominal analysis were assigned as the systematic uncertainty due to the tracking efficiency.

To assess the model dependence of the analysis on the generated spectra used to compute the RM (based on PYTHIA~8 with Monash 2013 tune events propagated through GEANT~3), the analysis was repeated using RMs based on Herwig~7 (default tune) and PYTHIA~8 with Monash 2013 tune constructed using a fast simulation, and the difference between the two unfolded results was assigned as a systematic uncertainty. In a comparison based on PYTHIA~8 events, this fast simulation agrees with the GEANT~3-based simulation within 10\%.

The total systematic uncertainty is taken as the sum in quadrature of the contributions due to the unfolding, tracking efficiency, and choice of event generator.
Table~\ref{tab:syst_pT_40_60} summarizes the range of the systematic uncertainties in different $\DeltaR$ intervals for jets of $R=0.2$ and 0.4 in the transverse momentum range $40 < \pTchjet < 60 \ \GeVc$. 
Most often, the dominant systematic uncertainty originates from the uncertainty on the tracking efficiency.

\begin{table}[!ht]
\centering
\caption{Range of the estimated value of relative systematic uncertainties in $\DeltaR$ intervals in the transverse momentum range
$40 < \pTchjet < 60 \ \GeVc$.
The unfolding, tracking efficiency, and generator systematic uncertainties
can be found in the columns labeled Unf., Trk. Eff., and Gen., respectively.
In the case of the groomed observables, the grooming parameters are specified as $(\zcut,\beta)$.
The displayed uncertainties correspond to the lowest and highest values for a given setting.}
\label{tab:syst_pT_40_60}
\begin{tabular}{c|cccc|cccc}
 & \multicolumn{4}{c}{$R=0.2$} & \multicolumn{4}{c}{$R=0.4$} \\
\DeltaR & Unf. & Trk. Eff. & Gen. & Total & Unf. & Trk. Eff. & Gen. & Total \\
\hline
WTA\textendash{}Standard  & 0\textendash{}3\% & 0\textendash{}8\% & 0\textendash{}5\% & 1\textendash{}10\% & 1\textendash{}4\% & 0\textendash{}6\% & 0\textendash{}4\% & 1\textendash{}6\% \\
WTA\textendash{}SD (0.1, 0)  & 1\textendash{}4\% & 0\textendash{}4\% & 0\textendash{}3\% & 2\textendash{}6\% & 1\textendash{}4\% & 1\textendash{}4\% & 0\textendash{}4\% & 1\textendash{}6\% \\
WTA\textendash{}SD (0.1, 1)  & 0\textendash{}3\% & 0\textendash{}5\% & 0\textendash{}3\% & 1\textendash{}5\% & 1\textendash{}4\% & 0\textendash{}4\% & 0\textendash{}4\% & 2\textendash{}6\% \\
WTA\textendash{}SD (0.1, 2)  & 0\textendash{}2\% & 0\textendash{}5\% & 0\textendash{}3\% & 1\textendash{}5\% & 1\textendash{}4\% & 1\textendash{}5\% & 0\textendash{}4\% & 2\textendash{}6\% \\
WTA\textendash{}SD (0.1, 3)  & 0\textendash{}2\% & 0\textendash{}5\% & 0\textendash{}3\% & 1\textendash{}6\% & 1\textendash{}3\% & 0\textendash{}6\% & 0\textendash{}3\% & 1\textendash{}6\% \\
WTA\textendash{}SD (0.2, 1)  & 0\textendash{}4\% & 0\textendash{}5\% & 0\textendash{}4\% & 2\textendash{}6\% & 1\textendash{}4\% & 1\textendash{}4\% & 0\textendash{}4\% & 2\textendash{}6\% \\
WTA\textendash{}SD (0.3, 1)  & 1\textendash{}5\% & 1\textendash{}4\% & 0\textendash{}4\% & 2\textendash{}6\% & 0\textendash{}3\% & 1\textendash{}4\% & 0\textendash{}5\% & 1\textendash{}6\% \\
Standard\textendash{}SD (0.1, 0)  & 1\textendash{}6\% & 0\textendash{}3\% & 0\textendash{}2\% & 1\textendash{}7\% & 1\textendash{}3\% & 0\textendash{}4\% & 0\textendash{}2\% & 2\textendash{}4\% \\
Standard\textendash{}SD (0.1, 1)  & 0\textendash{}4\% & 0\textendash{}5\% & 0\textendash{}2\% & 1\textendash{}6\% & 1\textendash{}4\% & 0\textendash{}5\% & 1\textendash{}2\% & 2\textendash{}7\% \\
Standard\textendash{}SD (0.1, 2)  & 1\textendash{}4\% & 1\textendash{}5\% & 0\textendash{}2\% & 1\textendash{}7\% & 1\textendash{}4\% & 0\textendash{}5\% & 0\textendash{}2\% & 1\textendash{}6\% \\
Standard\textendash{}SD (0.1, 3)  & 1\textendash{}5\% & 1\textendash{}5\% & 0\textendash{}3\% & 1\textendash{}7\% & 1\textendash{}3\% & 1\textendash{}4\% & 0\textendash{}4\% & 2\textendash{}5\% \\
Standard\textendash{}SD (0.2, 1)  & 1\textendash{}4\% & 0\textendash{}4\% & 0\textendash{}2\% & 2\textendash{}6\% & 0\textendash{}2\% & 0\textendash{}4\% & 0\textendash{}3\% & 1\textendash{}5\% \\
Standard\textendash{}SD (0.3, 1)  & 2\textendash{}4\% & 1\textendash{}4\% & 0\textendash{}2\% & 2\textendash{}6\% & 0\textendash{}3\% & 0\textendash{}5\% & 0\textendash{}3\% & 1\textendash{}6\% \\
\end{tabular}
\end{table}
 
\section{Results and discussion}
\label{sec:results}

The $\DeltaR$ distributions are reported as normalized differential cross sections:

\begin{equation}
\frac{1}{\Sinc (\pTchjet)} \frac{{\rm d}\sigma}{{\rm d}\DeltaR} \left(\pTchjet\right)
\equiv
\frac{1}{\Ninc (\pTchjet)} \frac{{\rm d}N}{{\rm d}\DeltaR} \left(\pTchjet\right),
\end{equation}

for jets of $R=0.2$ and 0.4 in 20-\GeVc-wide \pTchjet{} intervals in the 20\textendash{}100 \GeVc{} range. Here, $\Ninc{}(\pTchjet)$ is the number of inclusive jets in a given \pTchjet{} interval.
In the groomed cases, the normalization factor \Ninc{} is obtained by including the number of untagged jets  (i.e.\ without any splitting in the jet that satisfies the SD condition, so that the grooming process fails) in the unfolding procedure. Therefore, the number of jets appearing in the $\DeltaR$ distributions differs from \Ninc{} by the number of jets that are untagged.

Sections~\ref{sect:mccomp} and~\ref{sect:calc} present the experimental results as well as comparisons to predictions from MC generators and analytical calculations, respectively.

\subsection{Comparison to MC generators}
\label{sect:mccomp}

\begin{figure}
    \centering
    \includegraphics[width=0.49\textwidth]{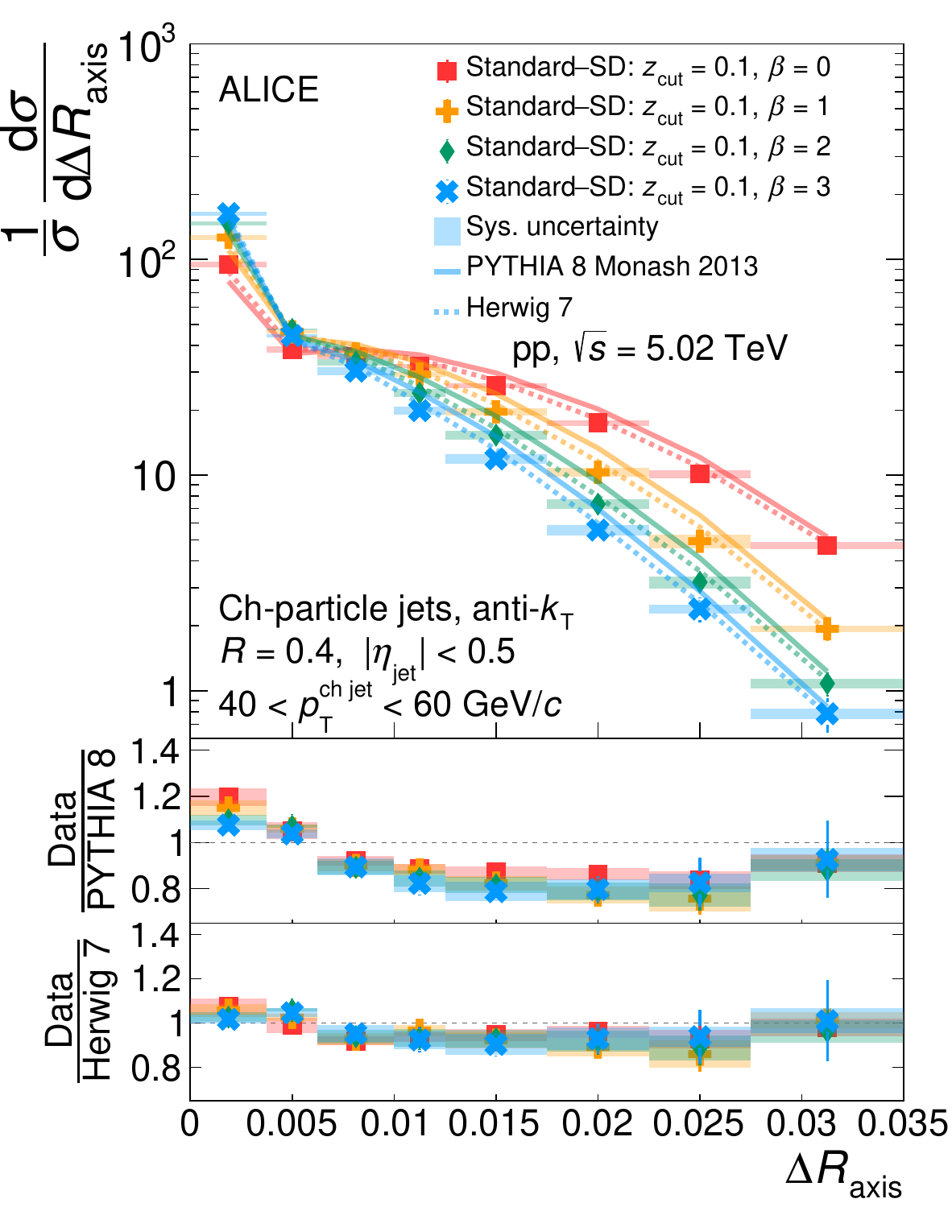}
    \includegraphics[width=0.49\textwidth]{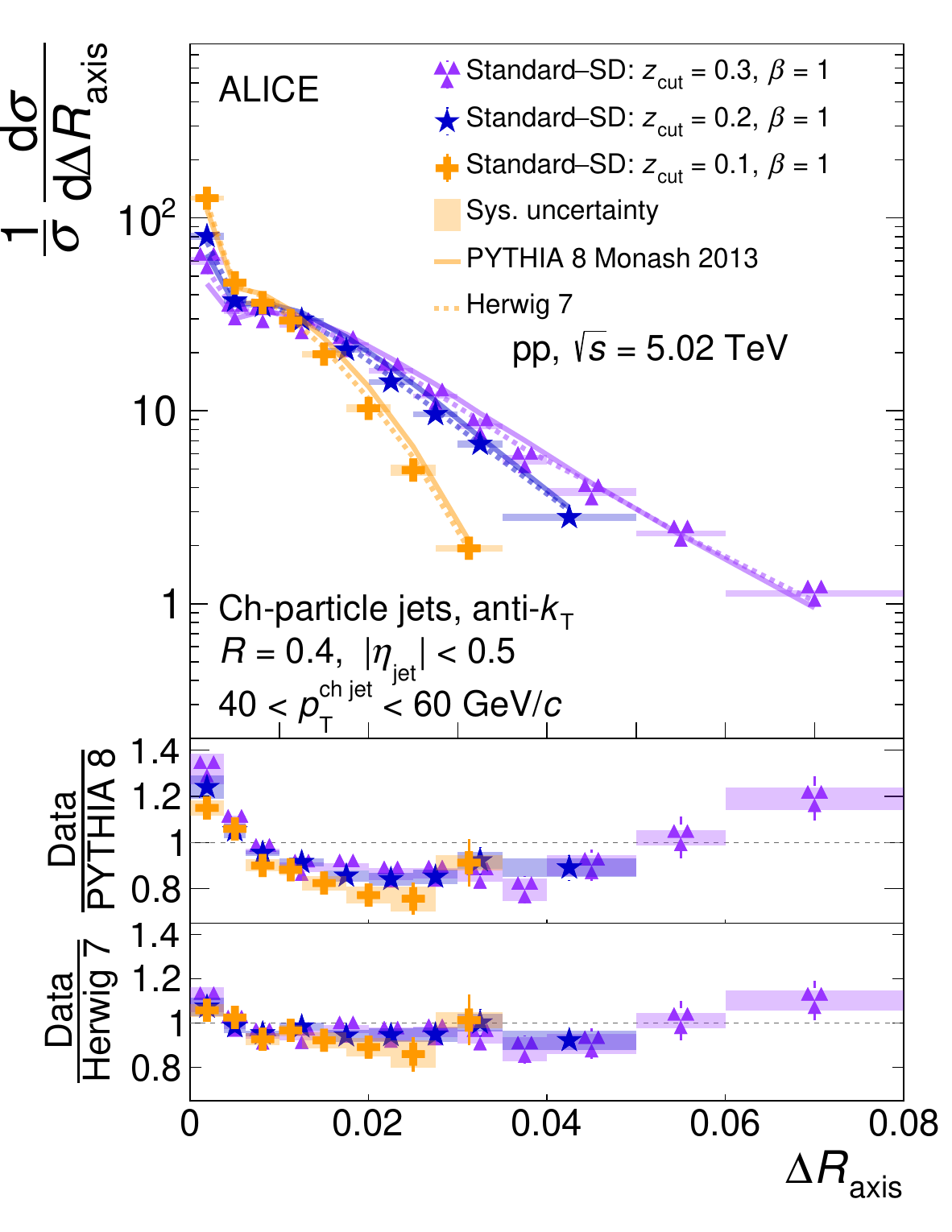}
    \includegraphics[width=0.49\textwidth]{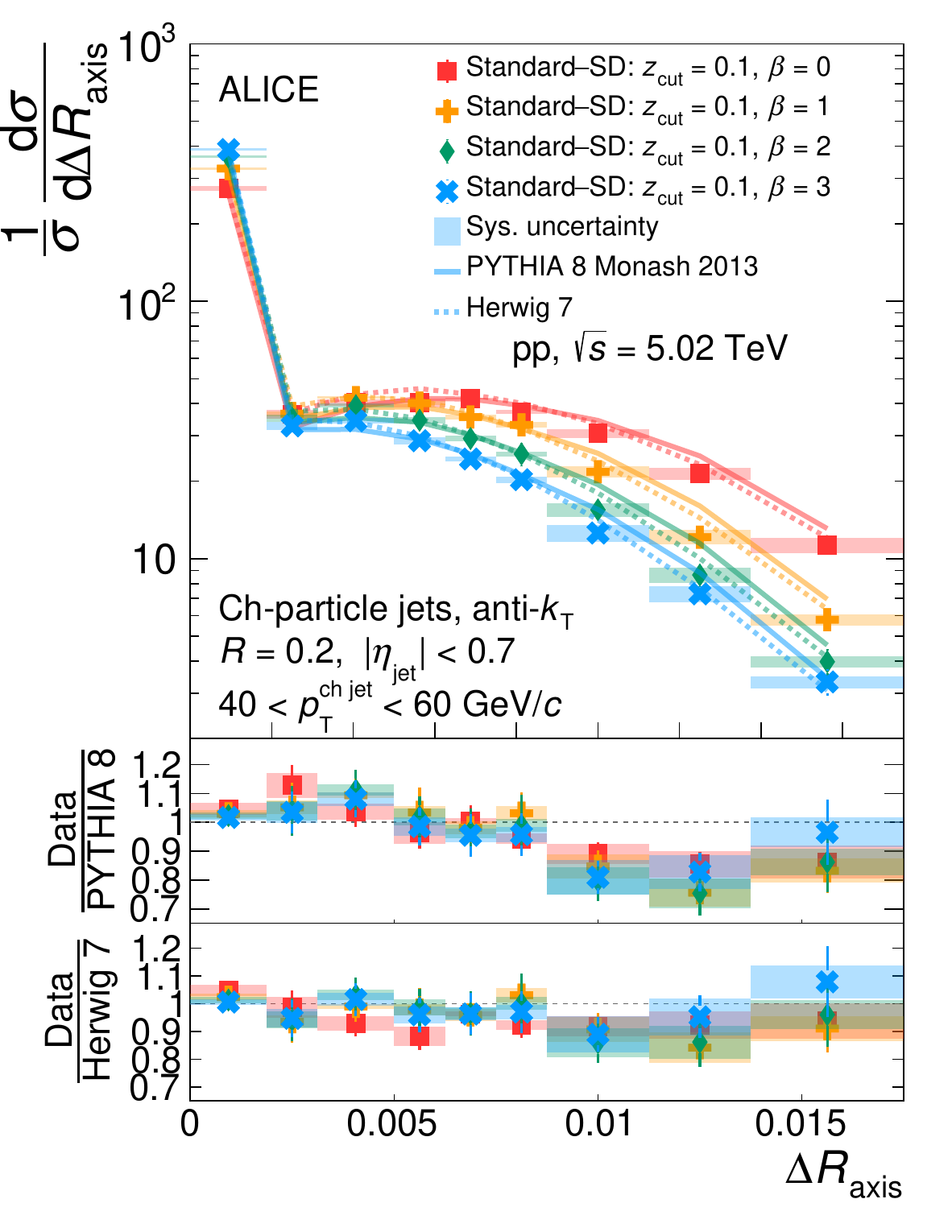}
    \includegraphics[width=0.49\textwidth]{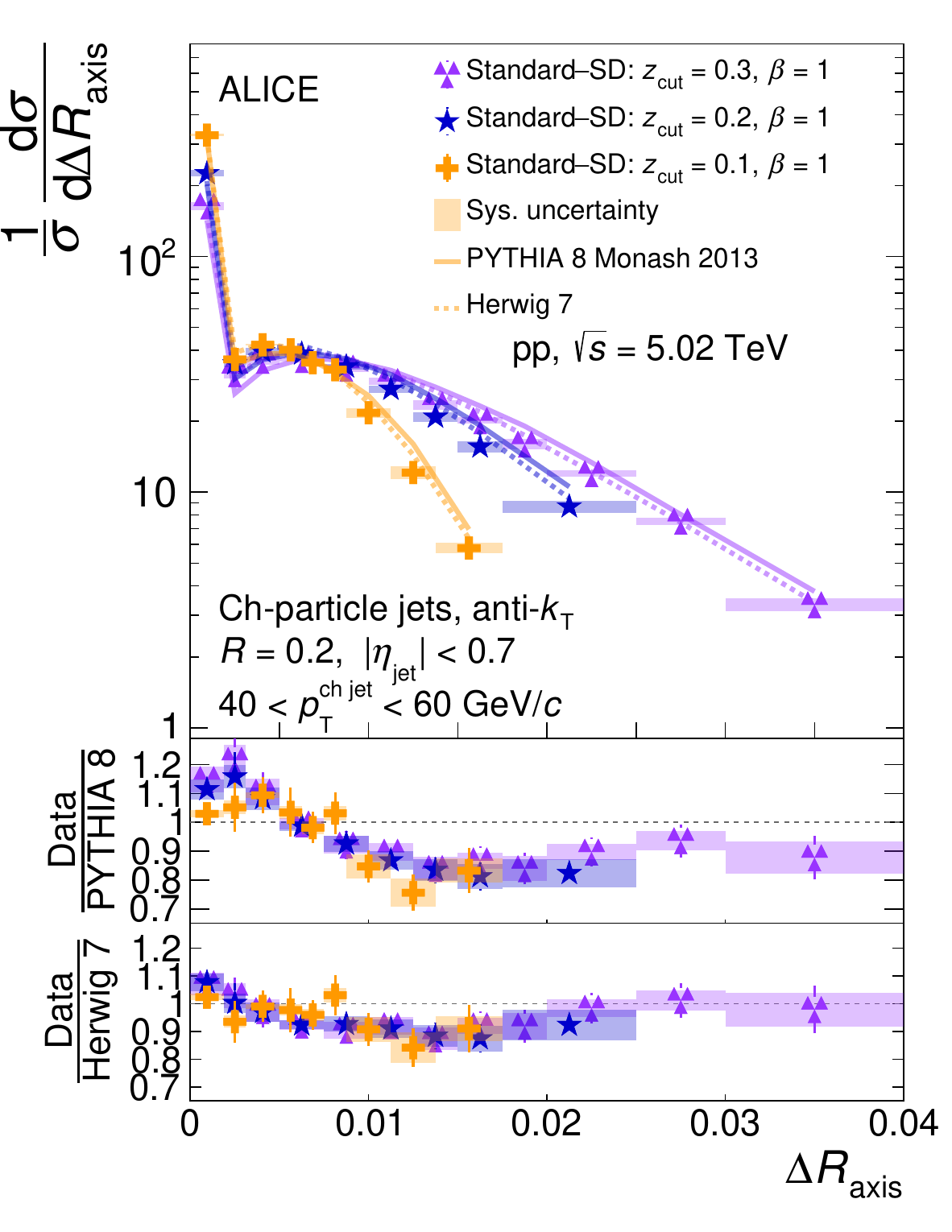}
    \caption{Comparison between the \DeltaR{} Standard\textendash{}SD measured distributions and Monte Carlo event generators for jets of $R=0.4$ (top) and 0.2 (bottom) in $40<\pTchjet<60$ \GeVc. Left: distributions with $\zcut=0.1$ and varying $\beta$. Right: distributions with $\beta=1$ and varying $\zcut$.}
    \label{fig:data_generators_Standard_SD}
\end{figure}

\begin{figure}
    \centering
    \includegraphics[width=0.49\textwidth]{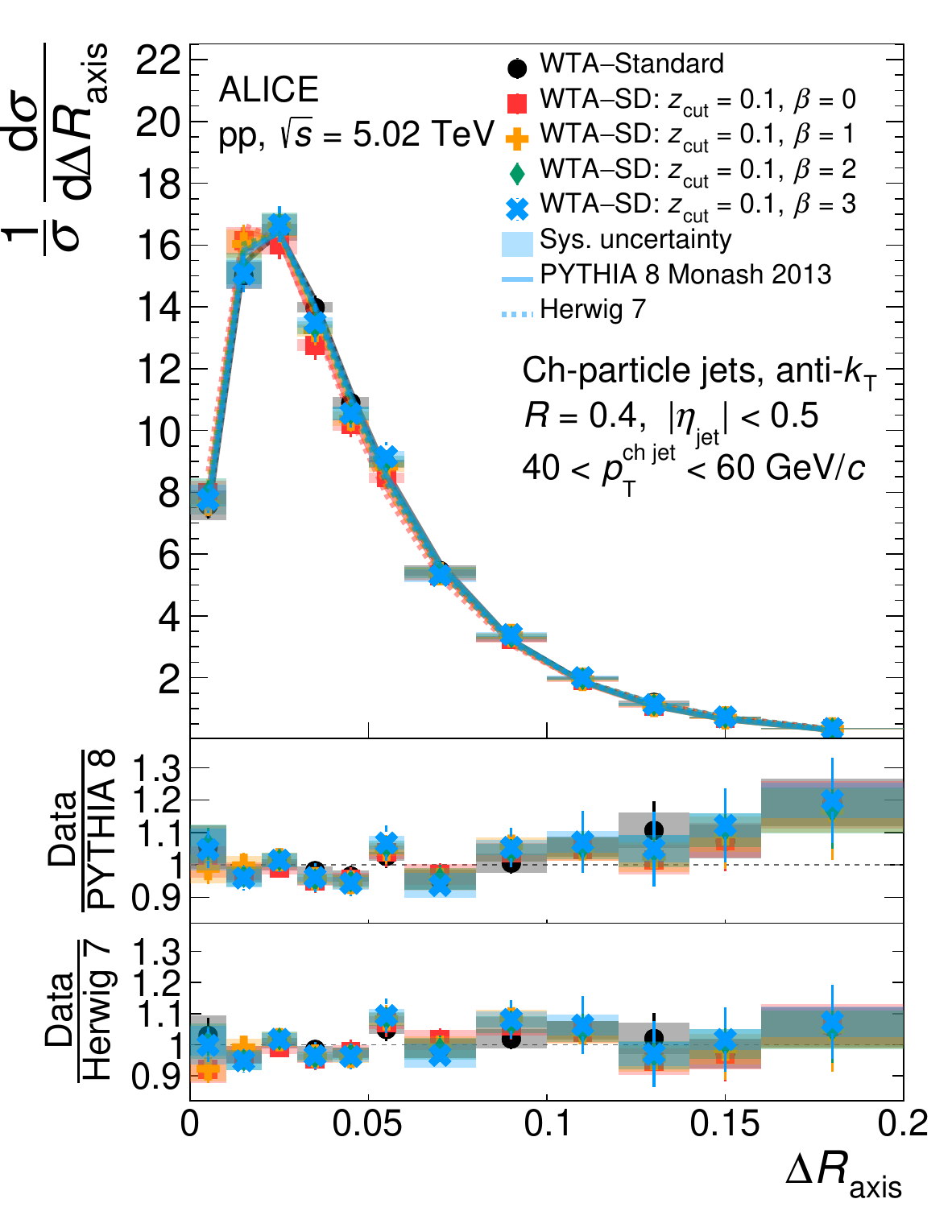}
    \includegraphics[width=0.49\textwidth]{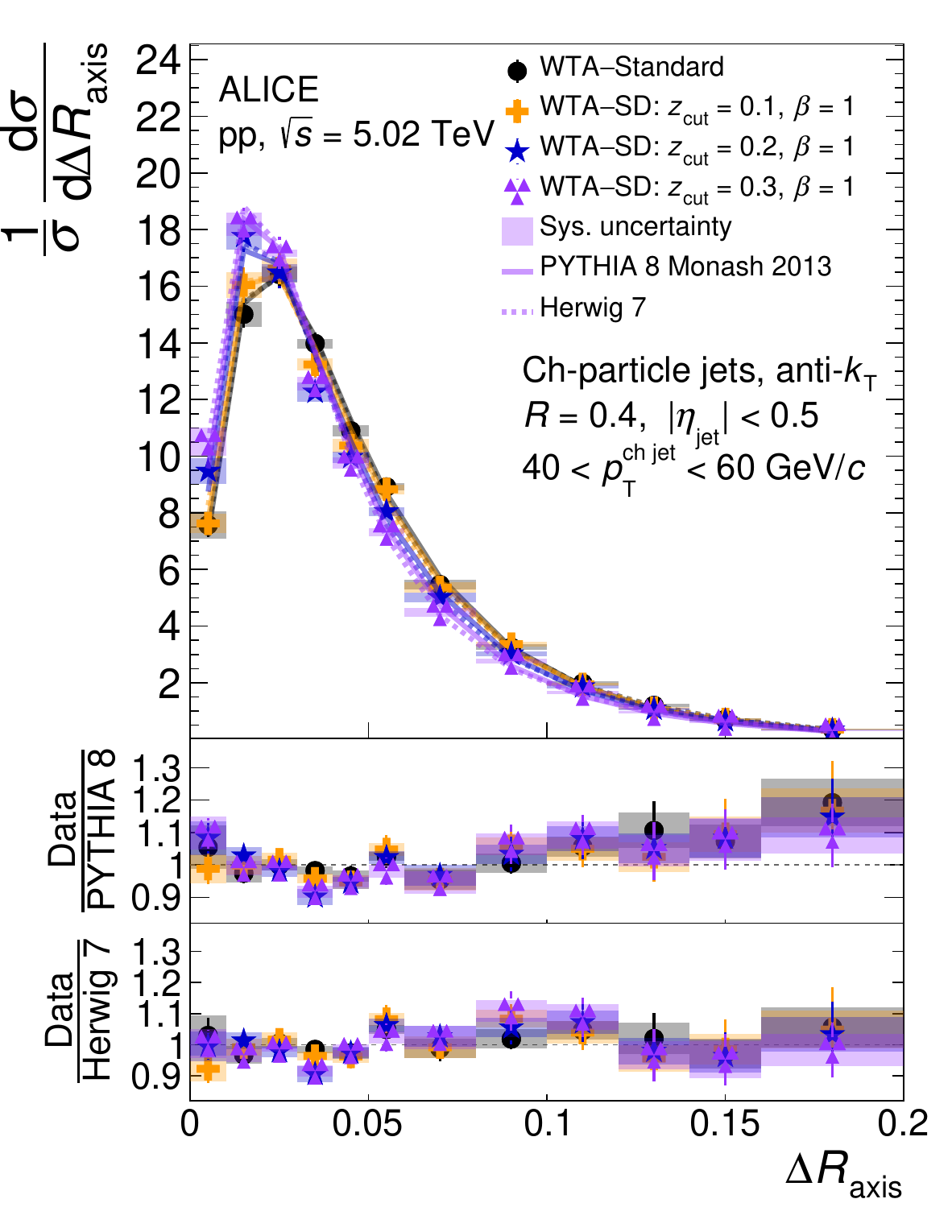}
    \includegraphics[width=0.49\textwidth]{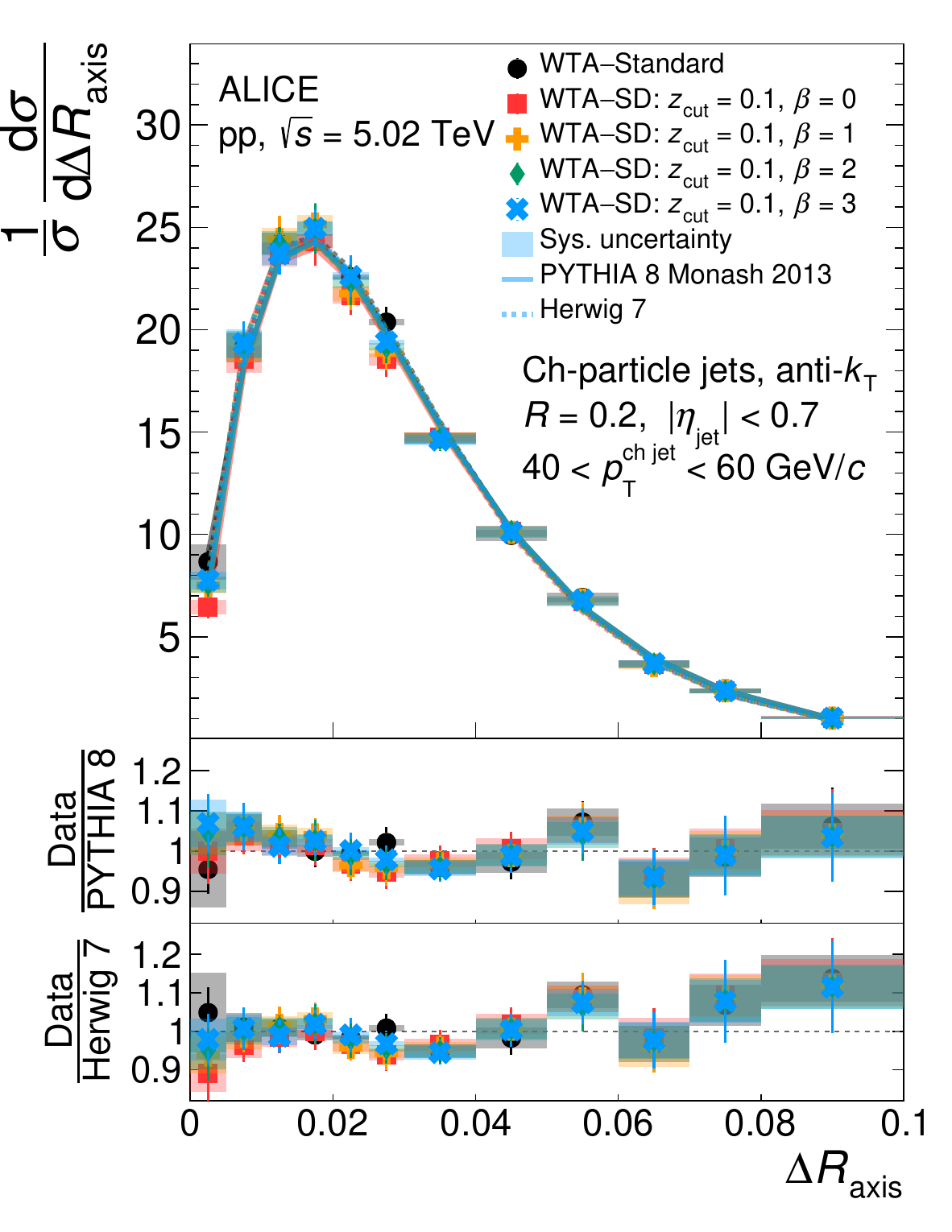}
    \includegraphics[width=0.49\textwidth]{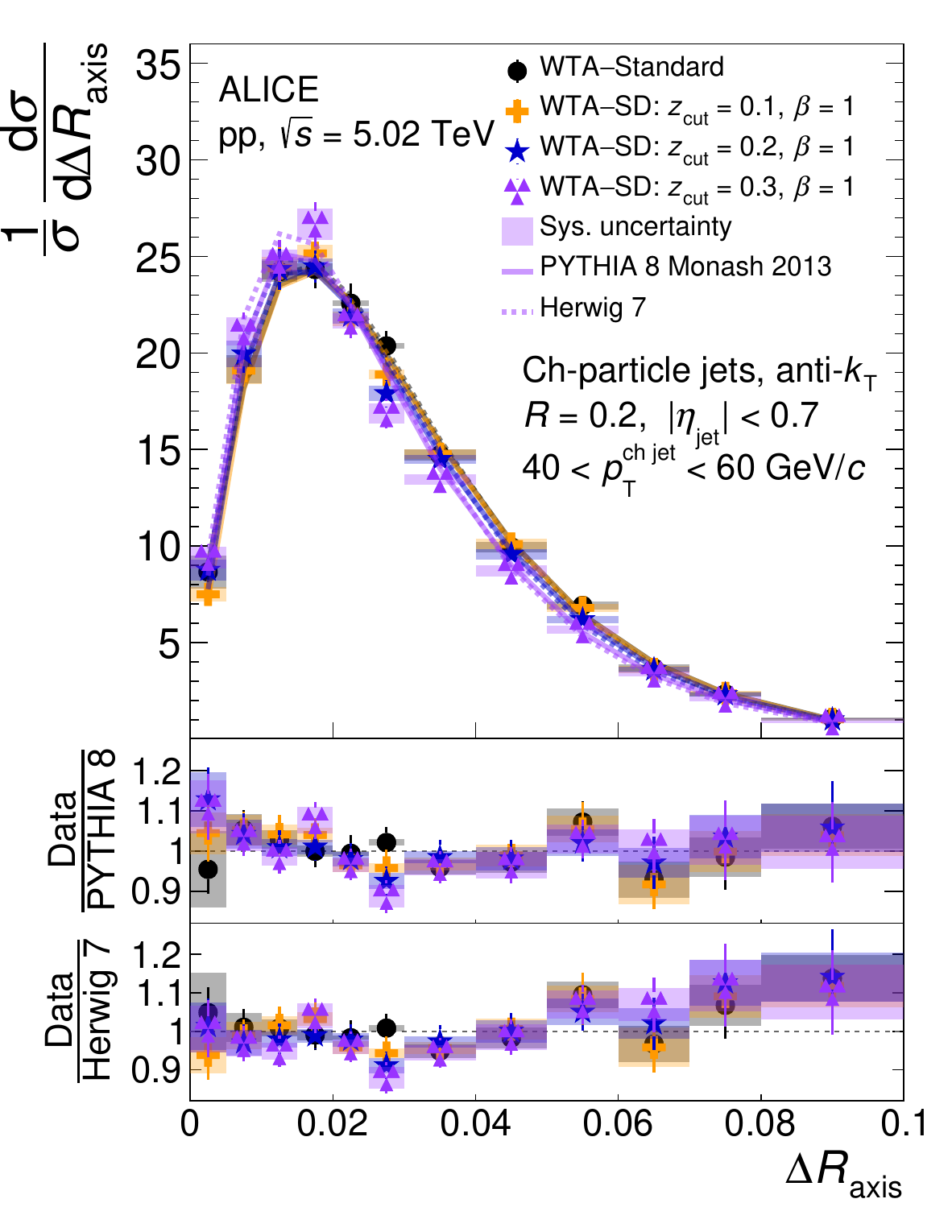}
    \caption{Comparison between the \DeltaR{} WTA\textendash{}Standard and WTA\textendash{}SD measured distributions and Monte Carlo event generators for jets of $R=0.4$ (top) and 0.2 (bottom) in $40<\pTchjet<60$ \GeVc. Left: distributions with $\zcut=0.1$ and varying $\beta$. Right: distributions with $\beta=1$ and varying $\zcut$.}
    \label{fig:data_generators_WTA}
\end{figure}

Figure~\ref{fig:data_generators_Standard_SD} 
compares the measured Standard\textendash{}SD distributions 
with predictions from Monte Carlo event generators, for jets of $R=0.4$ (top) and 0.2 (bottom) in $40<\pTchjet<60$ \GeVc. 
Results for different \pTchjet{} intervals are presented in Appendix~\ref{app:A}.
The vertical error bars 
are statistical uncertainties, and the rectangles 
indicate the total systematic uncertainties.
The solid (dashed) lines show results 
from PYTHIA~8 (Herwig~7).
The bottom two panels in these figures correspond to the ratios of data to PYTHIA~8 and Herwig~7 distributions. The left panel of Fig.~\ref{fig:data_generators_Standard_SD} shows the case in which \zcut{} is fixed at 0.1 and $\beta$ is varied from 0 to 3. In the right panel, 
$\beta$ is fixed at 1 and \zcut{} is varied from 0.1 to 0.3.
These spectra are plotted with a logarithmic $y$-axis 
to better exhibit the entire distribution.
Figure~\ref{fig:data_generators_WTA} shows the equivalent comparison for the case of the WTA\textendash{}Standard and WTA\textendash{}SD distributions.

Overall, the Standard\textendash{}SD distributions are narrow and  peaked at very small values.
This implies that the Standard and SD axes are aligned and grooming does not significantly impact the jet direction. However, as the grooming becomes more aggressive (i.e.\ higher \zcut{} or smaller $\beta$), the alignment between the Standard and SD axes worsens somewhat. This trend is present for both $R = 0.2$ and R = 0.4. There is a maximum at $\DeltaR^{\rm Standard-SD}=0$ that corresponds to jets for which the first splitting after reclustering the jet with the C/A algorithm
already satisfies the SD condition. As a result,
the difference
with respect to the Standard axis is exactly 0.

The shape of the $\DeltaR^{\rm Standard-SD}$ spectra is better described by Herwig~7 than by PYTHIA~8. In the $40<\pTchjet<60$ \GeVc{} range, bin-by-bin deviations with respect to PYTHIA~8 (Herwig~7) reach values up to $\approx24\%$ ($\approx7\%$).
Given that this observable is particularly sensitive to soft-radiation effects, our data can be used to further constrain the hadronization models in these generators.

The \DeltaR{} distributions for the WTA\textendash{}Standard and WTA\textendash{}SD cases are broader
($0<\DeltaR\lesssim R/2$) and peak at larger $\DeltaR$, showing substantial deviation between the WTA and Standard/SD jet axes.
Additionally, these distributions show very low sensitivity to the parameters chosen in the Soft Drop grooming procedure.

\begin{figure}
    \centering
    \includegraphics[width=0.6\textwidth]{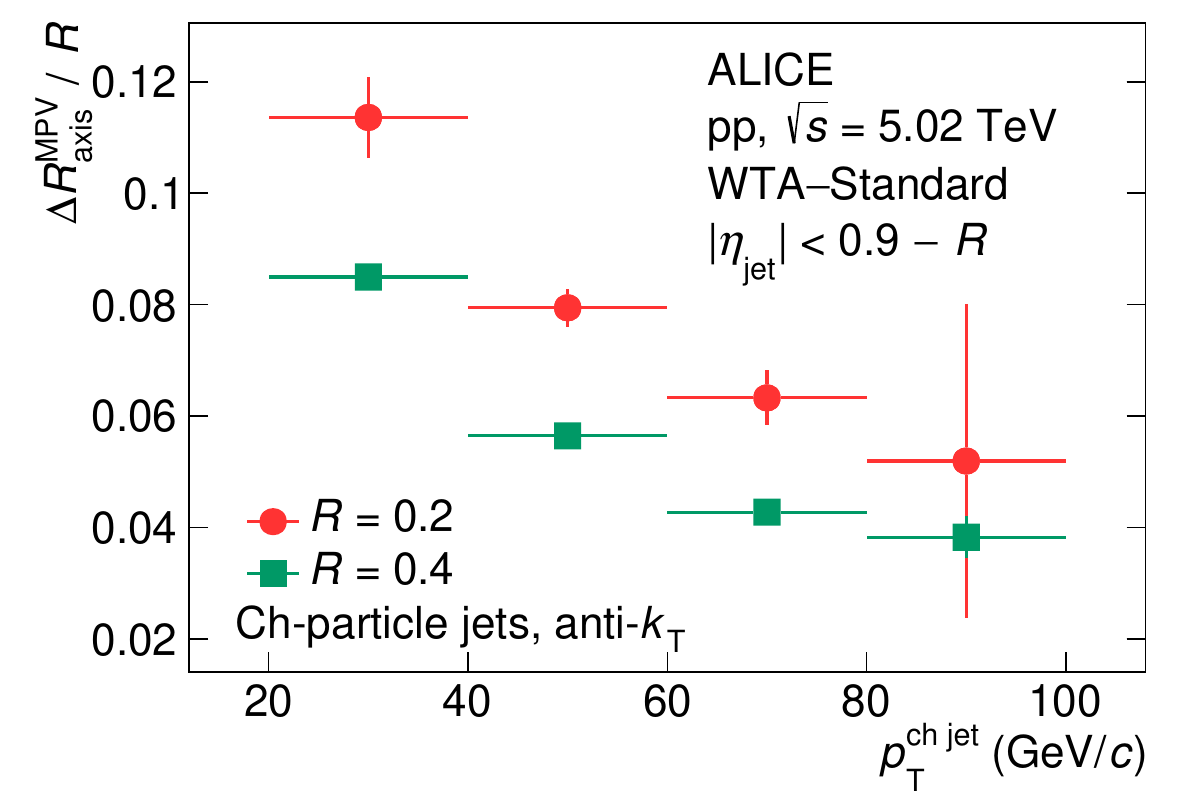}
    \caption{Distribution of the most-probable value of the WTA\textendash{}Standard spectra as a function
    of \pTchjet{} for $R=0.2$ and $R=0.4$.}
    \label{fig:peak}
\end{figure}

The $\DeltaR^{\rm WTA-Standard}$ and $\DeltaR^{\rm WTA-SD}$ predictions from PYTHIA~8 and Herwig~7  generally agree with the data, deviating 
by up to $\approx10\%$ in the $40<\pTchjet<60$ \GeVc{} interval and
up to $\approx40\%$ in several other \pTchjet{} intervals. The largest deviations are at low values of $\DeltaR$, corresponding to the region in which non-perturbative effects are significant. 
It should be noted, though, that the normalization convention implies that conclusions can only be made about the overall shape and not about a discrepancy in a specific $\DeltaR$ interval.

The axis differences with respect to WTA shift to lower \DeltaR{} values at higher \pTchjet, implying that the WTA and Standard or SD axes are more aligned in more energetic jets. This is summarized in Fig.~\ref{fig:peak}, where the most-probable value (MPV) of the $\DeltaR^{\rm WTA-Standard}$ distribution normalized to the jet resolution parameter $R$ is shown as a function of \pTchjet. 
The values in this figure are determined by repeatedly fitting a Landau distribution to the $\DeltaR^{\rm WTA-Standard}$ spectra changing each time the fit range. A histogram is then filled with the peak-position values extracted from each fit. From this histogram, the mean and standard deviation are extracted as the central value and uncertainty, respectively.
The values decrease from $\approx 11 \%$ (8.5\%) at low \pTchjet{} to $\approx 5 \%$ (4\%) at high \pTchjet{} for $R=0.2$ (0.4).

\subsection{Comparison to analytical calculations}
\label{sect:calc}

The \DeltaR{} observable has been calculated in the Soft Collinear Effective Theory (SCET) framework~\cite{Bauer:2001yt}.
In this framework, the jet-production cross section is factorized into parton distribution functions (PDF), ``hard'', and ``jet'' contributions in order to separate physics processes at different scales.
The PDFs encode the probability of finding a parton with a given flavor and momentum fraction from a proton and are non-perturbative objects extracted from global fits to measured data (see, e.g.\ Refs.~\citenum{Lai:2010vv,Gao:2013xoa}).
The SCET calculations presented here use the CT14 NLO PDF set~\cite{Dulat:2015mca}.
The hard contribution encodes the short-distance physics, i.e.\ the hard scattering of one parton from each colliding proton, and the distribution of the resulting partons.
Finally, the jet function describes the evolution of a final-state parton from the hard scattering into a collimated jet.
Large logarithms are resummed to Next-to-Leading Logarithmic (NLL) accuracy for the Standard\textendash{}SD \DeltaR, and NLL$^{\prime}$ for the WTA\textendash{}Standard and WTA\textendash{}SD \DeltaR,
including the contribution from non-global logarithms~\cite{Dasgupta:2001sh}.
NLL$^{\prime}$ refers to the inclusion in the resummation of terms that formally only enter at Next-to-Next-to-Leading Logarithmic (NNLL) accuracy, but their inclusion in the NLL calculation improves the theoretical uncertainty.
Specifically, logarithms of the grooming parameter \zcut{} and the observable \DeltaR{} are resummed. Logarithms of the jet resolution parameter $R$ are also resummed, which makes the calculations valid down to arbitrarily low values of $R$ and allows for tests of such resummations.
The resummed result is presented without matching to the fixed-order calculation in the high-\DeltaR{} region.

Unlike other TMD-sensitive jet substructure observables, such as hadron-in-jet fragmentation~\cite{Bain:2016rrv,Kang:2017glf}, \DeltaR{} does not depend on collinear fragmentation functions. Thus, a significant source of uncertainty is removed from these calculations.

The leading hadronization correction for these observables includes terms of the form $\exp[-g_K(b_\perp;b_{\perp}^{\rm max})]$~\cite{Cal:2019gxa}, where $b_\perp$ is the Fourier conjugate of
$k_{\rm T}$, the projection of the transverse momentum of a jet axis transverse to the other axis in \DeltaR{}.
The function $g_K(b_\perp;b_{\perp}^{\rm max})$ is the non-perturbative component of the Collins\textendash{}Soper\textendash{}Sterman (CSS) evolution kernel or rapidity anomalous dimension.
In the so-called $b_*$ prescription~\cite{Collins:1984kg,Makris:2017arq} it is often parametrized as

\begin{equation}
g_K(b_\perp;b_{\perp}^{\rm max}) = g_2(b_{\perp}^{\rm max})b^2_{\perp}.
\label{eq:css}
\end{equation}
In this work, the universality of the CSS kernel is tested by verifying that the \DeltaR{} observables are well described across two resolution parameters $R$, and several grooming settings and \pTchjet{} intervals
with the same parameters in $g_K(b_\perp;b_{\perp}^{\rm max})$ corresponding to $b_{\perp}^{\rm max} = 1.5 \, ({\rm GeV}/c)^{-1}$ and $g_2(b_{\perp}^{\rm max}) = 0.18$ from a global data analysis to Drell\textendash{}Yan lepton pair and Z$^0$-boson production~\cite{Konychev:2005iy}.

The analytic predictions are provided by the authors of Ref.~\citenum{Cal:2019gxa} for the kinematics of our measurement
at hadron level (i.e.\ after hadronization, in contrast to parton level) for full jets (i.e.\ including both charged and neutral hadrons), and without including multi-parton interaction (MPI) effects.
To do a comparison to the measured distributions, the analytic predictions are corrected using data from Monte Carlo event generators. Final-state particles from pp events are clustered into full and charged-particle jets
following the same procedure as in the data analysis (i.e.\ using the anti-$k_{\rm T}$ algorithm with the $E$ recombination scheme for a given $R$),
and the resulting jets are required to satisfy $|\eta_{\rm jet}|<0.9-R$ and $\pTjet>5$ \GeVc. The full and charged-particle jets are then geometrically matched following the matching procedure of the RM from the data analysis.
This sample is used to construct a 4D response matrix that maps the \pTjet{} and \DeltaR{} dependence from full- to charged-particle-jet levels for each observable.

The analytic predictions are provided as normalized densities $(1/\sigma_{\rm jet})({\rm d}\sigma_{\rm jet}/{\rm d}\DeltaR)$ for the \DeltaR{} observable in different \pTjet{} intervals (of 5 \GeVc{} width) for the nominal case, as well as for the results obtained by systematically varying the scales that appear in the calculation to account for theoretical uncertainties~\cite{Cal:2019gxa}. The first step in the correction corresponds to multiplying the spectrum by the average value of the inclusive cross section in the considered \pTjet{} interval to obtain the distribution for ${\rm d}\sigma_{\rm jet}/{\rm d}\DeltaR$. The cross section used was calculated at Next-to-Leading Order (NLO) with resummation of logarithms of the jet radius at NLL~\cite{Kang:2016mcy}. The scaled distributions are stored in 2D histograms in $p_{\rm T}$ and \DeltaR{} and multiplied by the 4D response matrices described in the previous paragraph to obtain the analytic predictions for charged-particle jets.
Subsequently, the resulting 2D histograms are corrected to account for MPI effects by multiplying bin-by-bin by the ratio of the simulated distribution without and with MPI effects included. The result is projected onto the observable axis for a range in \pTchjet{} equal to that of the measured distribution.
The final theory prediction corresponds to the curve obtained from the nominal calculation.
Additionally, the other results obtained from the systematic scale variations are equally corrected, and the envelope of all resulting distributions is taken as the theoretical uncertainty on the calculations.

The use of a Monte Carlo event generator in these corrections introduces a model dependence, the significance of which is explored by applying the correction procedure with two different generators: PYTHIA~8 with the Monash 2013 tune~\cite{Skands:2014pea} and Herwig~7 with the default tune~\cite{Gieseke:2012ft}.

\begin{figure}
    \centering
    \includegraphics[width=\textwidth]{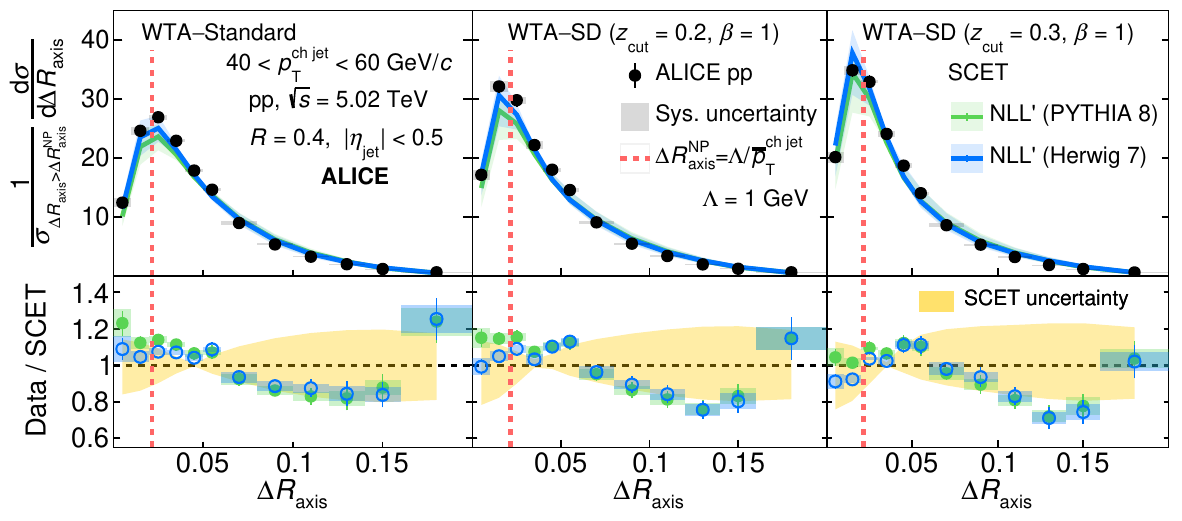}
    \includegraphics[width=\textwidth]{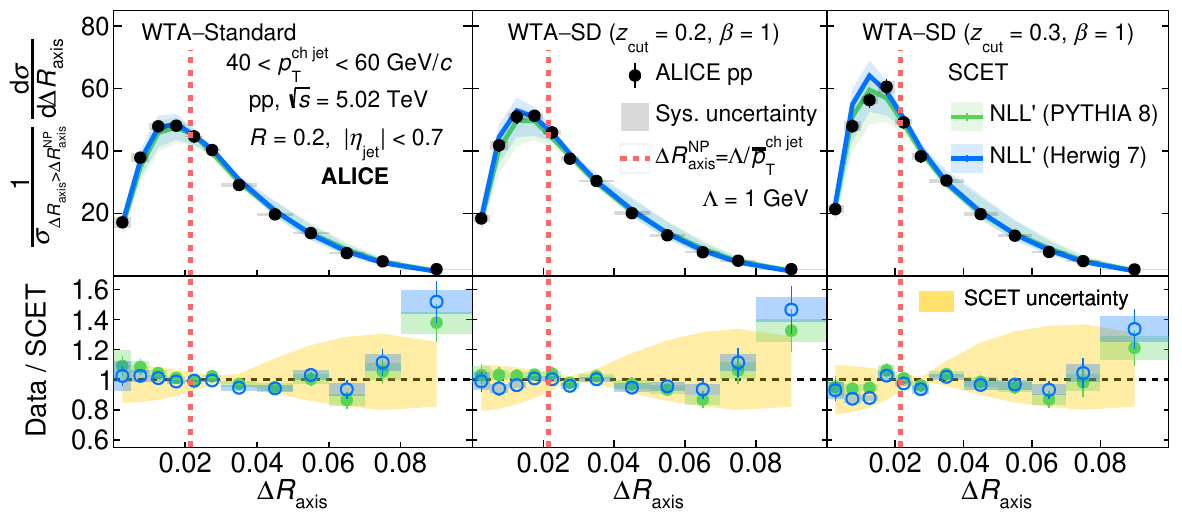}
    \caption{Comparison between measured distributions and analytic predictions for the difference between the WTA axis and the Standard (left), SD ($\zcut=0.2,\beta=1$) (center), and SD ($\zcut=0.3,\beta=1$) (right) axes for jets of $R=0.4$ (top) and 0.2 (bottom) in the transverse momentum range $40 < \pTchjet < 60$ \GeVc. The black markers correspond to the distributions determined from measured data. The vertical error bars correspond to the statistical uncertainties, and the rectangles correspond to the total systematic uncertainties. The colored curves correspond to the SCET-based analytic predictions corrected for charge and MPI effects using two event generators (PYTHIA~8 and Herwig~7). The vertical dashed line defines the approximate boundary between the non-perturbative and perturbative regions. Both the measured and analytic predictions are normalized so that the integral $\int_{\DeltaR^{\rm NP}}^{R/2}{\rm d}\DeltaR ({\rm d}\sigma/{\rm d}\DeltaR)=1$. The bottom panels show the data/SCET ratios. The colored rectangles correspond to the systematic uncertainty of the measured distribution. The size of the theoretical uncertainty in the analytic predictions is shown as a yellow band.}
    \label{fig:comp_scet}
\end{figure}

Figure~\ref{fig:comp_scet} shows a subset of the comparison between the measured distributions and the corrected analytic predictions.
Equivalent comparisons for the rest of all available predictions are presented in Appendix~\ref{app:B}.
The black markers correspond to the distributions determined from measured data.
These distributions are identical to those in Fig.~\ref{fig:data_generators_WTA} up to a normalization factor defined below.
The vertical error bars correspond to the statistical uncertainties, and the rectangles correspond to the total systematic uncertainties. The colored curves correspond to the SCET-based analytic predictions corrected for charge and MPI effects using two event generators (PYTHIA~8 and Herwig~7).

Differences between the SCET predictions corrected with either Monte Carlo event generator are very small. This is due to the fact that, since the input calculations are provided at hadron level, the most significant correction is done to the \pT{} scale of the jet, and this correction is well modeled by both generators. Thus, the resulting distributions are not significantly model dependent.

The analytic calculations are only expected to describe the measured distributions in the perturbative region. The predictions presented here become non-perturbative approximately at $\DeltaR \lesssim \DeltaR^{\rm NP} = \Lambda/\pTchjet$~\cite{Cal:2019gxa}, where $\Lambda$ corresponds to the scale at which the strong coupling constant becomes non-perturbative. 
The red vertical line in Fig.~\ref{fig:comp_scet} corresponds to this value calculated using $\Lambda = 1$ GeV and $\pTchjet=\overline{p}_{\rm T}^{\rm ch \, jet}$, the average jet-transverse-momentum value in the interval. The $\overline{p}_{\rm T}^{\rm ch \, jet}$ value for a given \pTchjet{} interval was determined by fitting the measured spectra with a power law and subsequently analytically calculating the mean value in the interval.
The comparison is done by normalizing both the measured distributions and analytic predictions so that $\int_{\DeltaR^{\rm NP}}^{R/2}{\rm d}\DeltaR ({\rm d}\sigma/{\rm d}\DeltaR)=1$. The $\DeltaR$ interval that contains the value $\DeltaR^{\rm NP}$ is defined to belong to the non-perturbative region and excluded from the integral.
In reality, there is no sharp boundary between the perturbative and non-perturbative regimes, but a continuous transition. This vertical line is therefore an indicative value defined for the sake of the comparison.

The Standard\textendash{}SD variable is also IRC safe and therefore calculable in the SCET framework. However, these calculations are computationally expensive and the results are not available at the time of this article. These distributions are particularly sensitive to soft effects, and thus also to the hadronization corrections.

\subsection{Discussion of analytic calculations}

The analytic predictions describe the measured distributions in the perturbative regime within uncertainties for all variations of the observable (i.e.\ jet-resolution parameter, grooming setting and \pTchjet) considered in the analysis. 
The agreement is excellent (largest deviations observed are within 10 to 20\% in the lower \pT{} bins, where the smaller systematic uncertainties allow us to draw more precise conclusions) given the low \pTchjet{} values from this measurement.
Additionally, even though the distributions are normalized to their integral in the perturbative region, significant agreement is also found in the non-perturbative region, where the calculations are most sensitive to the non-perturbative correction. This agreement persists independently of jet-resolution parameter, grooming setting, and \pTchjet. There are larger shape differences between the data and analytic predictions for larger $R$.

The agreement at lower \DeltaR{} values is increasingly 
due to the inclusion of the non-perturbative correction from Eq.~\ref{eq:css}.
The measured distributions are compatible with the universal behavior of the non-perturbative component of the CSS evolution kernel to the extent this hadronization correction is well modeled. This is the first time this has been verified in jet substructure observables. The measurements presented in this article should be included in future global fits to further increase the precision of the experimental extractions of the non-perturbative component of the CSS kernel.

The theoretical uncertainties in the analytic predictions are significant, and reducing them can lead to a higher-precision test. Specifically, as shown in the ratio plots from Fig.~\ref{fig:comp_scet}, while the measured distributions and the calculations agree within uncertainties, there are systematic shape differences between them. For instance, there is a sudden rise in the interval at the highest $\DeltaR$ which can be related to the lack of matching to a fixed-order calculation or power corrections in the observable in the SCET prediction. However, because the distributions are self-normalized, the disagreement cannot be pinned to a specific \DeltaR{} value.
 
\section{Conclusions}
\label{sec:conclusions}

The first measurement of the angle
between the Standard, Soft Drop groomed, and Winner-Take-All charged-jet axes is reported.
The measurement was carried out
in pp collisions at $\sqrt{s} = 5.02$ TeV with the ALICE detector
for jets of $R=0.2$ and 0.4. This analysis focused on jets below 100 \GeVc, where non-perturbative effects are more important than in highly energetic jets.
Jet grooming does not substantially change the jet axis direction. The WTA and Standard jet axes differ substantially, though, showing that the Standard jet axis does not generally point in the direction of the highest-momentum jet fragment.
However, the WTA and Standard or SD axes become more aligned in more energetic jets.
The motivation to use the WTA axis arises from its insensitivity to soft radiation; it is amenable to perturbative calculations and is expected to be less modified in the medium created in ultrarelativistic nucleus\textendash{}nucleus collisions.

The distributions from PYTHIA~8 and Herwig~7 show overall good agreement with the data; both groomed and ungroomed jet axis differences with WTA are generally described within the uncertainties. The Standard\textendash{}SD jet axis difference is better described by Herwig~7 than by PYTHIA~8, suggesting that Herwig better reproduces the soft splittings which are groomed away by Soft Drop. 
There is also
good agreement between the analytic predictions and the measured spectra, even in the region of \DeltaR{} where non-perturbative physics is important. 
Within the precision of the measurement and analytic calculations, our result is compatible with the universality of the non-perturbative component of the Collins\textendash{}Soper\textendash{}Sterman (CSS) TMD evolution kernel. This study constitutes the first experimental verification of such compatibility in the context of jet-substructure measurements.

These measurements shed light on the interplay between perturbative and non-perturbative effects. 
The majority of jets produced in \pp{} collisions at this center-of-mass energy are initiated by gluons. It will be illuminating to compare these distributions to those from quark-initiated jets. This can be done by measuring \DeltaR{} in, for example, heavy-flavor jets or photon\textendash{}jet coincidences.

The angle between different jet axes may also reflect medium  modification of jet angular substructure in heavy-ion collisions.
The results presented here will provide a valuable baseline for \PbPb{} studies.

\clearpage

\newenvironment{acknowledgement}{\relax}{\relax}
\begin{acknowledgement}
\section*{Acknowledgements}
The ALICE Collaboration gratefully acknowledges Felix Ringer for providing theoretical predictions.

The ALICE Collaboration would like to thank all its engineers and technicians for their invaluable contributions to the construction of the experiment and the CERN accelerator teams for the outstanding performance of the LHC complex.
The ALICE Collaboration gratefully acknowledges the resources and support provided by all Grid centres and the Worldwide LHC Computing Grid (WLCG) collaboration.
The ALICE Collaboration acknowledges the following funding agencies for their support in building and running the ALICE detector:
A. I. Alikhanyan National Science Laboratory (Yerevan Physics Institute) Foundation (ANSL), State Committee of Science and World Federation of Scientists (WFS), Armenia;
Austrian Academy of Sciences, Austrian Science Fund (FWF): [M 2467-N36] and Nationalstiftung f\"{u}r Forschung, Technologie und Entwicklung, Austria;
Ministry of Communications and High Technologies, National Nuclear Research Center, Azerbaijan;
Conselho Nacional de Desenvolvimento Cient\'{\i}fico e Tecnol\'{o}gico (CNPq), Financiadora de Estudos e Projetos (Finep), Funda\c{c}\~{a}o de Amparo \`{a} Pesquisa do Estado de S\~{a}o Paulo (FAPESP) and Universidade Federal do Rio Grande do Sul (UFRGS), Brazil;
Bulgarian Ministry of Education and Science, within the National Roadmap for Research Infrastructures 2020-2027 (object CERN), Bulgaria;
Ministry of Education of China (MOEC) , Ministry of Science \& Technology of China (MSTC) and National Natural Science Foundation of China (NSFC), China;
Ministry of Science and Education and Croatian Science Foundation, Croatia;
Centro de Aplicaciones Tecnol\'{o}gicas y Desarrollo Nuclear (CEADEN), Cubaenerg\'{\i}a, Cuba;
Ministry of Education, Youth and Sports of the Czech Republic, Czech Republic;
The Danish Council for Independent Research | Natural Sciences, the VILLUM FONDEN and Danish National Research Foundation (DNRF), Denmark;
Helsinki Institute of Physics (HIP), Finland;
Commissariat \`{a} l'Energie Atomique (CEA) and Institut National de Physique Nucl\'{e}aire et de Physique des Particules (IN2P3) and Centre National de la Recherche Scientifique (CNRS), France;
Bundesministerium f\"{u}r Bildung und Forschung (BMBF) and GSI Helmholtzzentrum f\"{u}r Schwerionenforschung GmbH, Germany;
General Secretariat for Research and Technology, Ministry of Education, Research and Religions, Greece;
National Research, Development and Innovation Office, Hungary;
Department of Atomic Energy Government of India (DAE), Department of Science and Technology, Government of India (DST), University Grants Commission, Government of India (UGC) and Council of Scientific and Industrial Research (CSIR), India;
National Research and Innovation Agency - BRIN, Indonesia;
Istituto Nazionale di Fisica Nucleare (INFN), Italy;
Japanese Ministry of Education, Culture, Sports, Science and Technology (MEXT) and Japan Society for the Promotion of Science (JSPS) KAKENHI, Japan;
Consejo Nacional de Ciencia (CONACYT) y Tecnolog\'{i}a, through Fondo de Cooperaci\'{o}n Internacional en Ciencia y Tecnolog\'{i}a (FONCICYT) and Direcci\'{o}n General de Asuntos del Personal Academico (DGAPA), Mexico;
Nederlandse Organisatie voor Wetenschappelijk Onderzoek (NWO), Netherlands;
The Research Council of Norway, Norway;
Commission on Science and Technology for Sustainable Development in the South (COMSATS), Pakistan;
Pontificia Universidad Cat\'{o}lica del Per\'{u}, Peru;
Ministry of Education and Science, National Science Centre and WUT ID-UB, Poland;
Korea Institute of Science and Technology Information and National Research Foundation of Korea (NRF), Republic of Korea;
Ministry of Education and Scientific Research, Institute of Atomic Physics, Ministry of Research and Innovation and Institute of Atomic Physics and University Politehnica of Bucharest, Romania;
Ministry of Education, Science, Research and Sport of the Slovak Republic, Slovakia;
National Research Foundation of South Africa, South Africa;
Swedish Research Council (VR) and Knut \& Alice Wallenberg Foundation (KAW), Sweden;
European Organization for Nuclear Research, Switzerland;
Suranaree University of Technology (SUT), National Science and Technology Development Agency (NSTDA), Thailand Science Research and Innovation (TSRI) and National Science, Research and Innovation Fund (NSRF), Thailand;
Turkish Energy, Nuclear and Mineral Research Agency (TENMAK), Turkey;
National Academy of  Sciences of Ukraine, Ukraine;
Science and Technology Facilities Council (STFC), United Kingdom;
National Science Foundation of the United States of America (NSF) and United States Department of Energy, Office of Nuclear Physics (DOE NP), United States of America.
In addition, individual groups or members have received support from:
Marie Sk\l{}odowska Curie, European Research Council, Strong 2020 - Horizon 2020 (grant nos. 950692, 824093, 896850), European Union;
Academy of Finland (Center of Excellence in Quark Matter) (grant nos. 346327, 346328), Finland;
Programa de Apoyos para la Superaci\'{o}n del Personal Acad\'{e}mico, UNAM, Mexico.

\end{acknowledgement}

\bibliographystyle{utphys}   
\bibliography{main}

\newpage
\appendix
\section{Comparison to Monte Carlo event generators} \label{app:A}

\begin{figure}[!htb]
    \centering
    \includegraphics[width=0.48\textwidth]{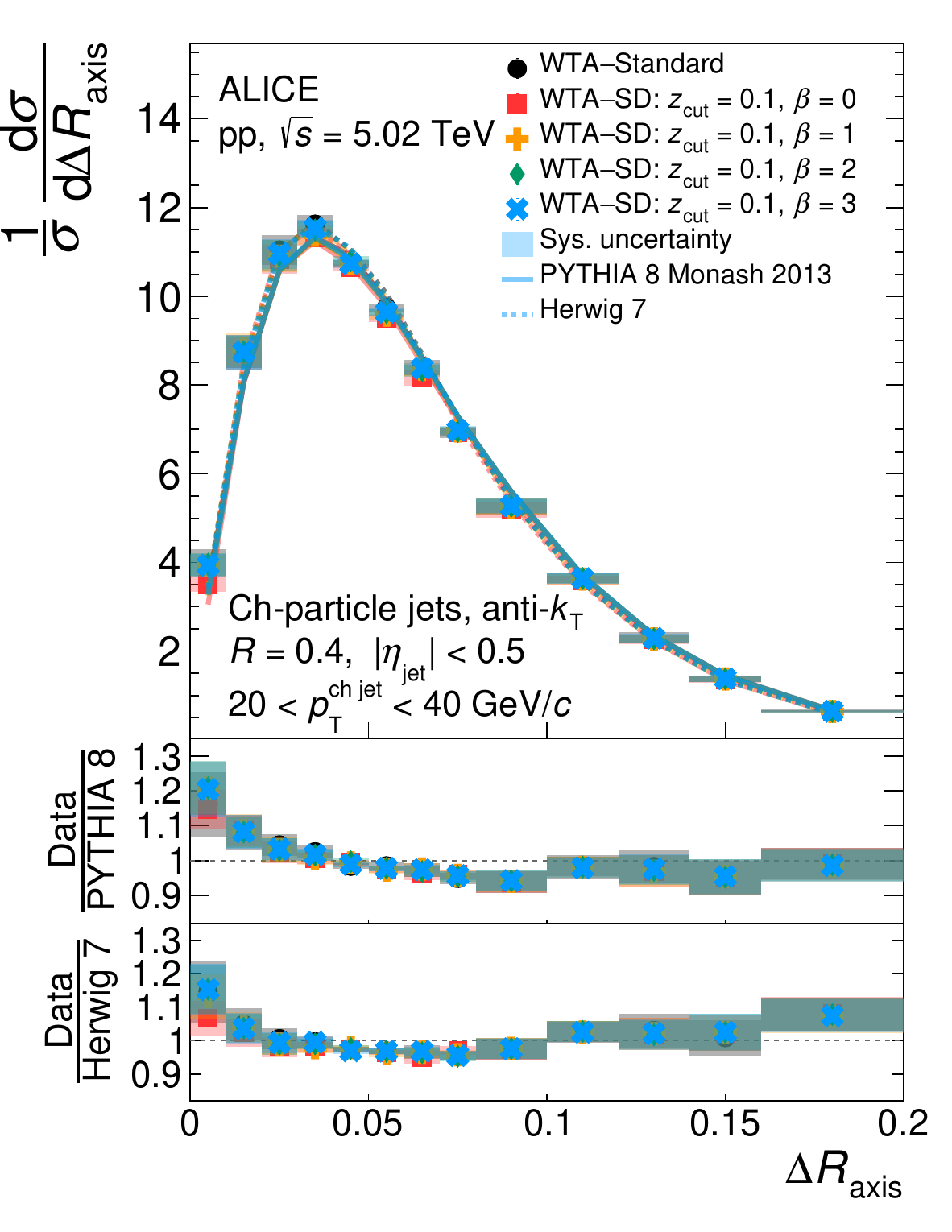}
    \includegraphics[width=0.48\textwidth]{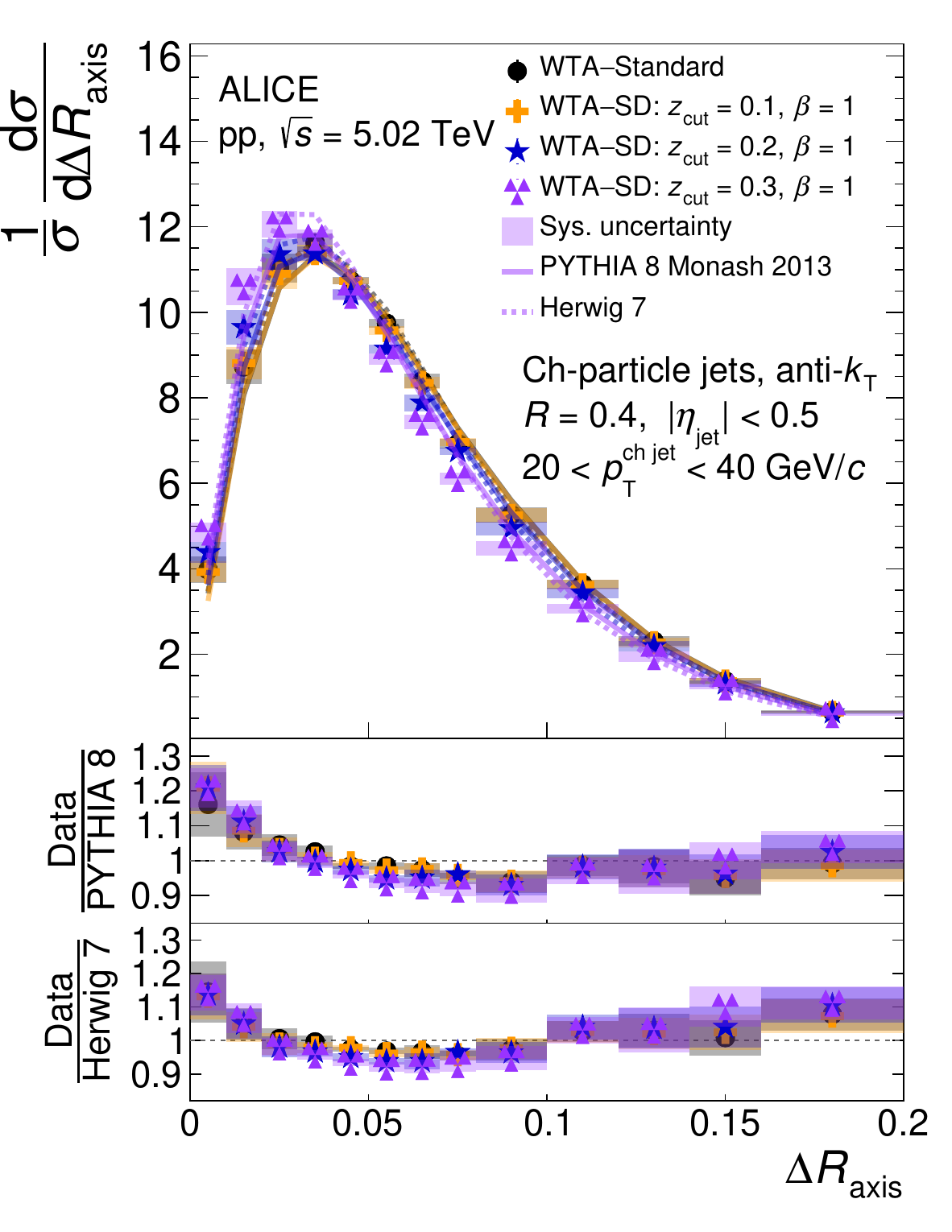}
    \caption{Comparison between \DeltaR{} WTA\textendash{}Standard and WTA\textendash{}SD measured distributions and Monte Carlo event generators for jets of $R=0.4$ in $20<\pTjet<40$ \GeVc. Left: distributions with $\zcut=0.1$ and varying $\beta$. Right: distributions with $\beta=1$ and varying $\zcut$.}
    \label{fig:data_generators_WTA_20_40_R04}
\end{figure}

\begin{figure}[!htb]
    \centering
    \includegraphics[width=0.48\textwidth]{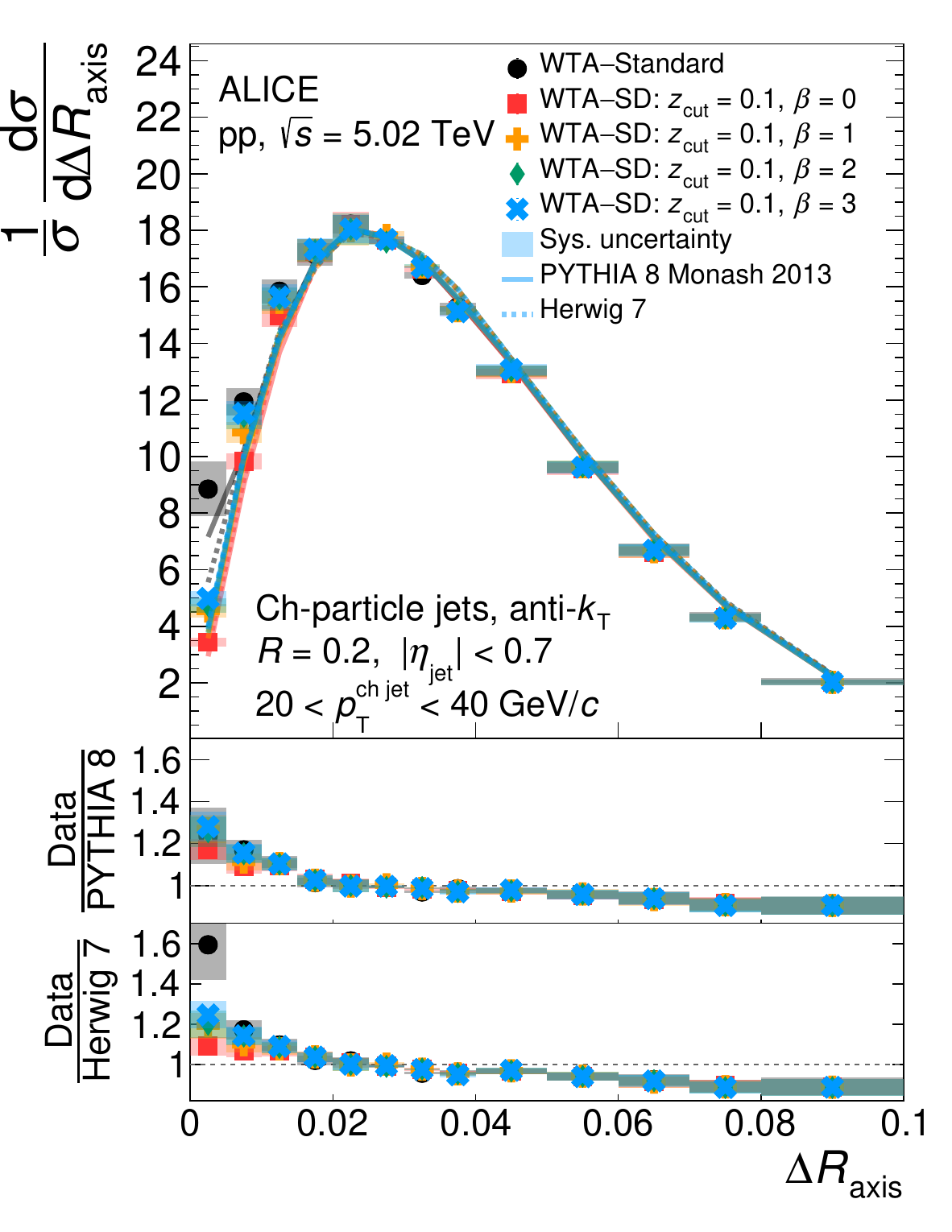}
    \includegraphics[width=0.48\textwidth]{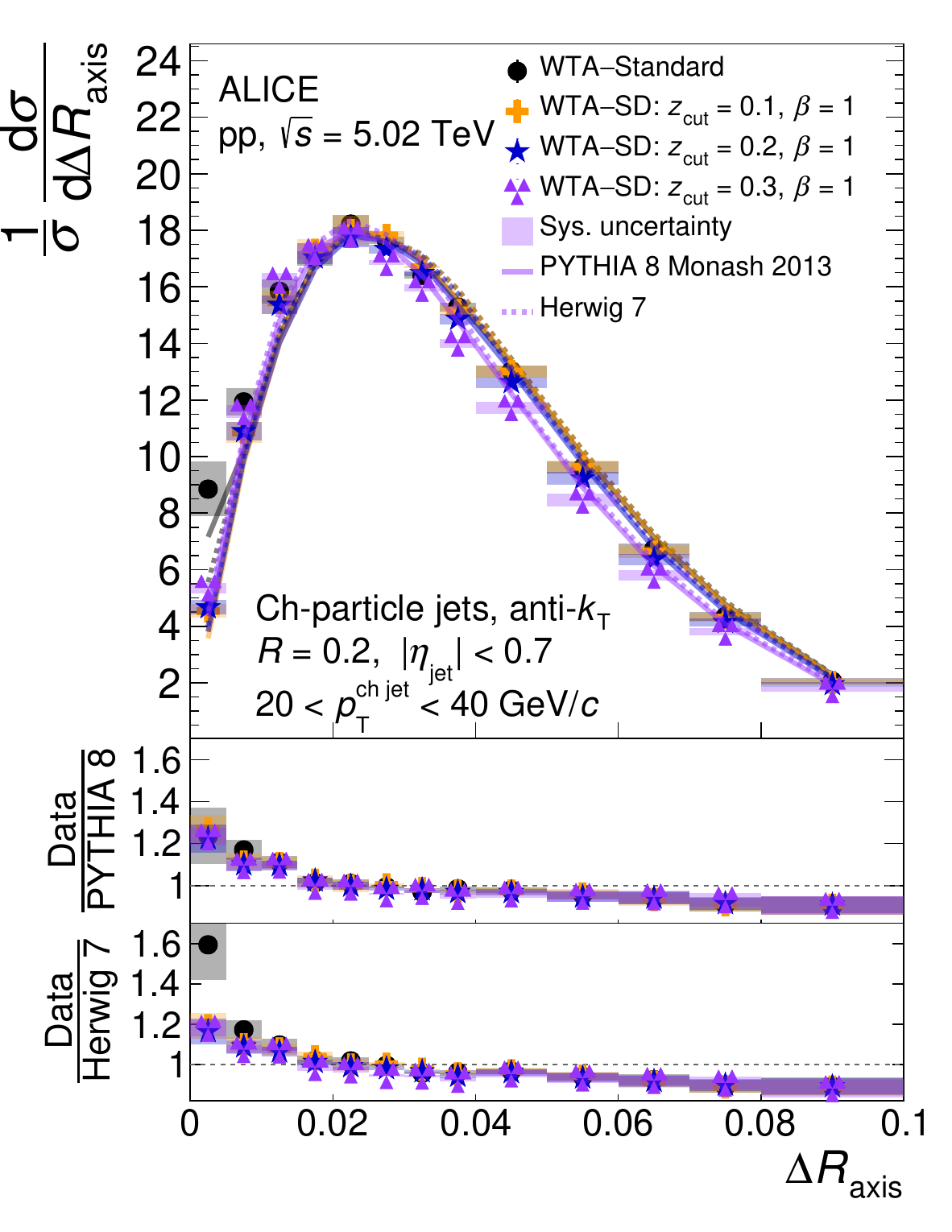}
    \caption{Same as Figure~\ref{fig:data_generators_WTA_20_40_R04} for $R=0.2$.}
    \label{fig:data_generators_WTA_20_40_R02}
\end{figure}

\begin{figure}[!htb]
    \centering
    \includegraphics[width=0.48\textwidth]{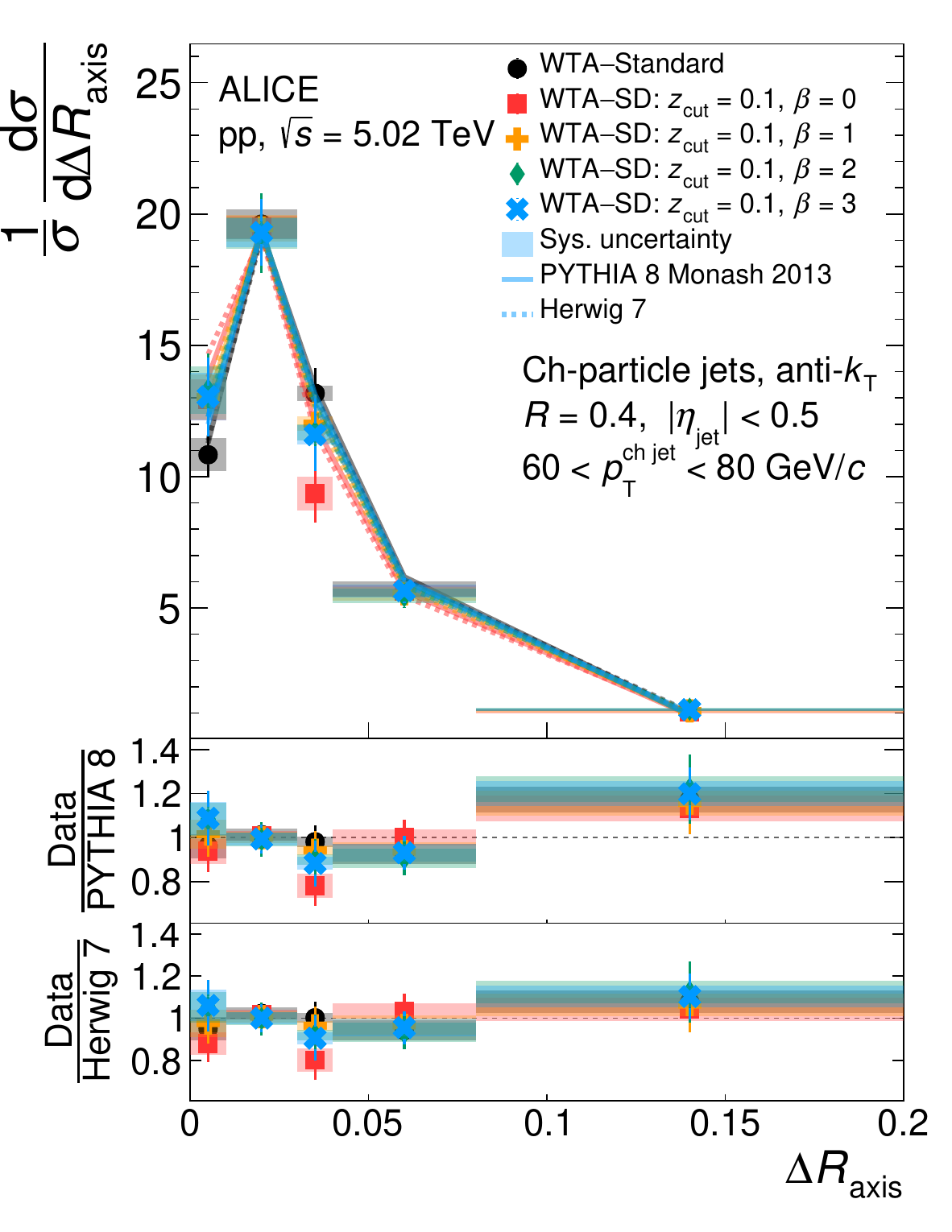}
    \includegraphics[width=0.48\textwidth]{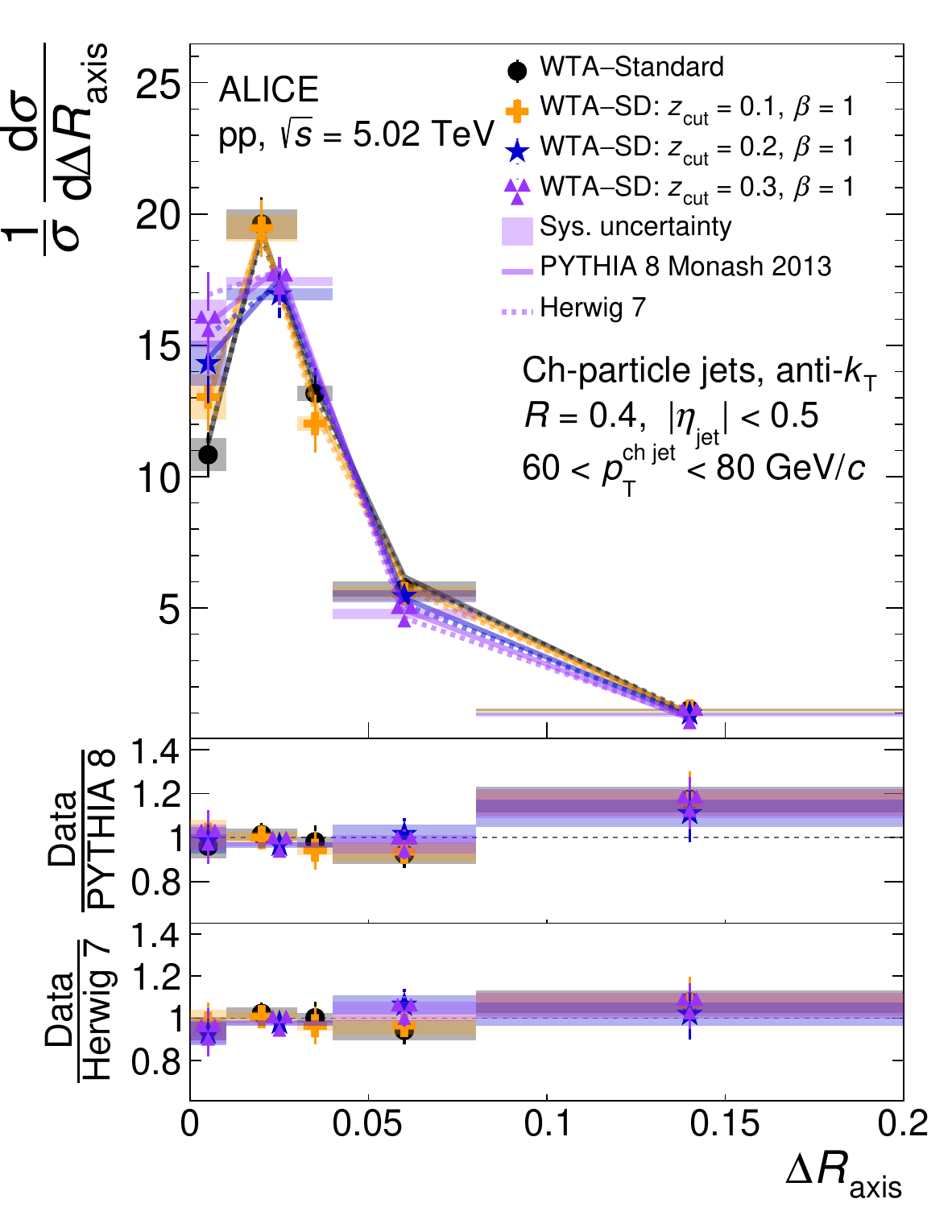}
    \caption{Same as Figure~\ref{fig:data_generators_WTA_20_40_R04} for $60<\pTjet<80$ \GeVc.}
    \label{fig:data_generators_WTA_60_80_R04}
\end{figure}

\begin{figure}[!htb]
    \centering
    \includegraphics[width=0.48\textwidth]{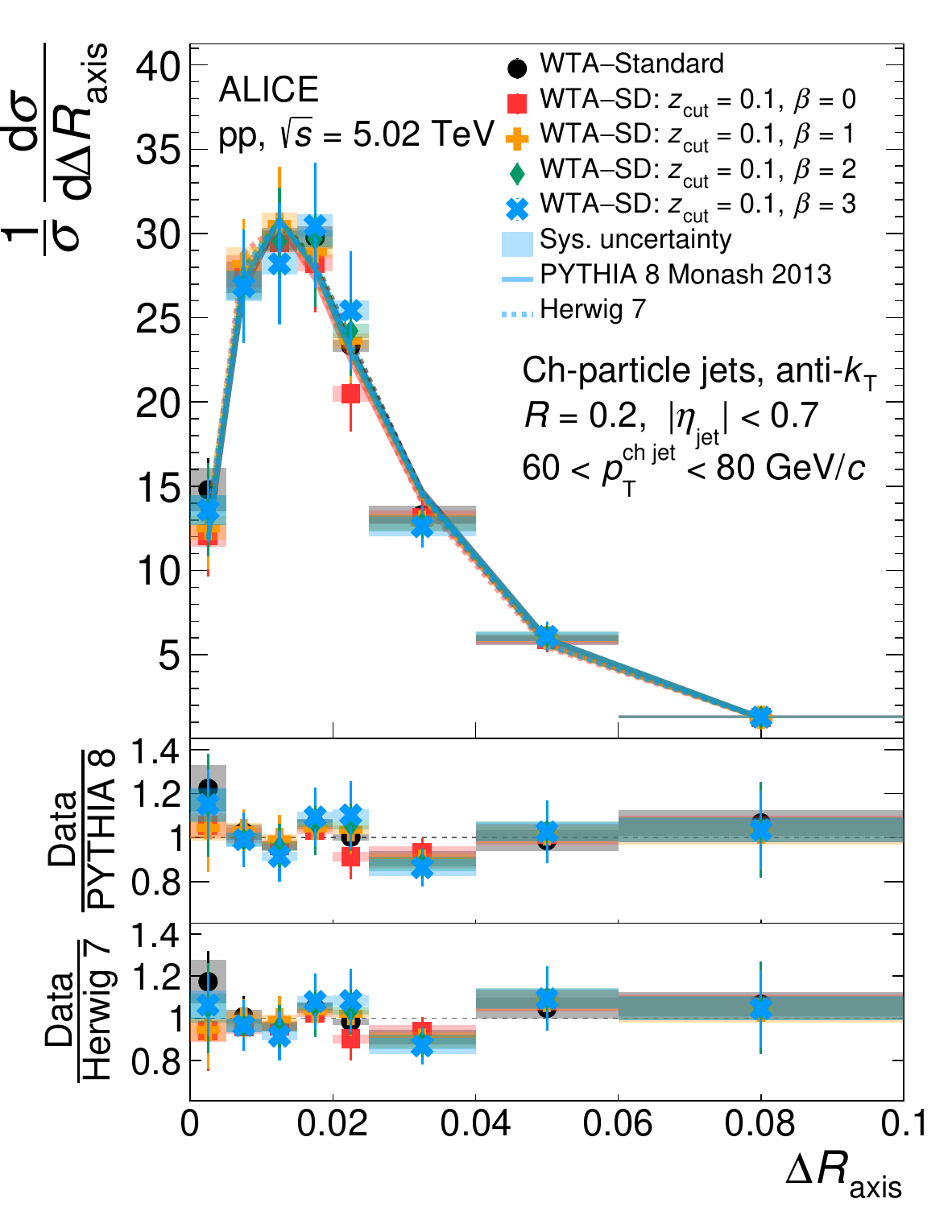}
    \includegraphics[width=0.48\textwidth]{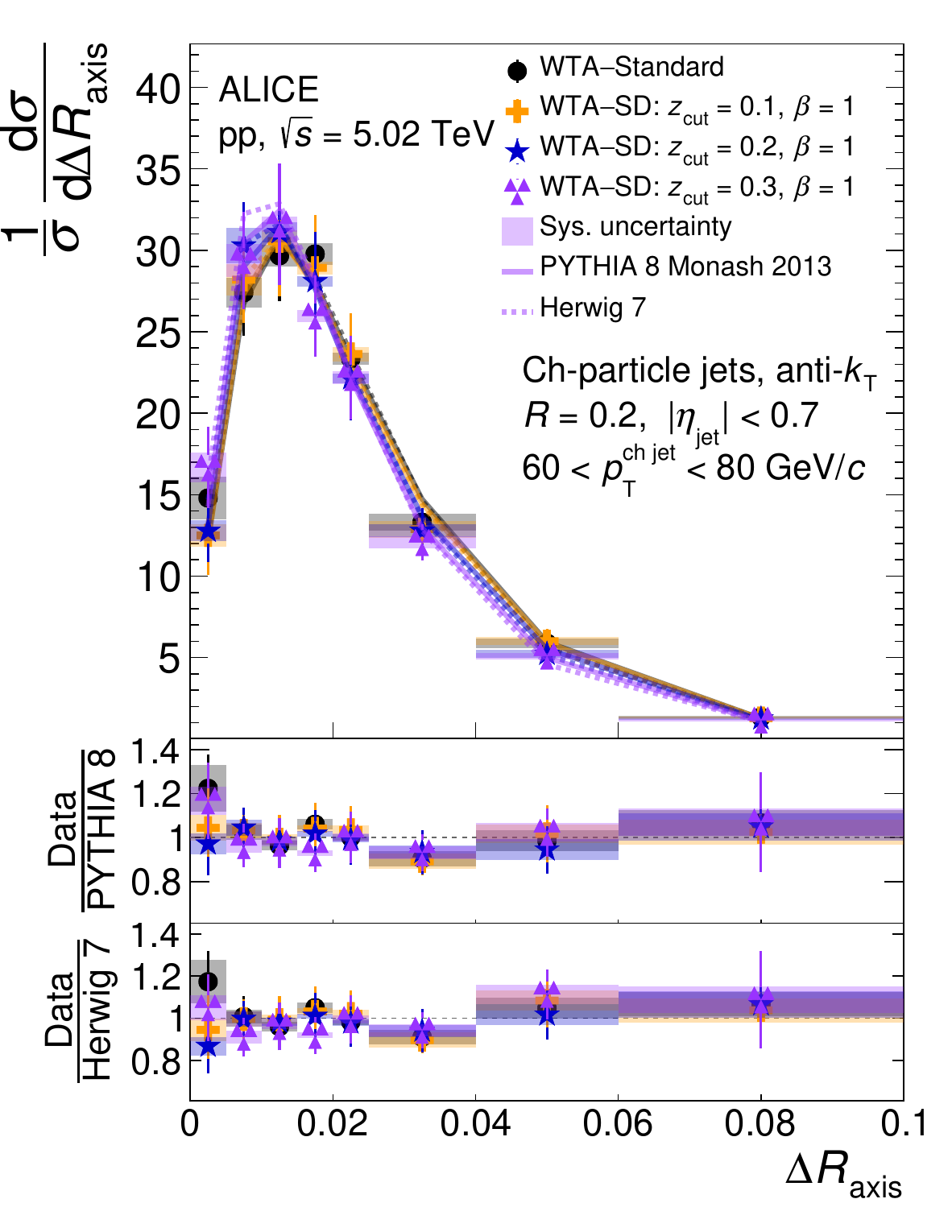}
    \caption{Same as Figure~\ref{fig:data_generators_WTA_20_40_R04} for $R=0.2$ for $60<\pTjet<80$ \GeVc.}
    \label{fig:data_generators_WTA_60_80_R02}
\end{figure}

\begin{figure}[!htb]
    \centering
    \includegraphics[width=0.48\textwidth]{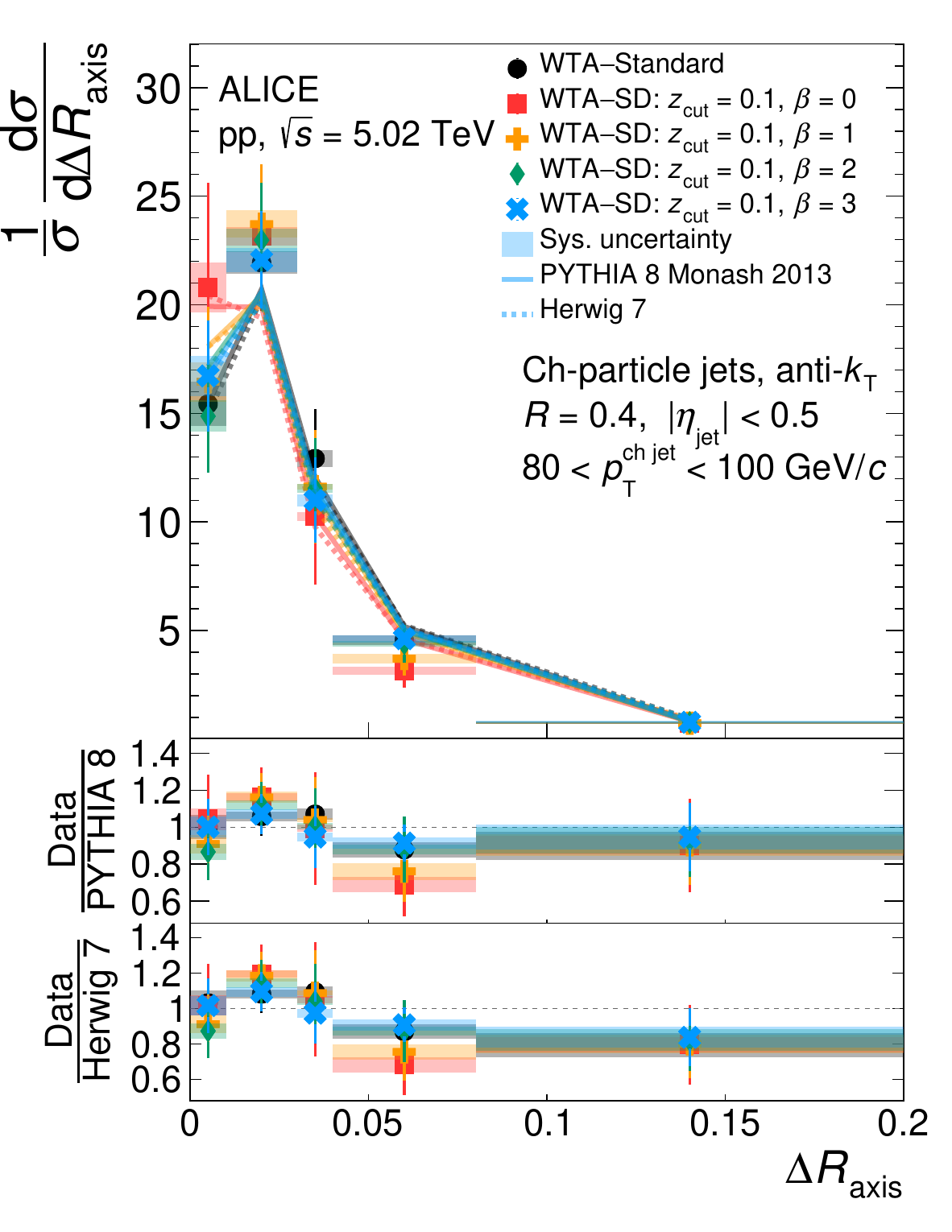}
    \includegraphics[width=0.48\textwidth]{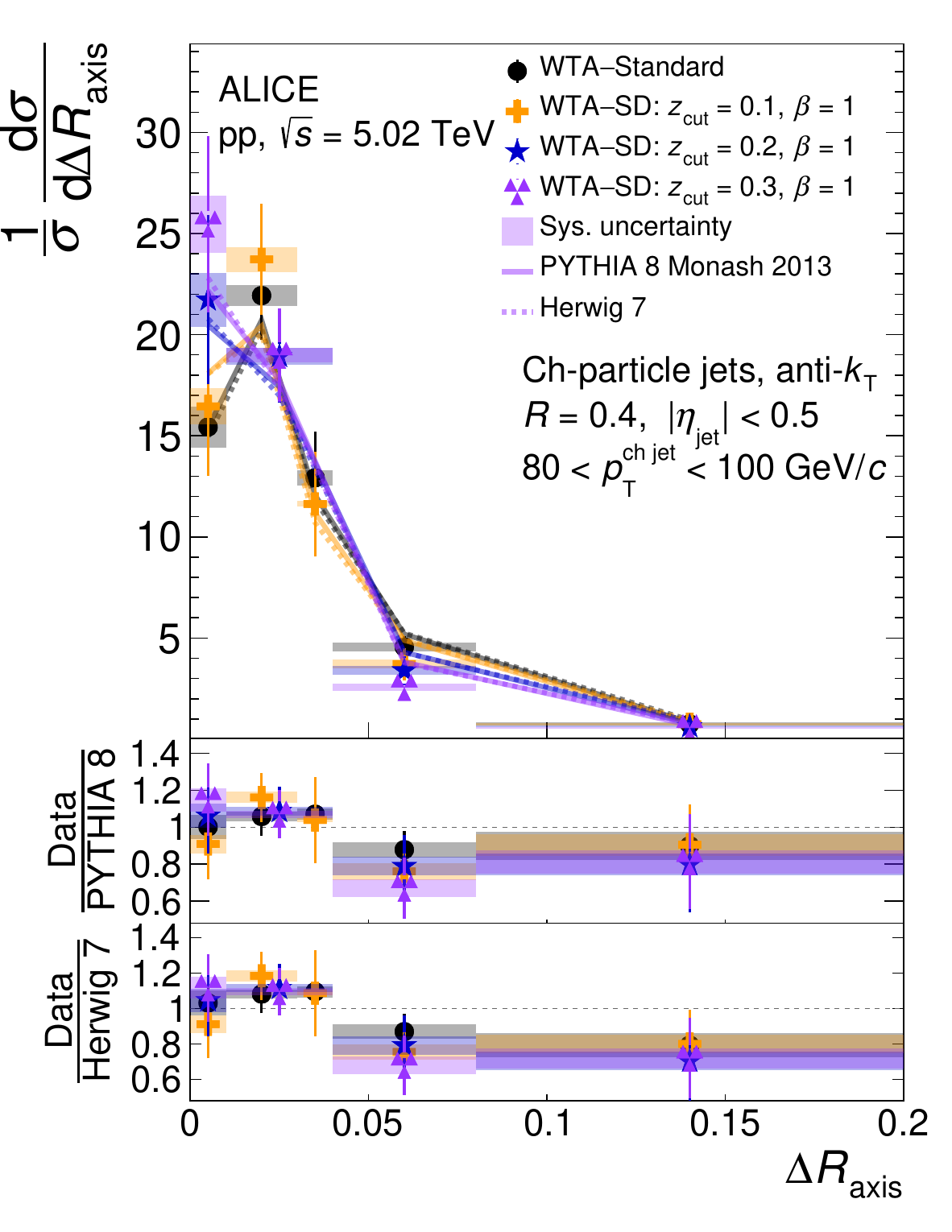}
    \caption{Same as Figure~\ref{fig:data_generators_WTA_20_40_R04} for $80<\pTjet<100$ \GeVc.}
    \label{fig:data_generators_WTA_80_100_R04}
\end{figure}

\begin{figure}[!htb]
    \centering
    \includegraphics[width=0.48\textwidth]{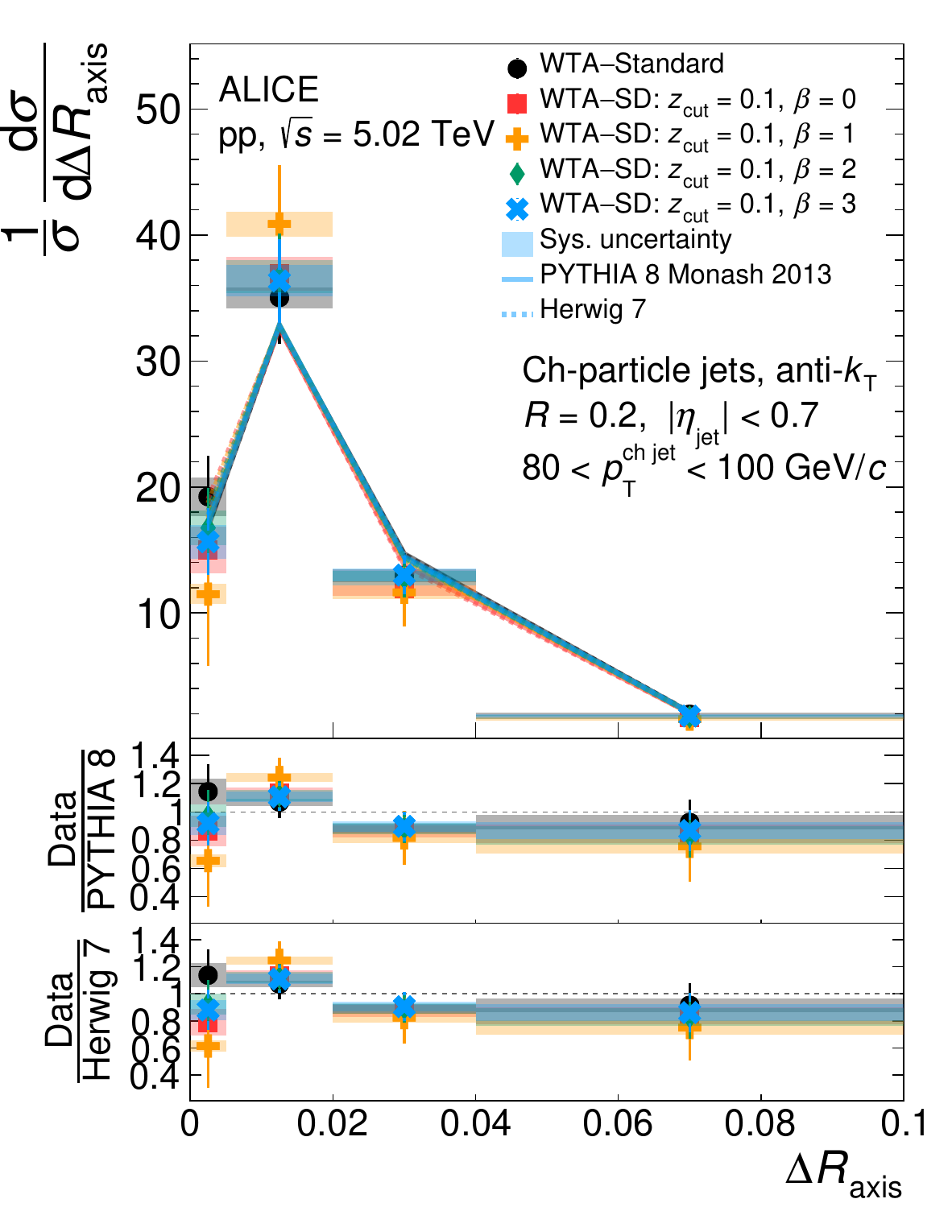}
    \includegraphics[width=0.48\textwidth]{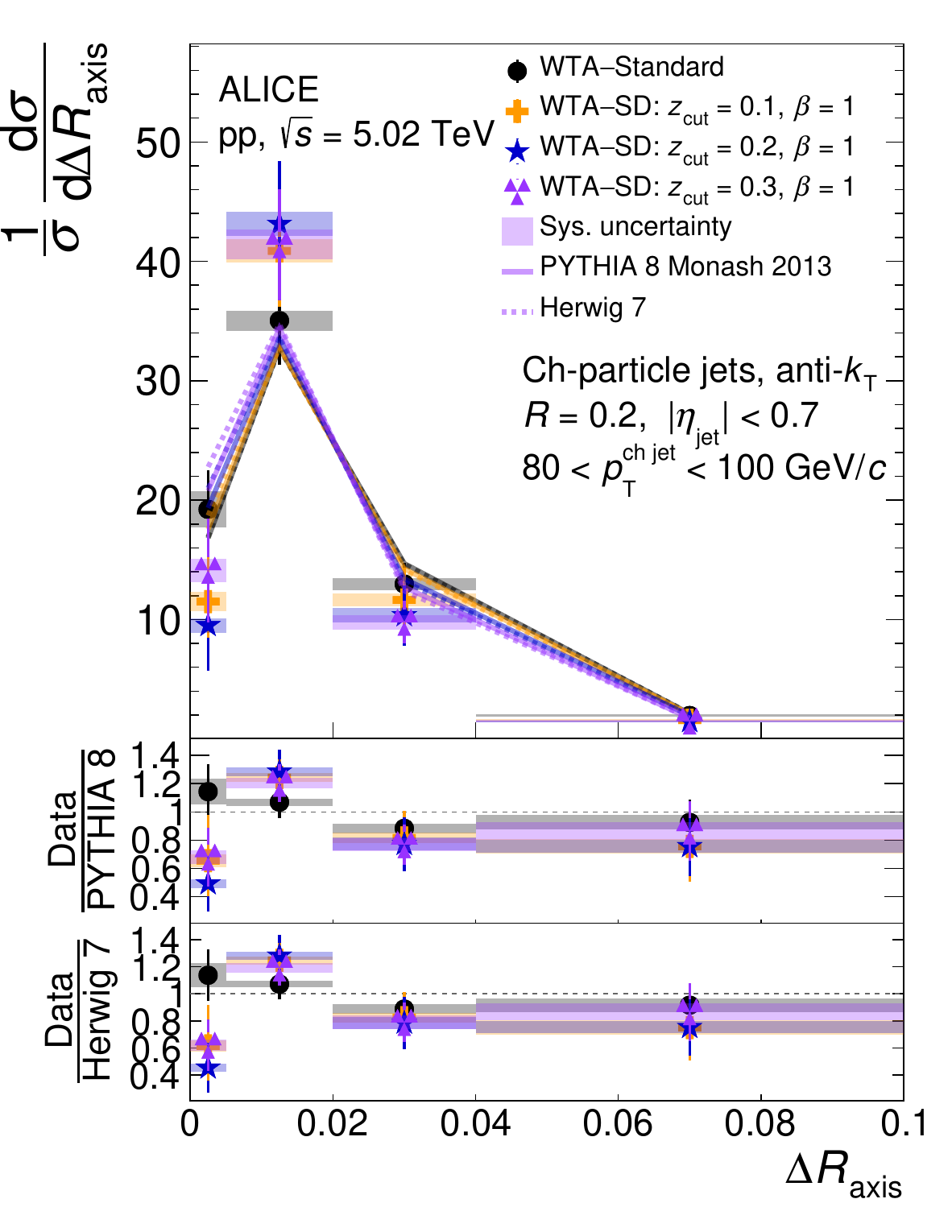}
    \caption{Same as Figure~\ref{fig:data_generators_WTA_20_40_R04} for $R=0.2$ for $80<\pTjet<100$ \GeVc.}
    \label{fig:data_generators_WTA_80_100_R02}
\end{figure}


\begin{figure}
    \centering
    \includegraphics[width=0.48\textwidth]{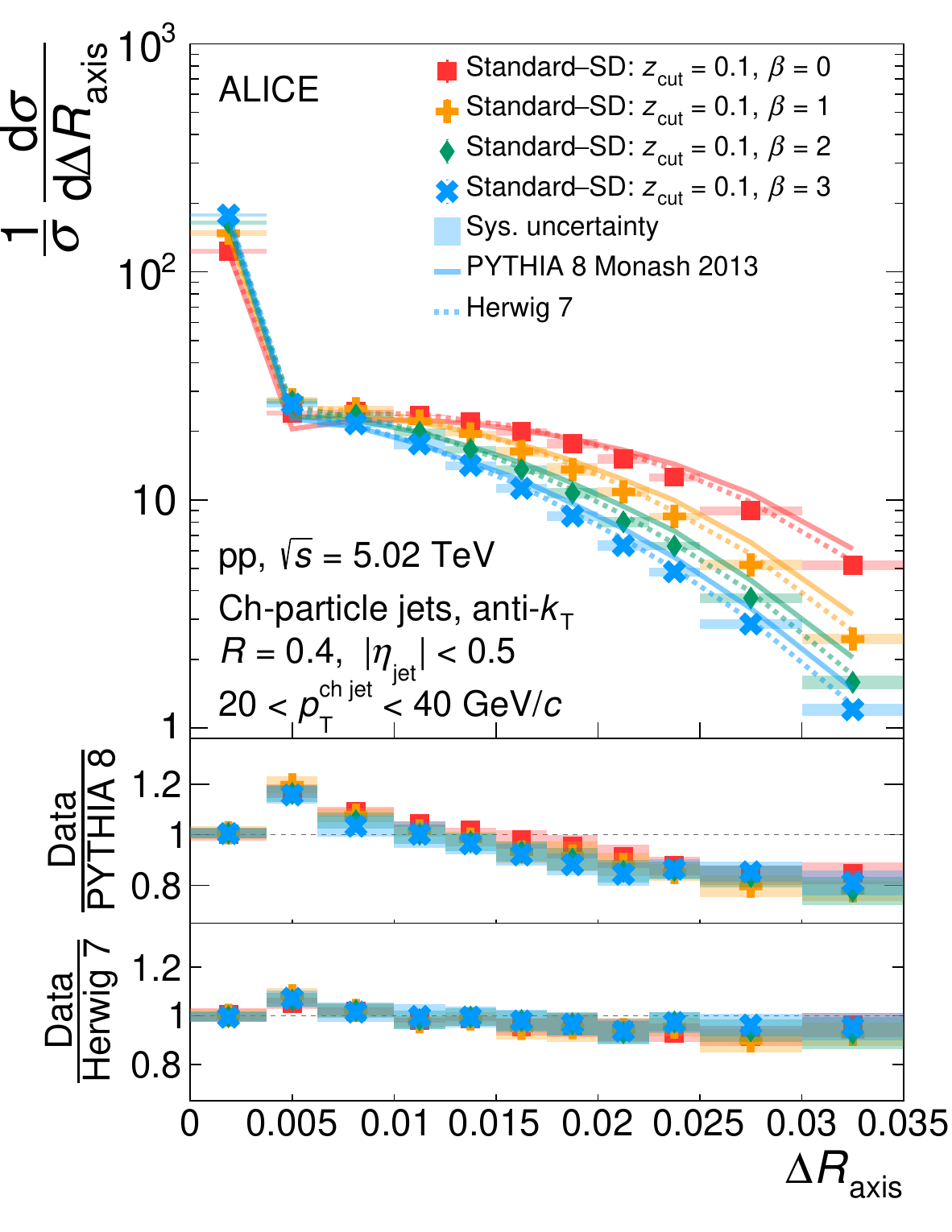}
    \includegraphics[width=0.48\textwidth]{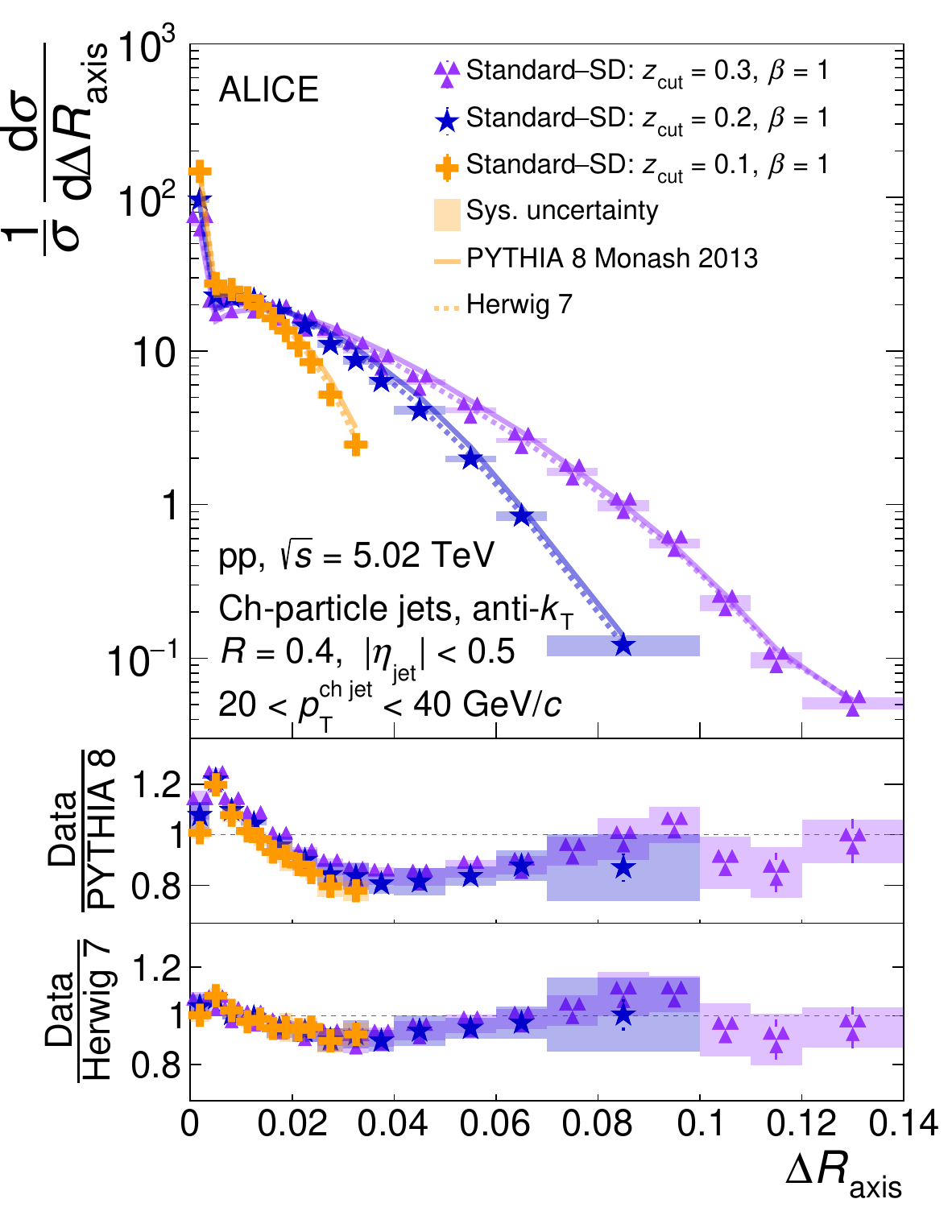}
    \caption{Comparison between \DeltaR{} Standard$-$SD measured distributions and Monte Carlo event generators for jets of $R=0.4$ in $20<\pTjet<40$ \GeVc. Left: distributions with $\zcut=0.1$ and varying $\beta$. Right: distributions with $\beta=1$ and varying $\zcut$.}
    \label{fig:data_generators_Standard_SD_20_40_R04}
\end{figure}

\begin{figure}
    \centering
    \includegraphics[width=0.48\textwidth]{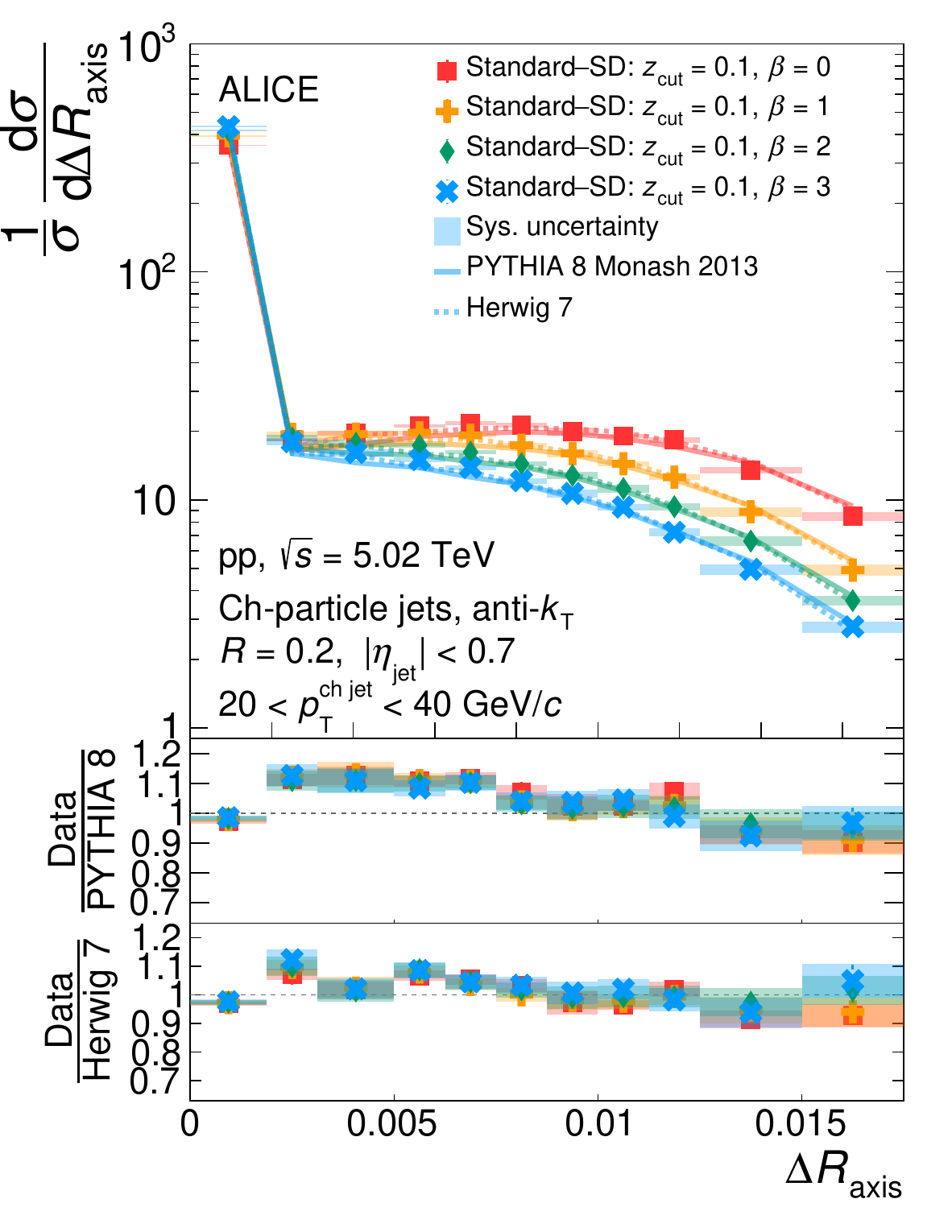}
    \includegraphics[width=0.48\textwidth]{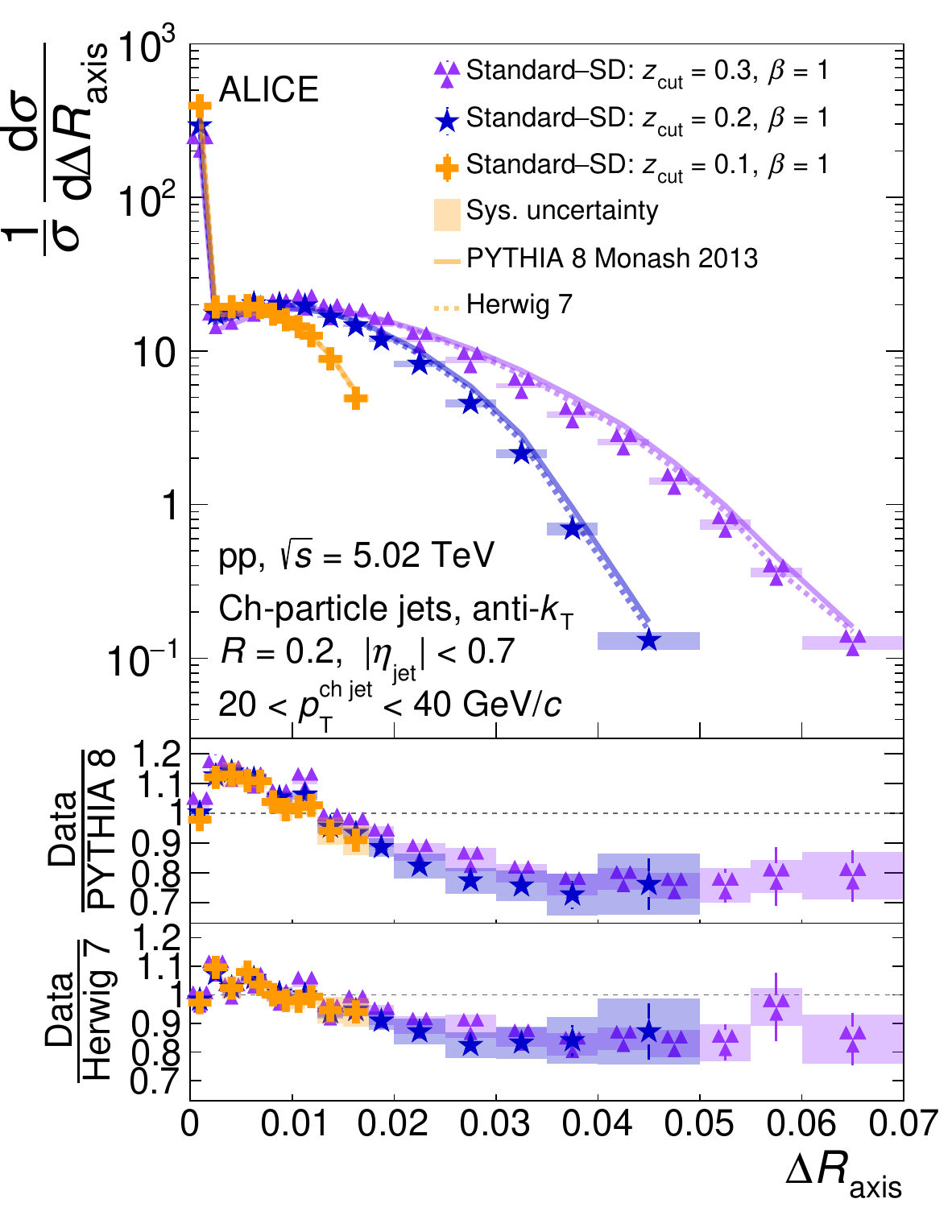}
    \caption{Same as Figure~\ref{fig:data_generators_Standard_SD_20_40_R04} for $R=0.2$.}
    \label{fig:data_generators_Standard_SD_20_40_R02}
\end{figure}

\begin{figure}
    \centering
    \includegraphics[width=0.48\textwidth]{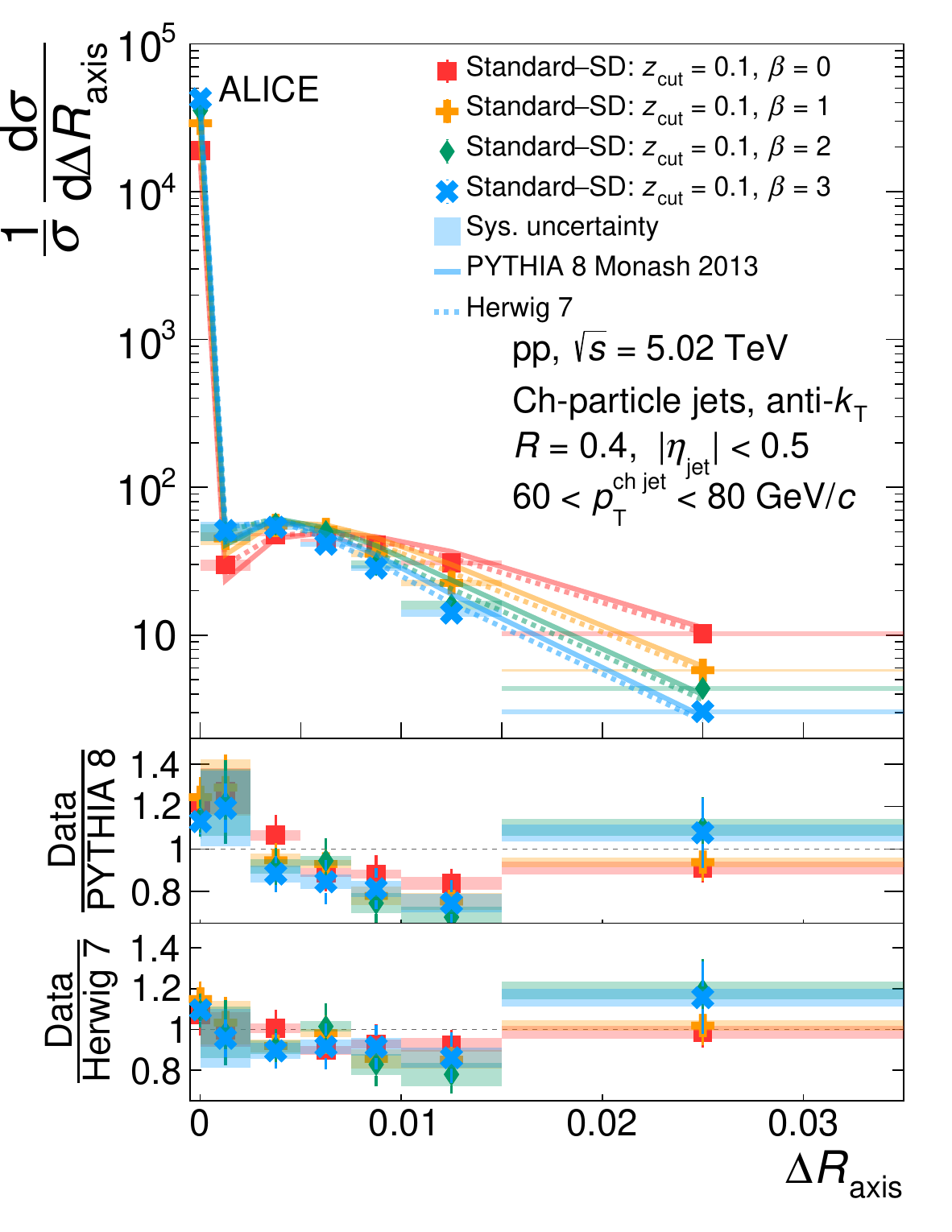}
    \includegraphics[width=0.48\textwidth]{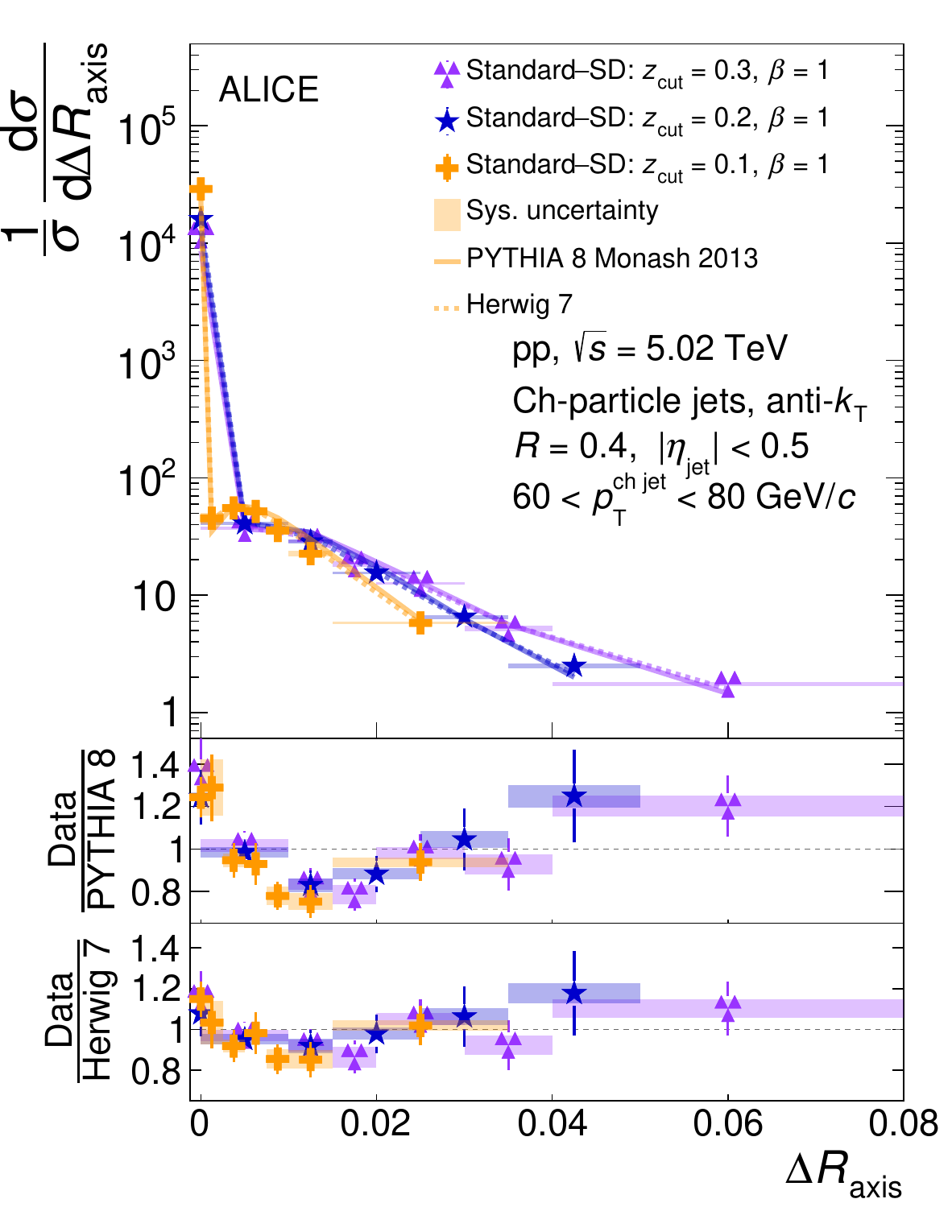}
    \caption{Same as Figure~\ref{fig:data_generators_Standard_SD_20_40_R04} for $60<\pTjet<80$ \GeVc.}
    \label{fig:data_generators_Standard_SD_60_80_R04}
\end{figure}

\begin{figure}
    \centering
    \includegraphics[width=0.48\textwidth]{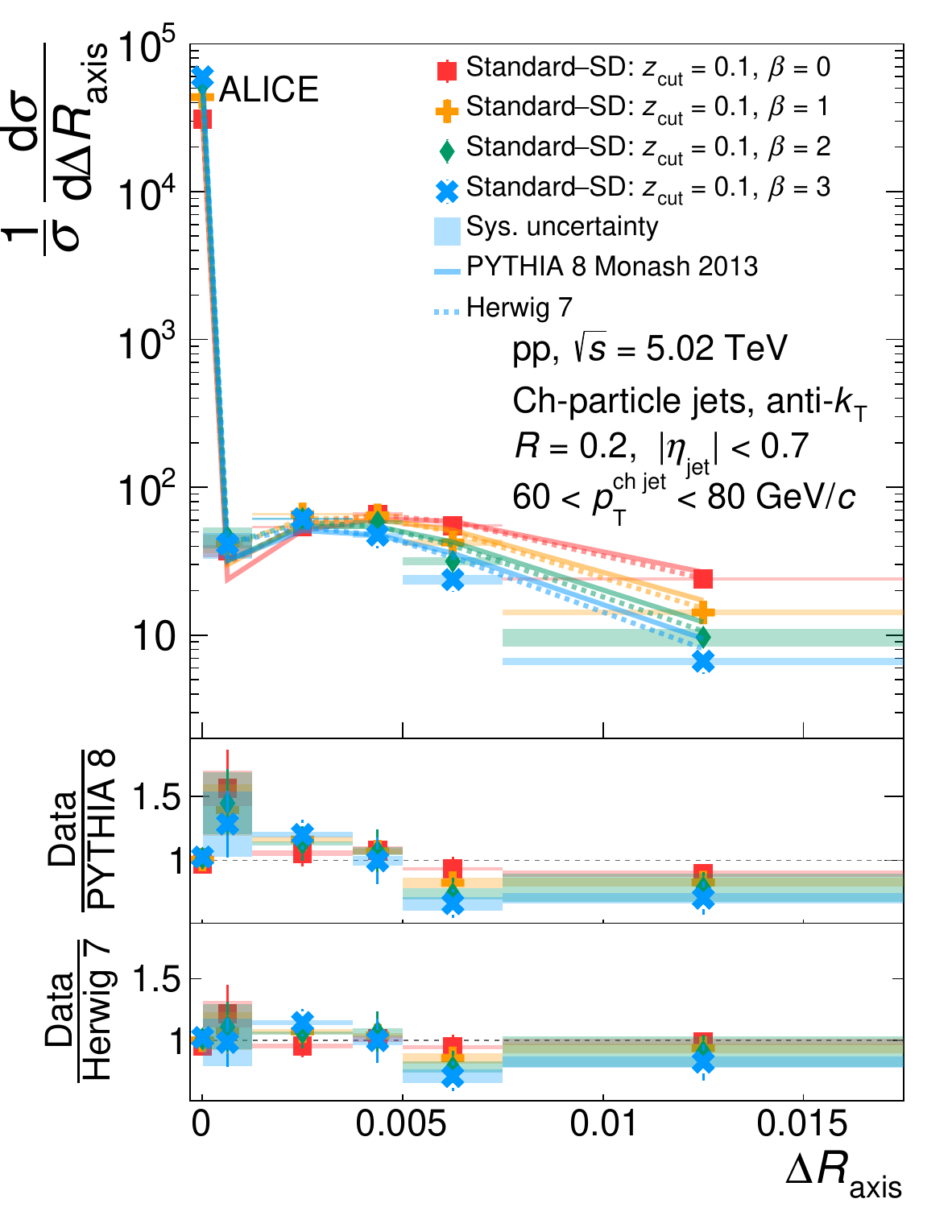}
    \includegraphics[width=0.48\textwidth]{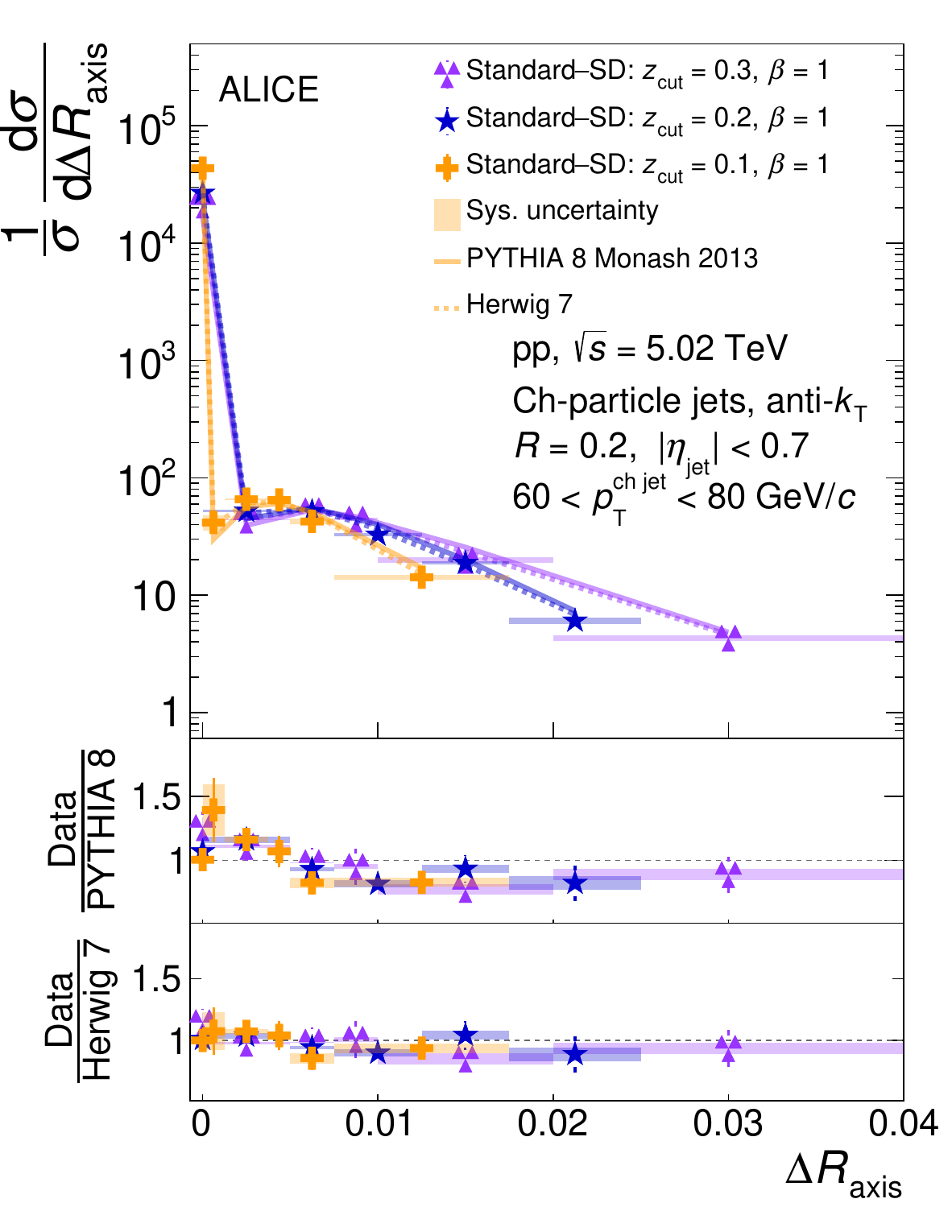}
    \caption{Same as Figure~\ref{fig:data_generators_Standard_SD_20_40_R04} for $R=0.2$ for $60<\pTjet<80$ \GeVc.}
    \label{fig:data_generators_Standard_SD_60_80_R02}
\end{figure}

\begin{figure}
    \centering
    \includegraphics[width=0.48\textwidth]{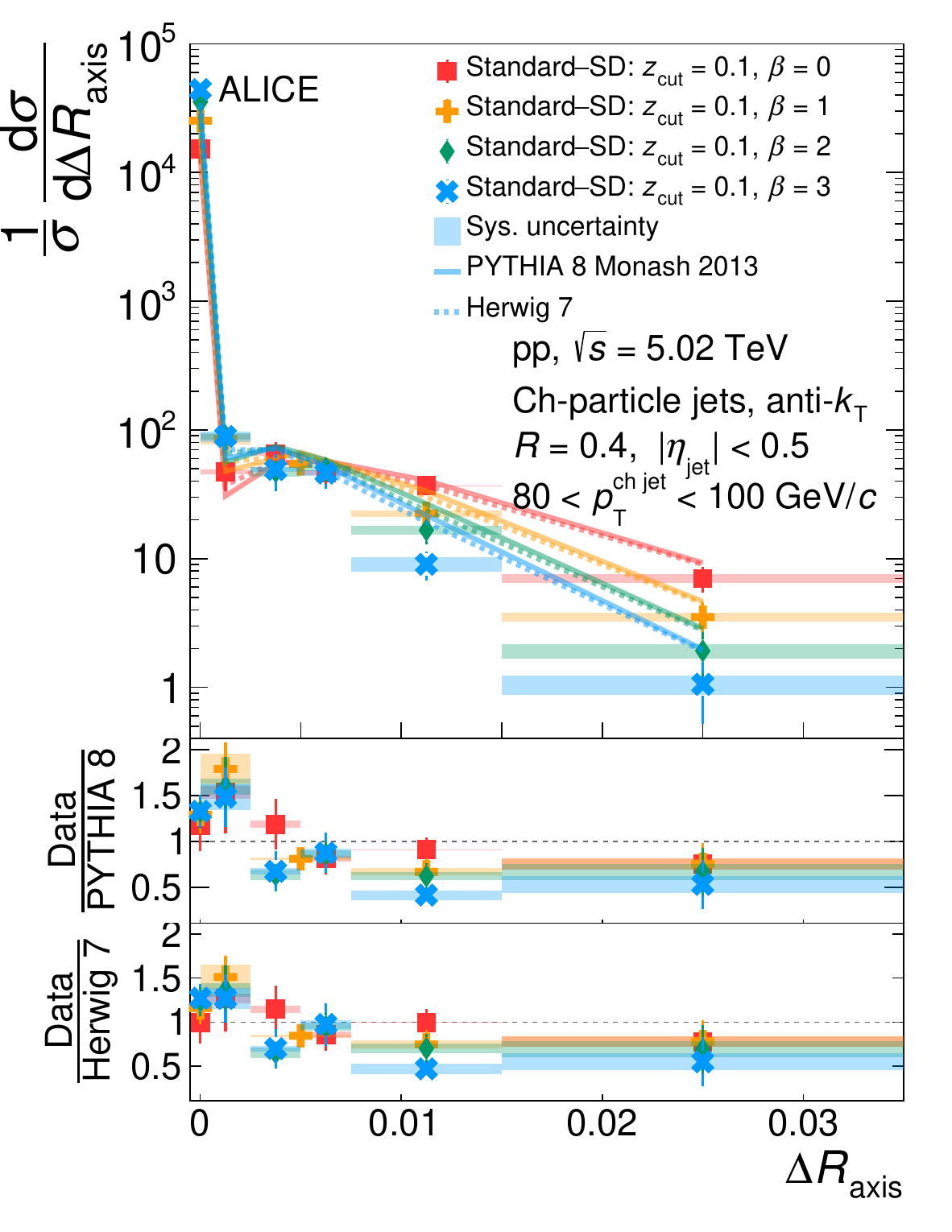}
    \includegraphics[width=0.48\textwidth]{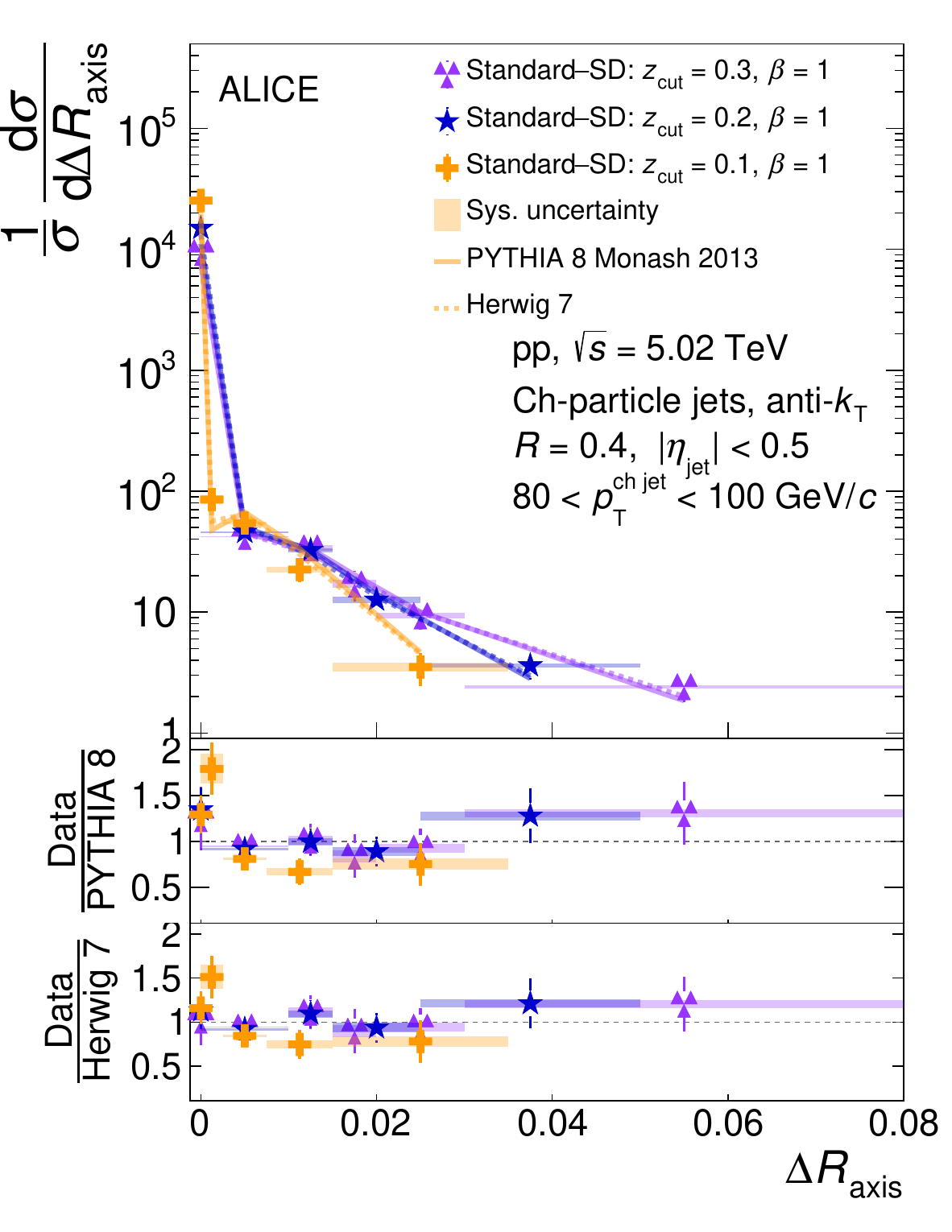}
    \caption{Same as Figure~\ref{fig:data_generators_Standard_SD_20_40_R04} for $80<\pTjet<100$ \GeVc.}
    \label{fig:data_generators_Standard_SD_80_100_R04}
\end{figure}

\begin{figure}
    \centering
    \includegraphics[width=0.48\textwidth]{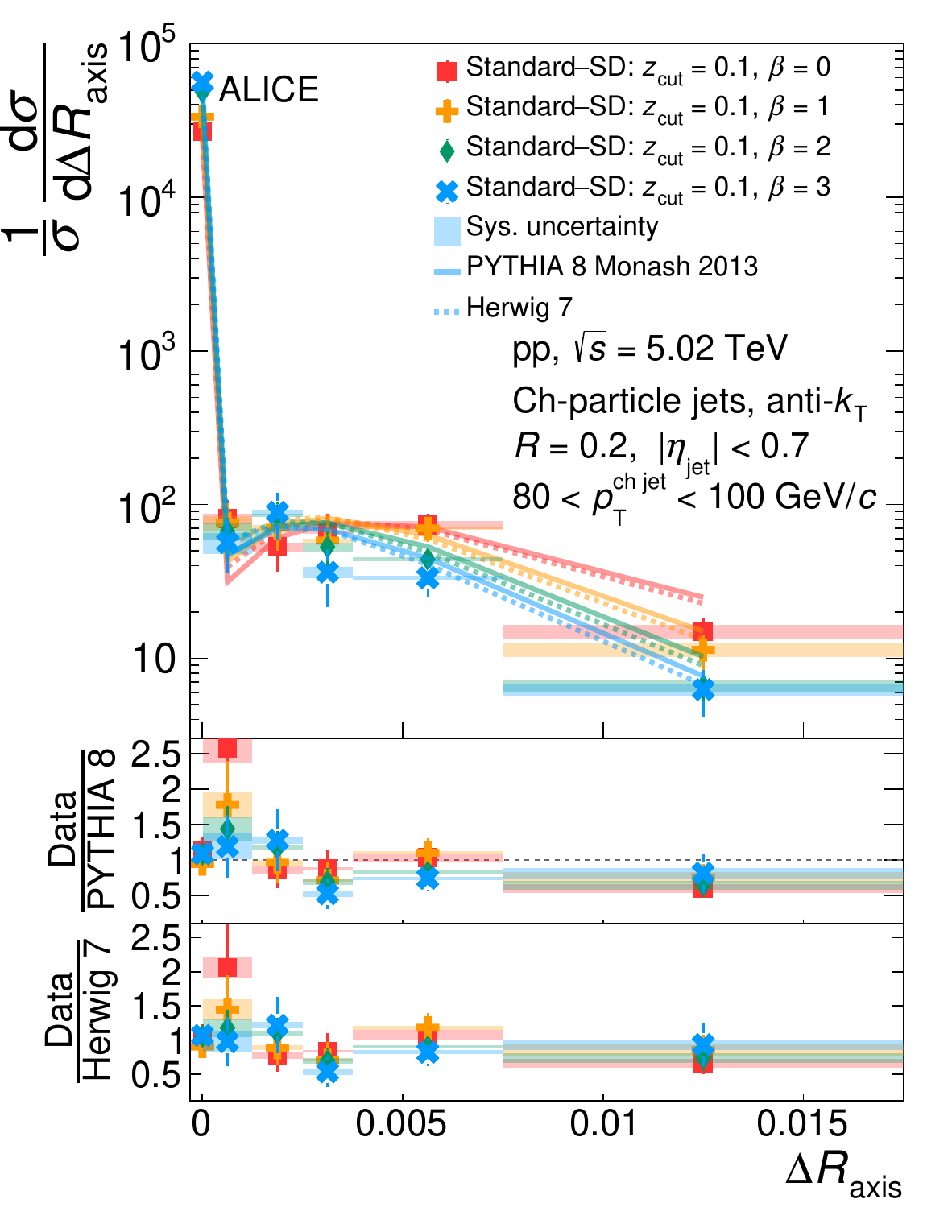}
    \includegraphics[width=0.48\textwidth]{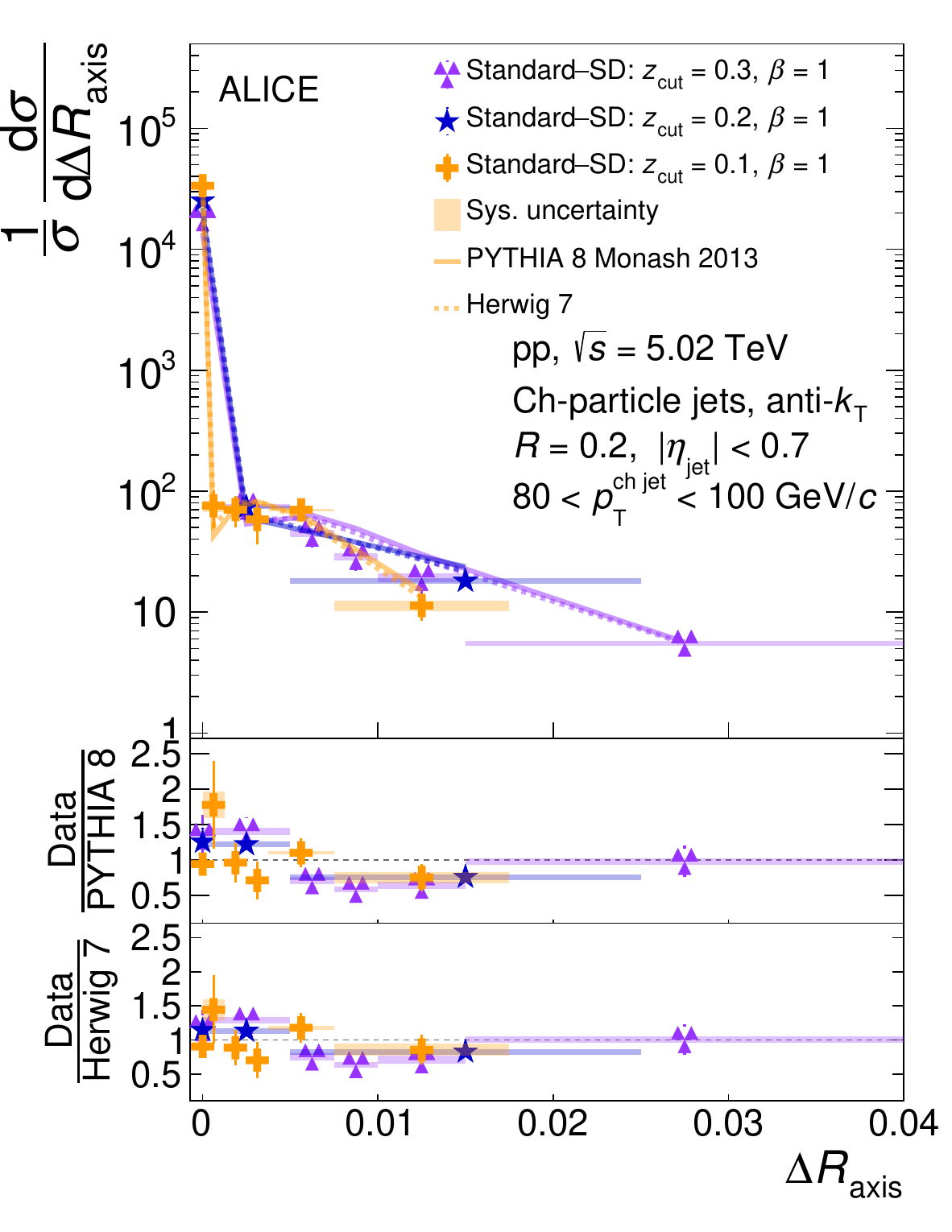}
    \caption{Same as Figure~\ref{fig:data_generators_Standard_SD_20_40_R04} for $R=0.2$ for $80<\pTjet<100$ \GeVc.}
    \label{fig:data_generators_Standard_SD_80_100_R02}
\end{figure}


\clearpage
\section{Comparison to analytic predictions} \label{app:B}

\begin{figure}[!htb]
\centering{}
\includegraphics[width=0.495\textwidth]{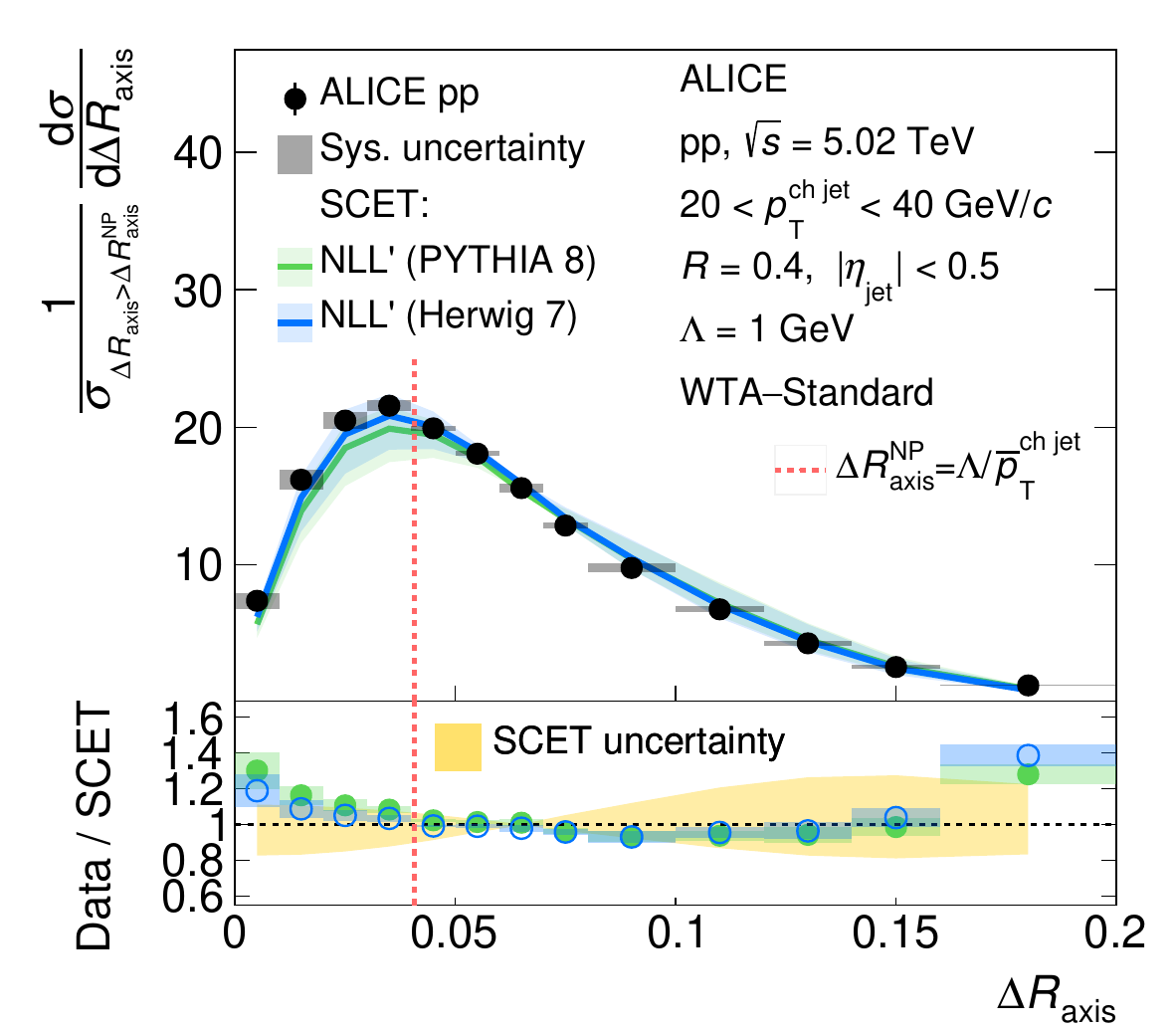}
\includegraphics[width=0.495\textwidth]{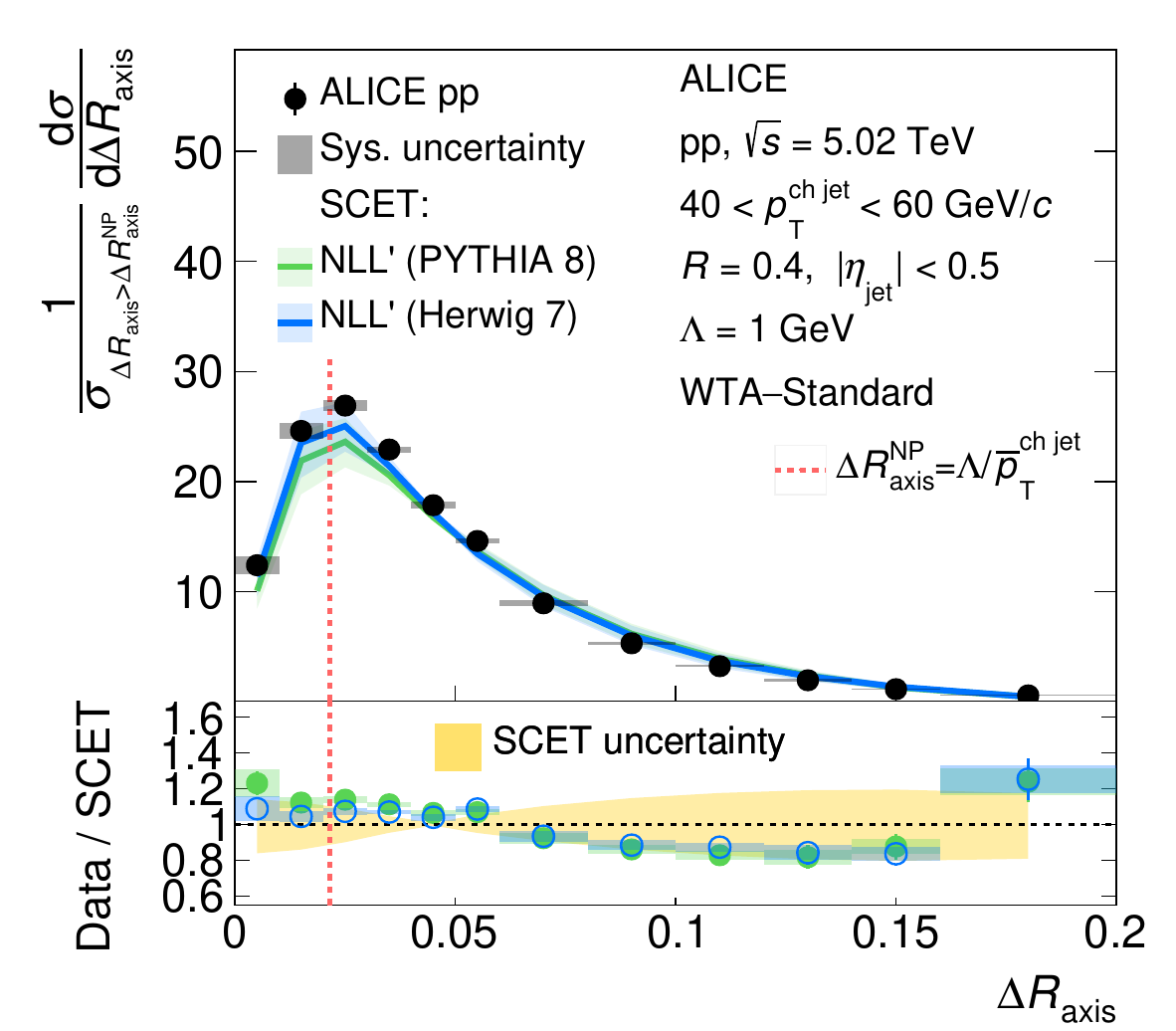}
\includegraphics[width=0.495\textwidth]{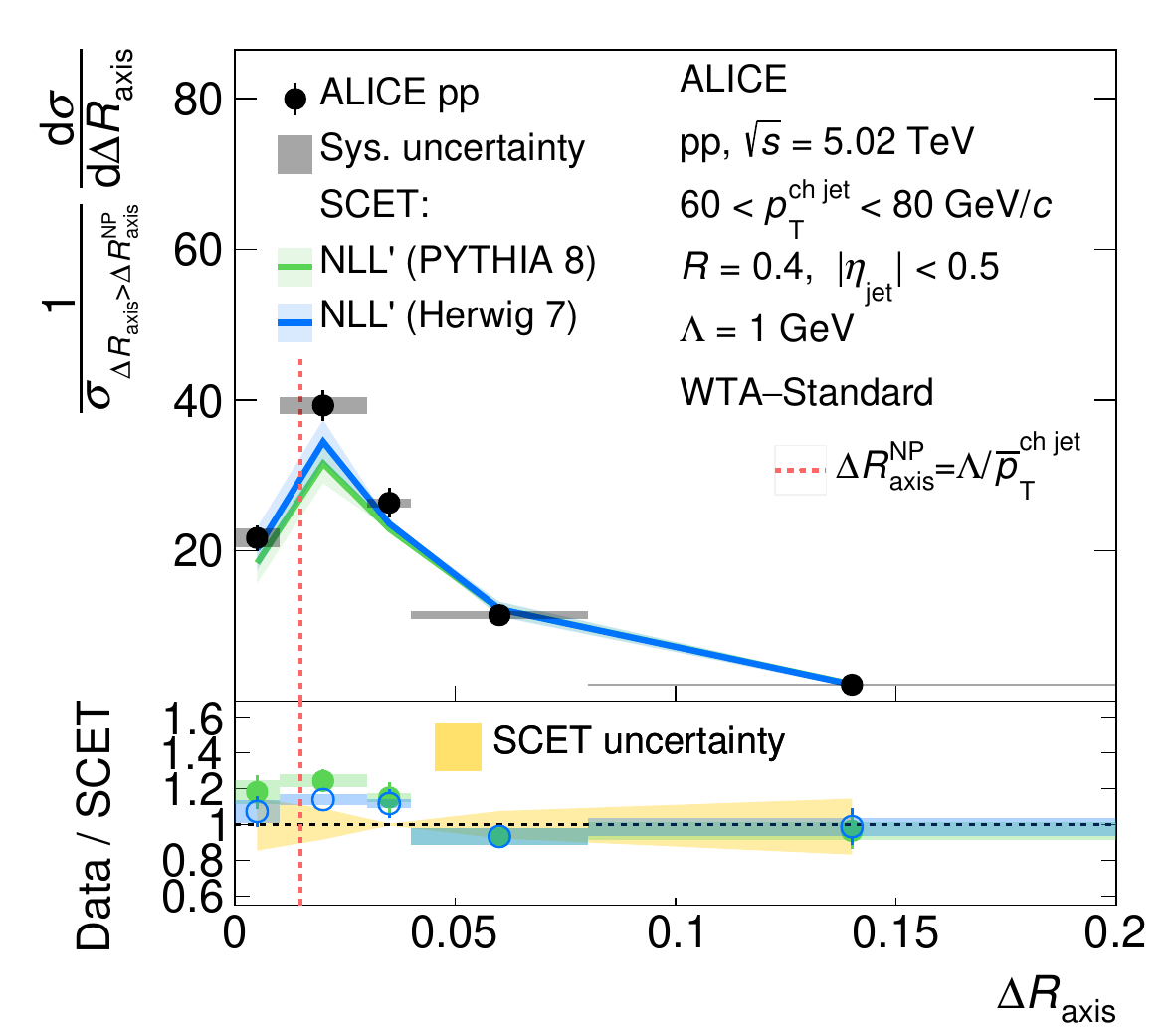}
\includegraphics[width=0.495\textwidth]{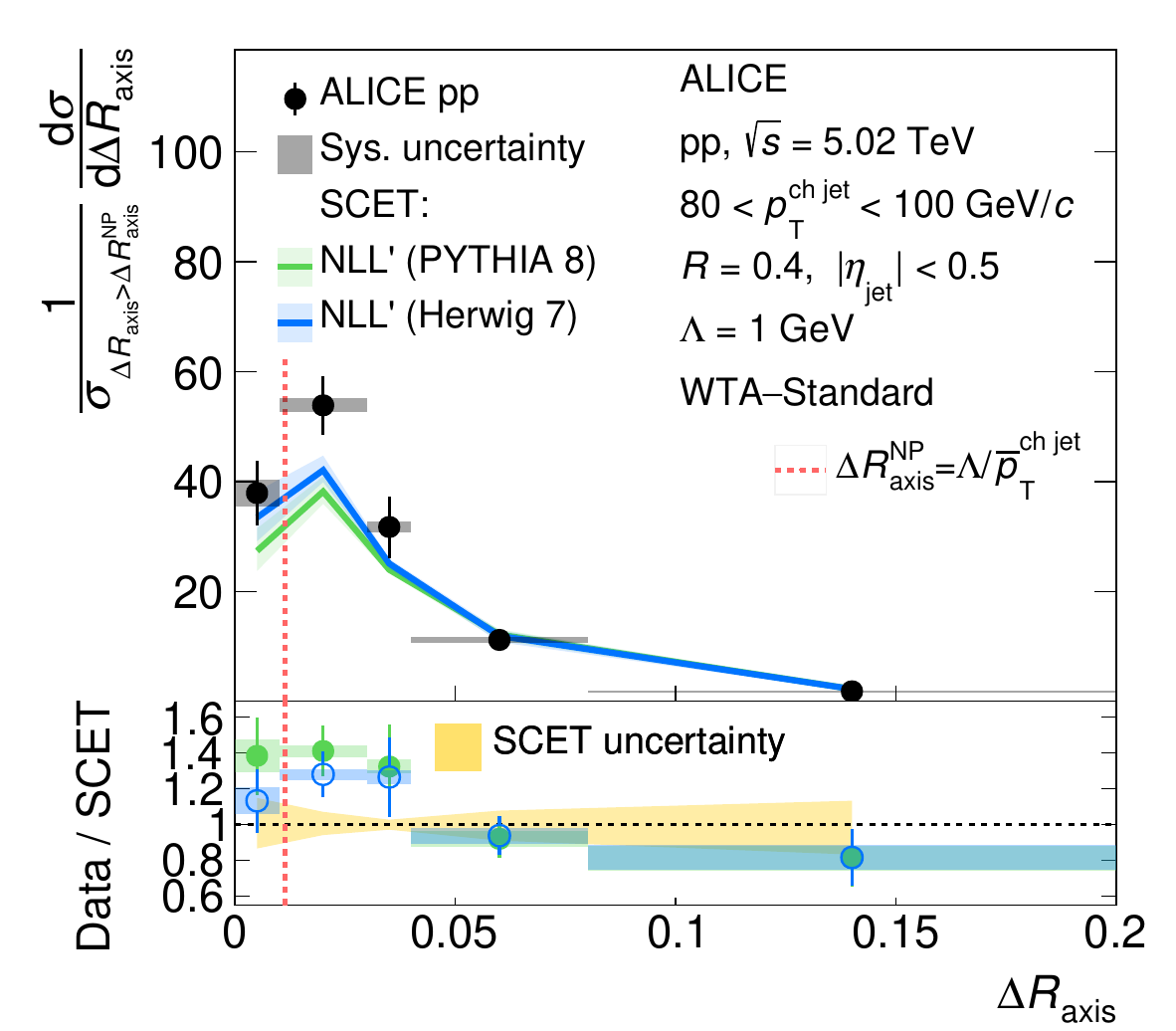}
\caption{Comparison between \DeltaR{} measured distributions and analytic predictions for the WTA\textendash{}Standard case with $R=0.4$.
Each panel corresponds to a different \pTchjet{} bin. The measured distribution is shown in black.
The colored distributions correspond to the SCET prediction corrected with different MC event generators.}
\label{fig:SCET_Standard_WTA_R04}
\end{figure}

\begin{figure}[!htb]
\centering{}
\includegraphics[width=0.495\textwidth]{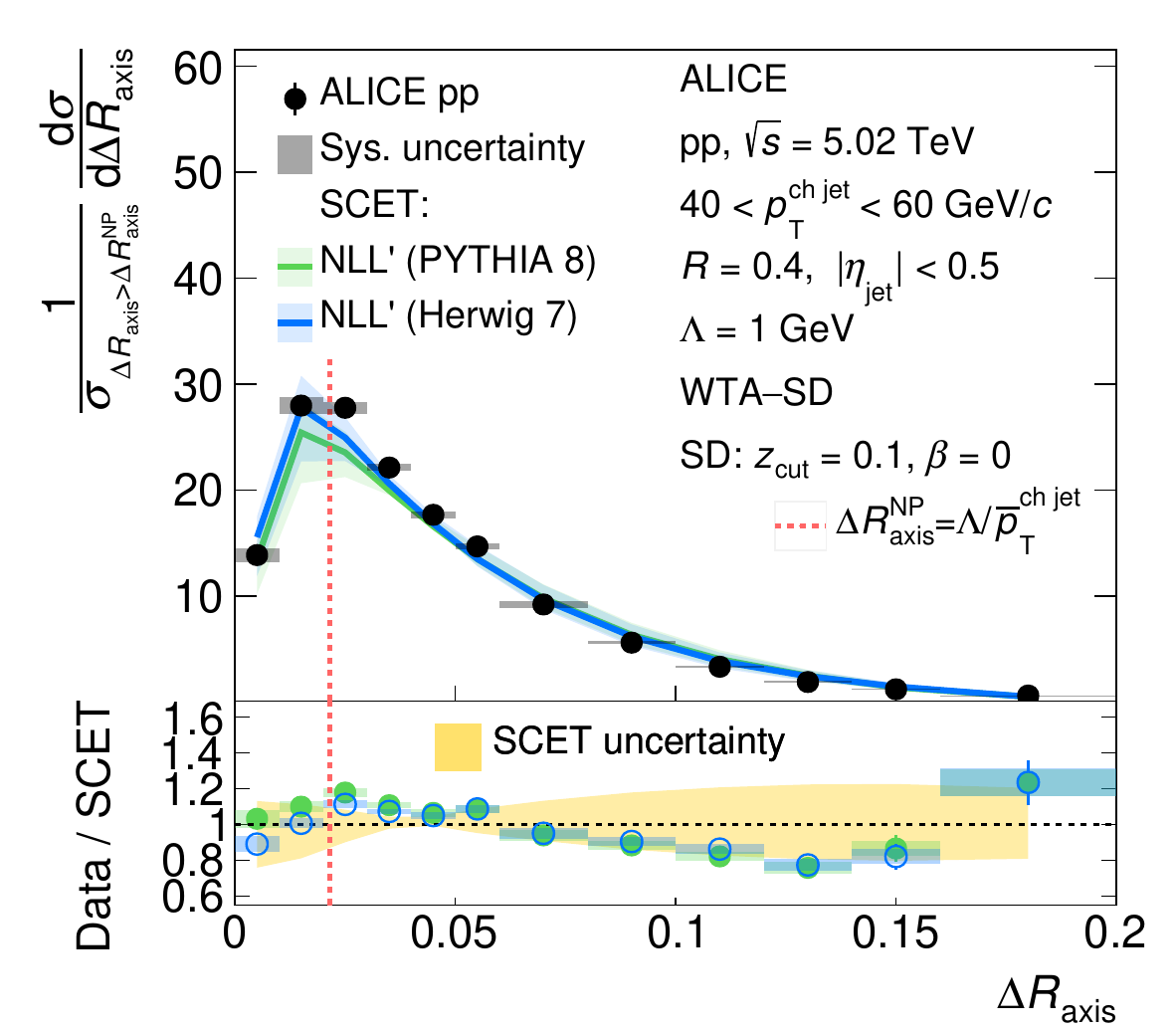}
\includegraphics[width=0.495\textwidth]{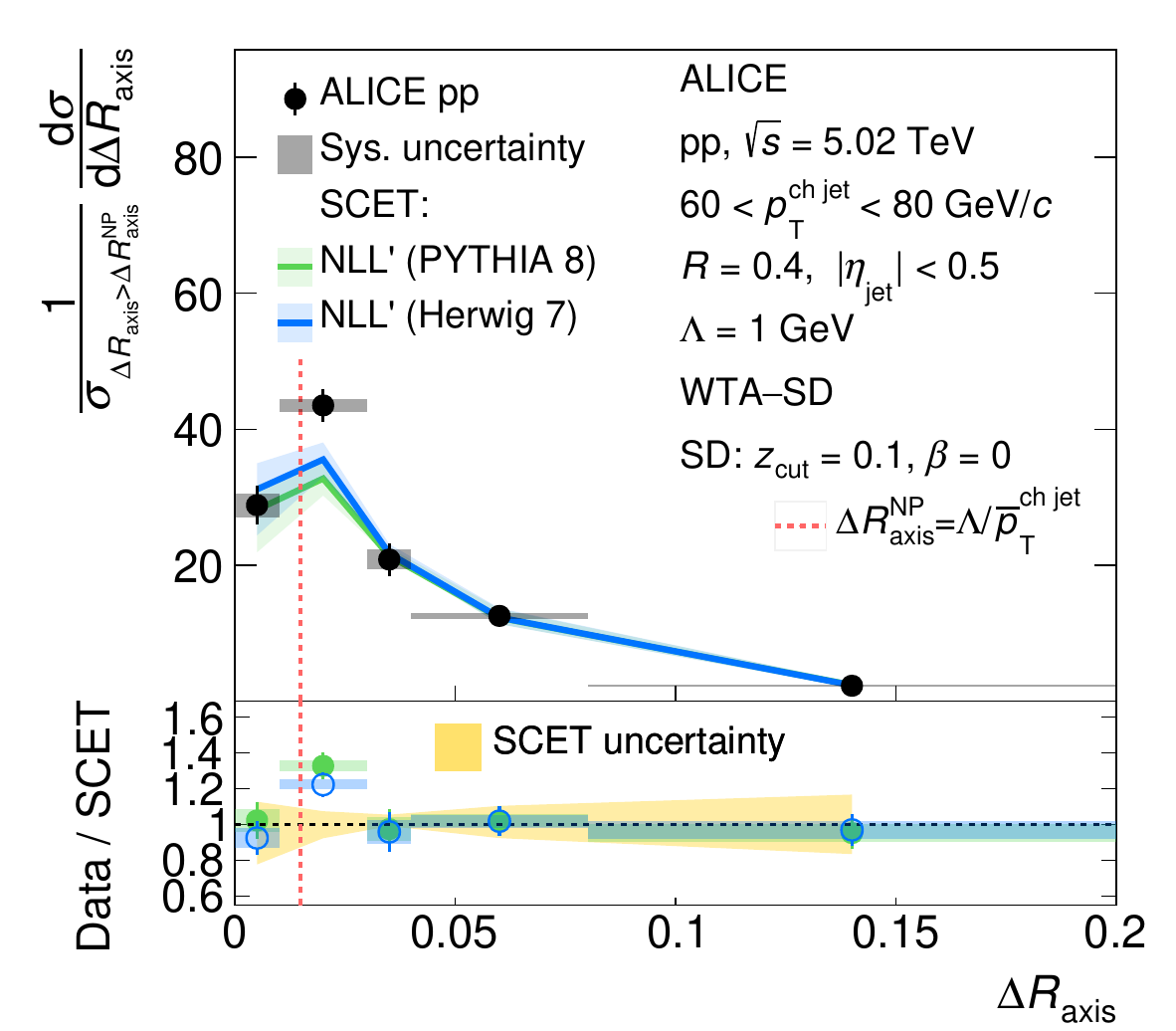}
\includegraphics[width=0.495\textwidth]{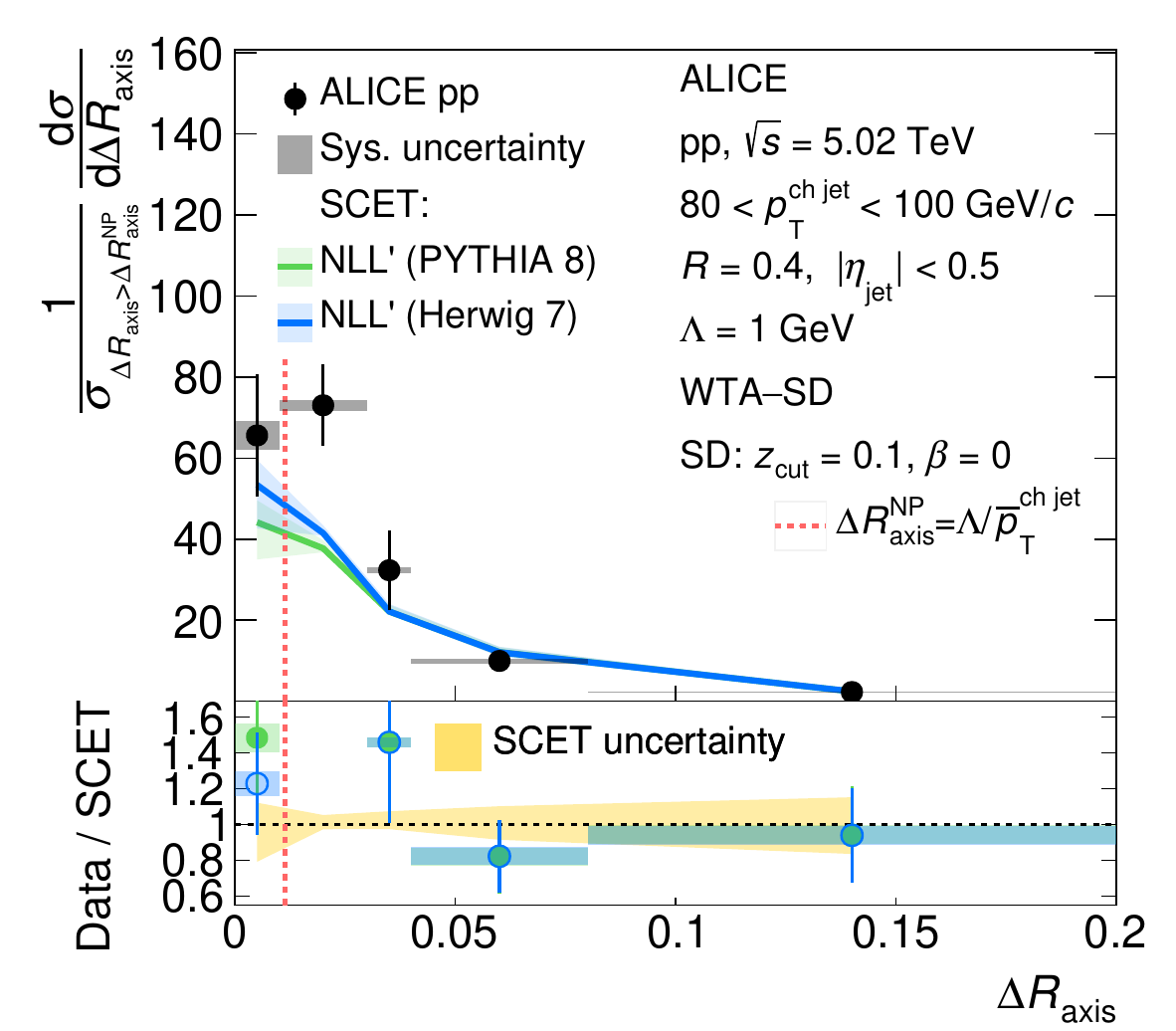}
\caption{Same as Figure~\ref{fig:SCET_Standard_WTA_R04}, for WTA$-$SD ($\zcut=0.1, \, \beta=0$) for $R=0.4$.}
\label{fig:SCET_WTA_SD_1_SD_zcut01_B0_R04}
\end{figure}

\begin{figure}[!htb]
\centering{}
\includegraphics[width=0.495\textwidth]{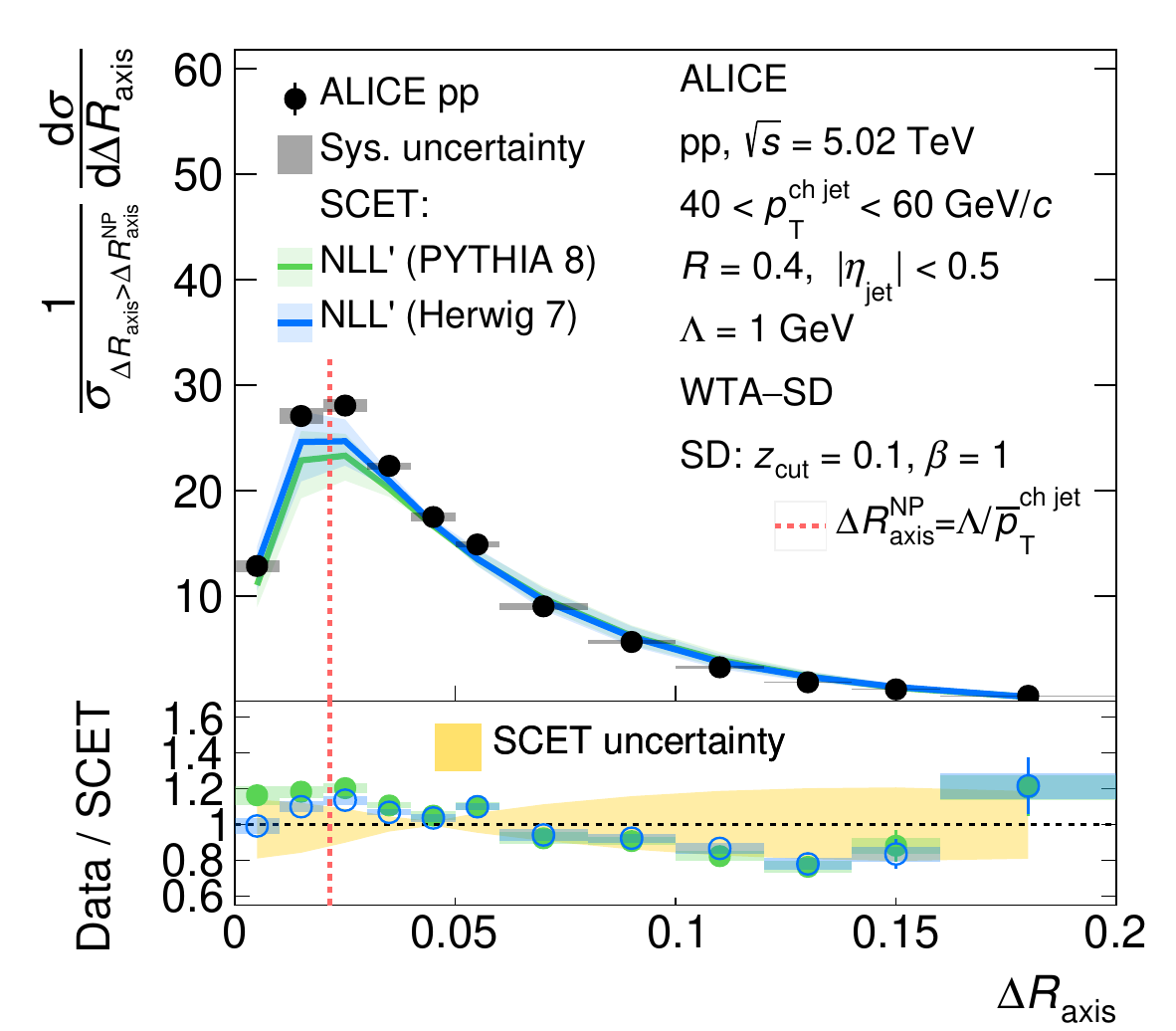}
\includegraphics[width=0.495\textwidth]{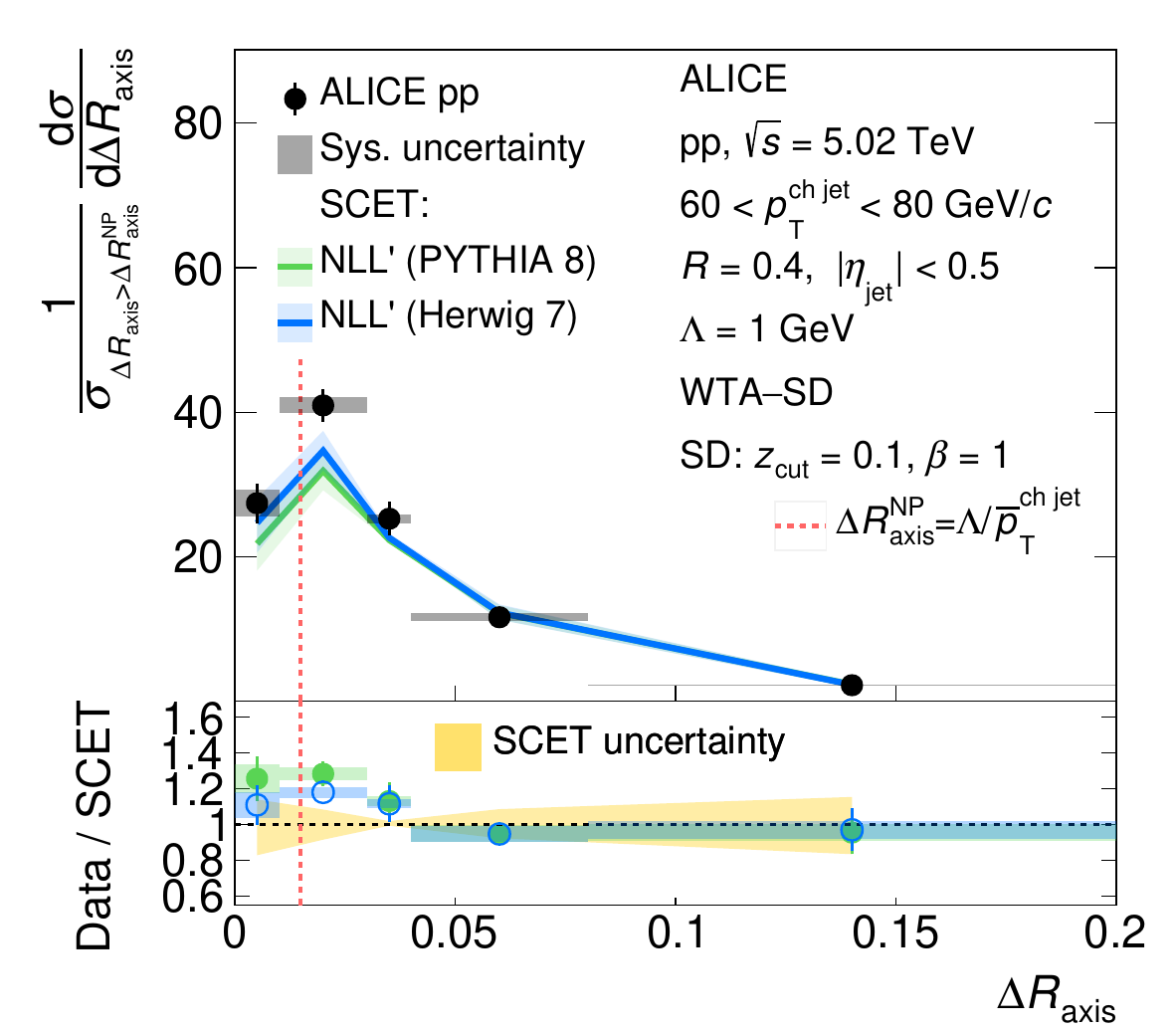}
\includegraphics[width=0.495\textwidth]{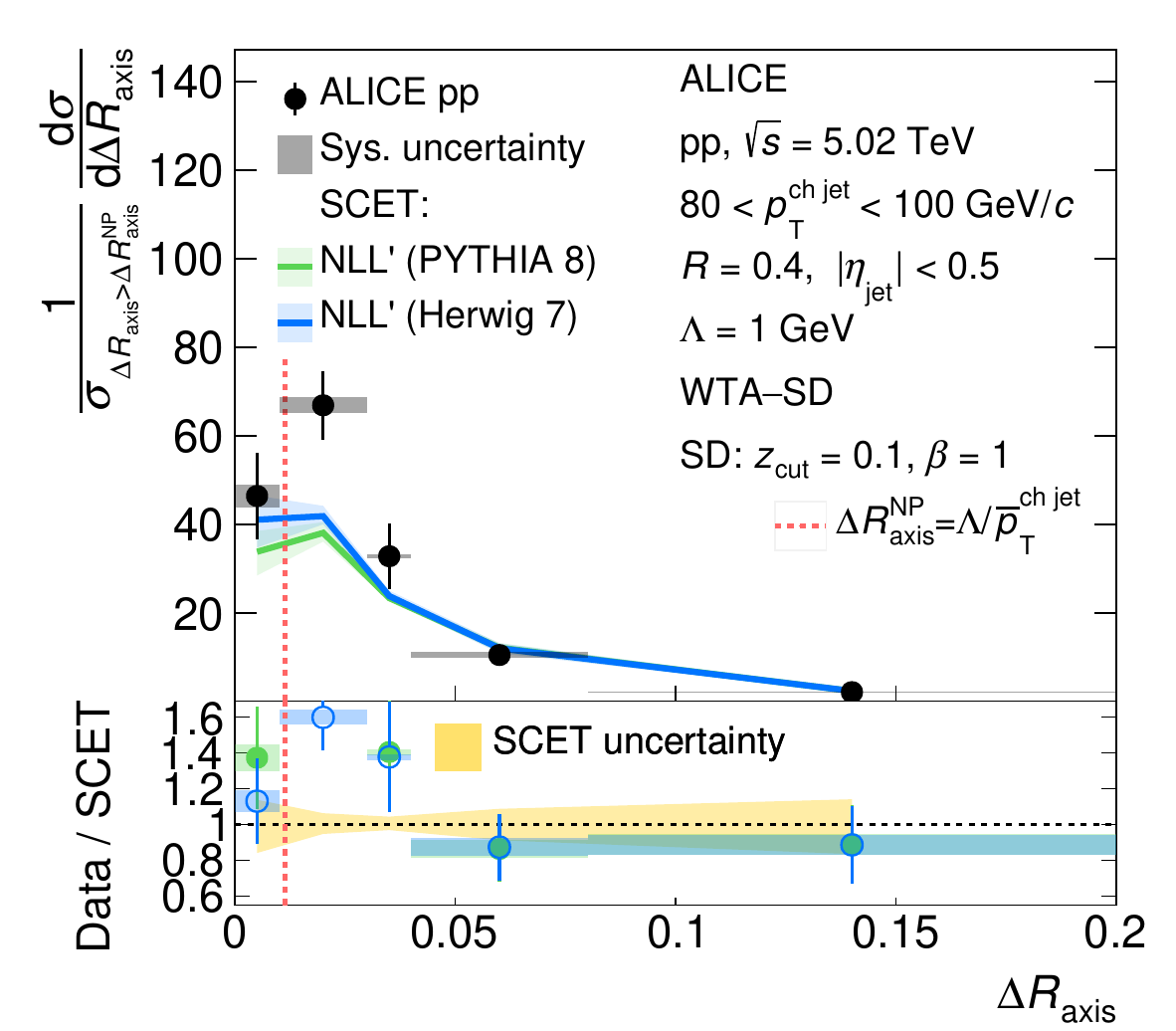}
\caption{Same as Figure~\ref{fig:SCET_Standard_WTA_R04}, for WTA$-$SD ($\zcut=0.1, \, \beta=1$) for $R=0.4$.}
\label{fig:SCET_WTA_SD_2_SD_zcut01_B1_R04}
\end{figure}

\begin{figure}[!htb]
\centering{}
\includegraphics[width=0.495\textwidth]{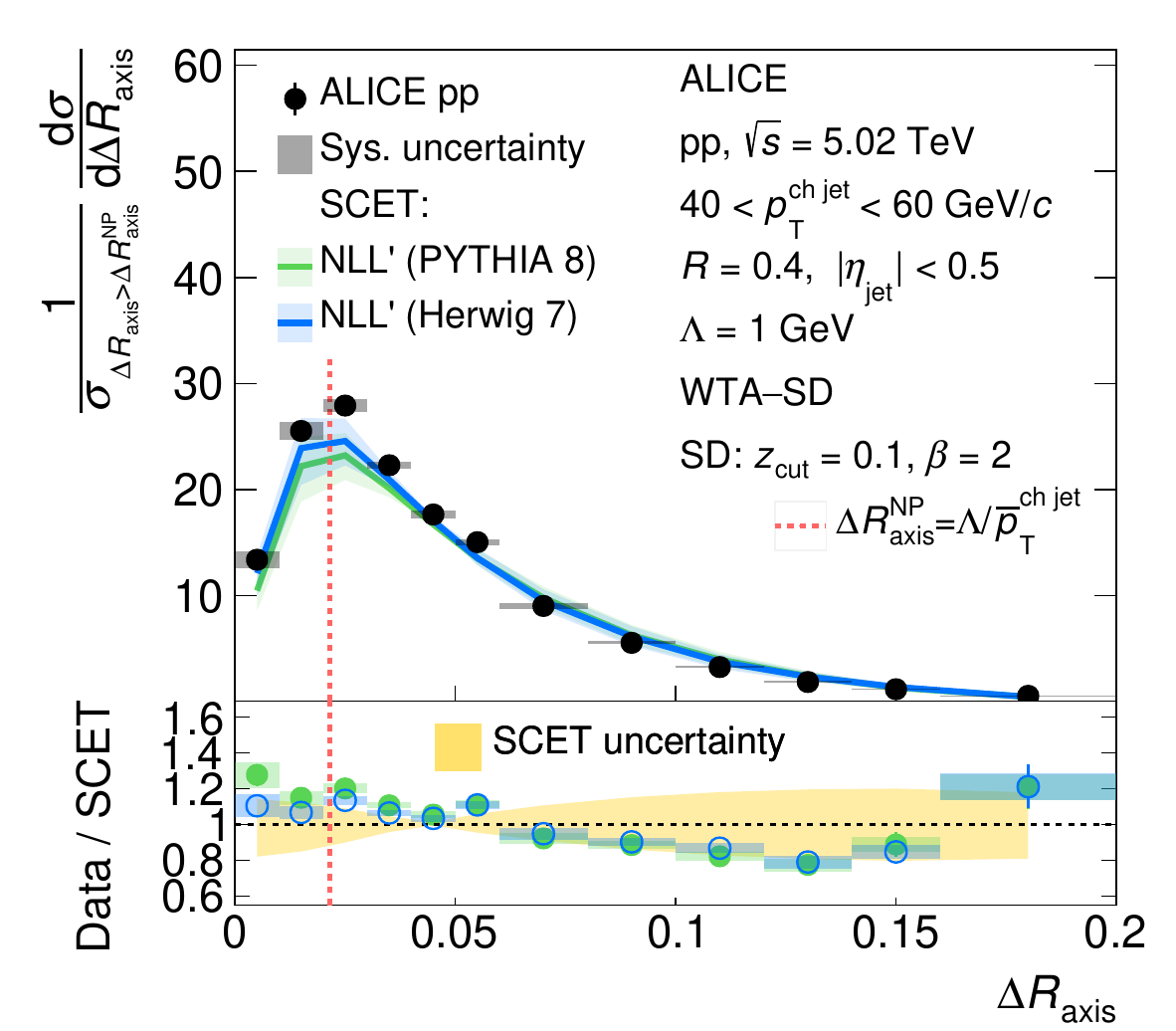}
\includegraphics[width=0.495\textwidth]{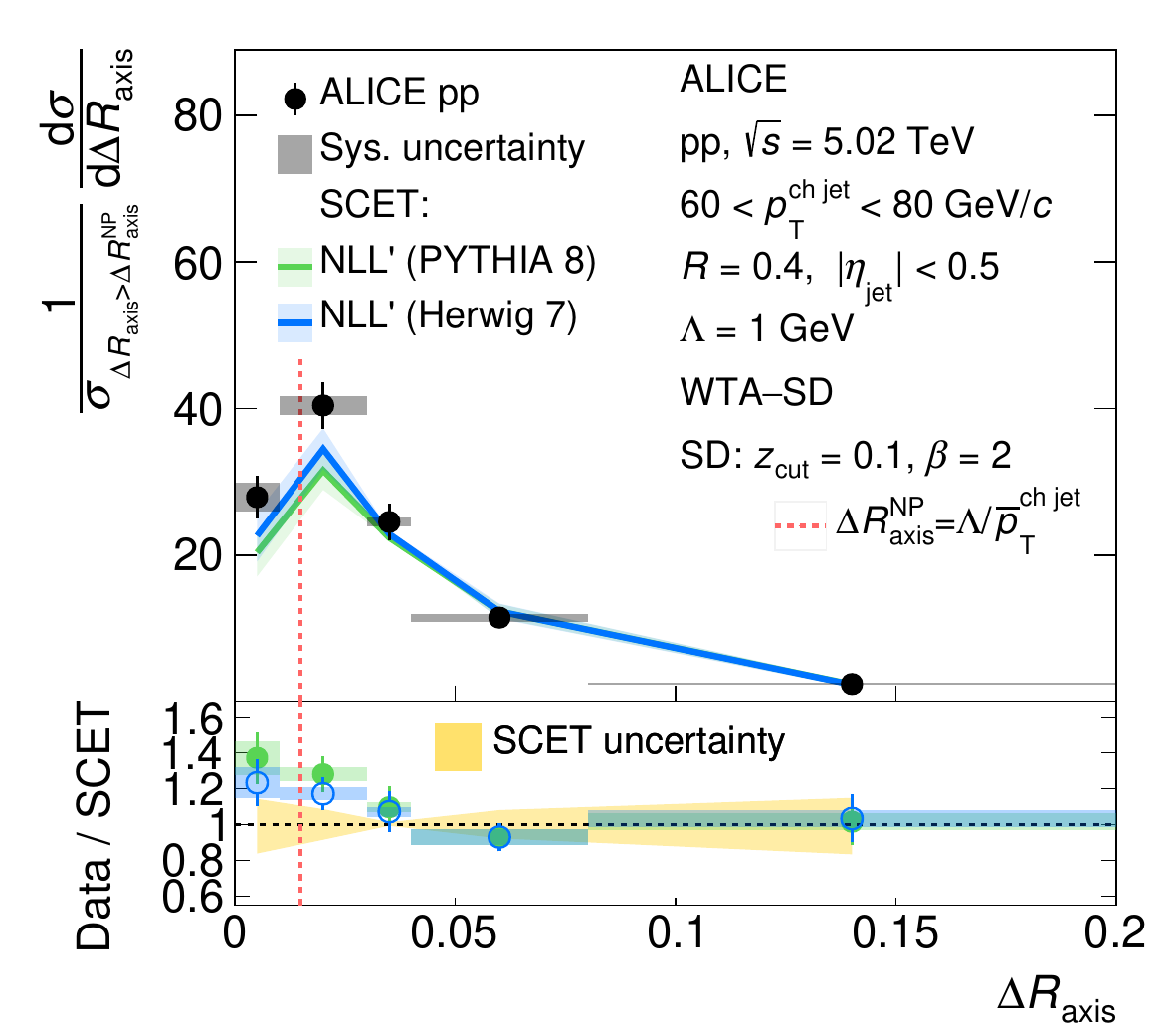}
\includegraphics[width=0.495\textwidth]{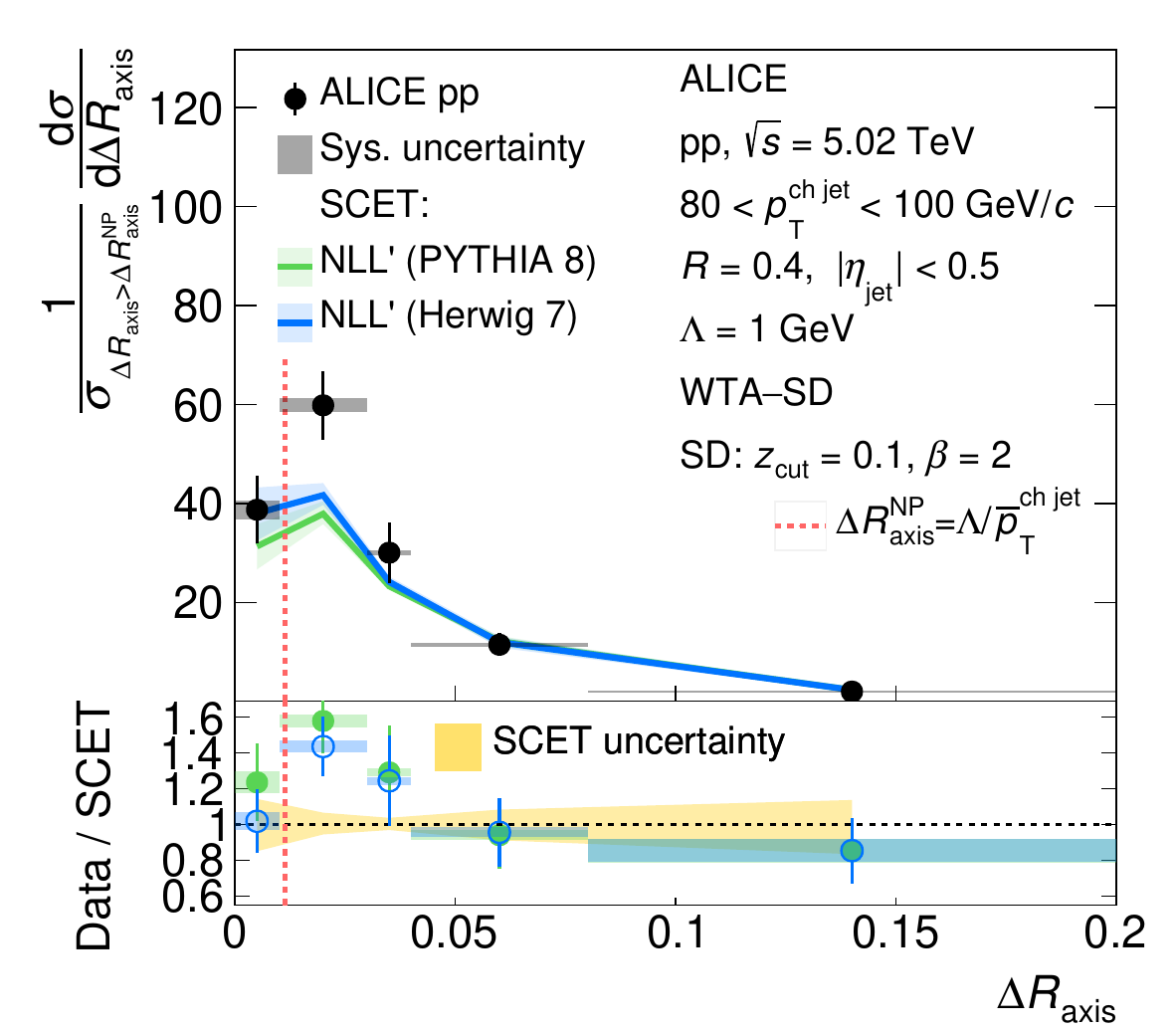}
\caption{Same as Figure~\ref{fig:SCET_Standard_WTA_R04}, for WTA$-$SD ($\zcut=0.1, \, \beta=2$) for $R=0.4$.}
\label{fig:SCET_WTA_SD_3_SD_zcut01_B2_R04}
\end{figure}

\begin{figure}[!htb]
\centering{}
\includegraphics[width=0.495\textwidth]{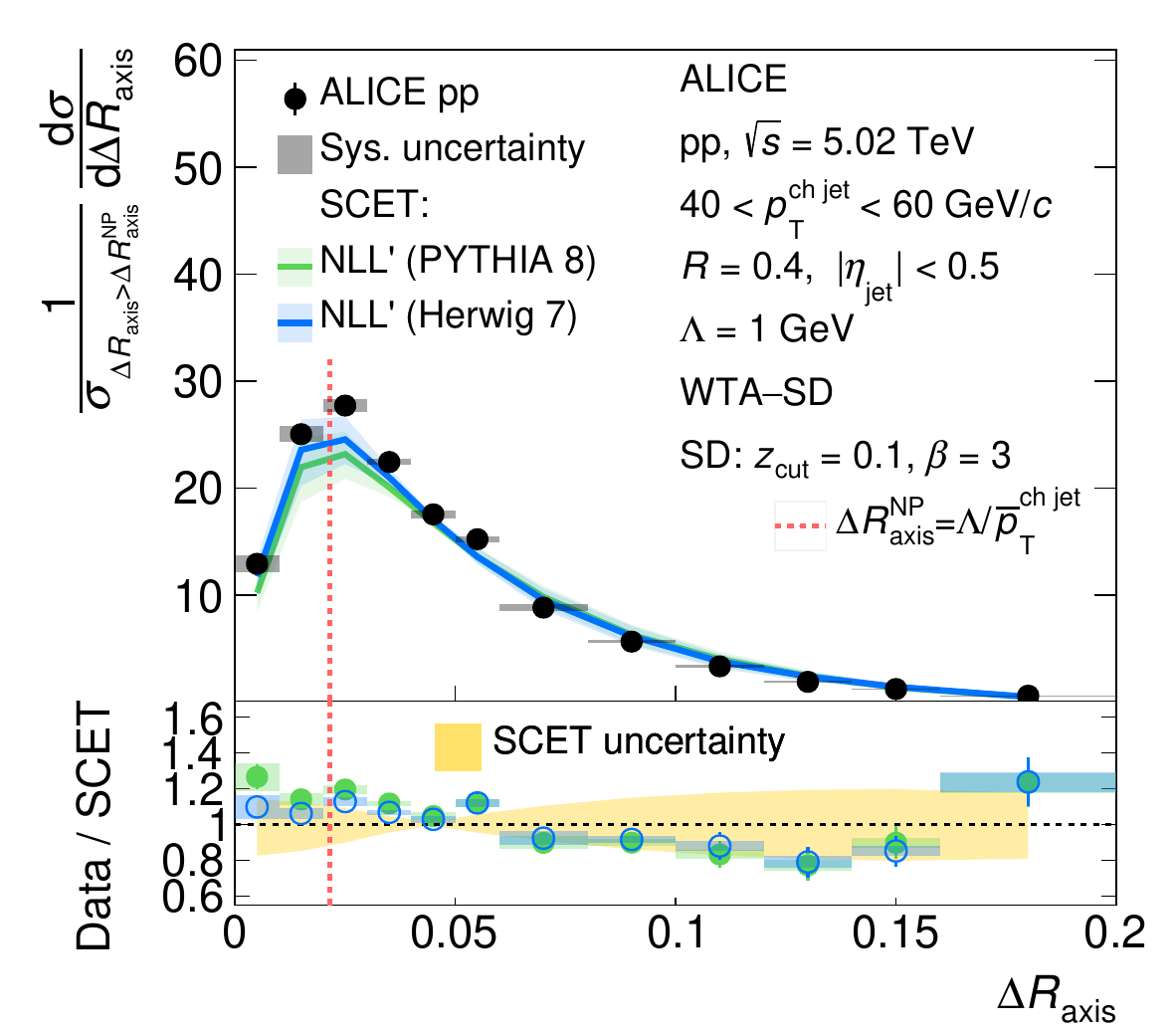}
\includegraphics[width=0.495\textwidth]{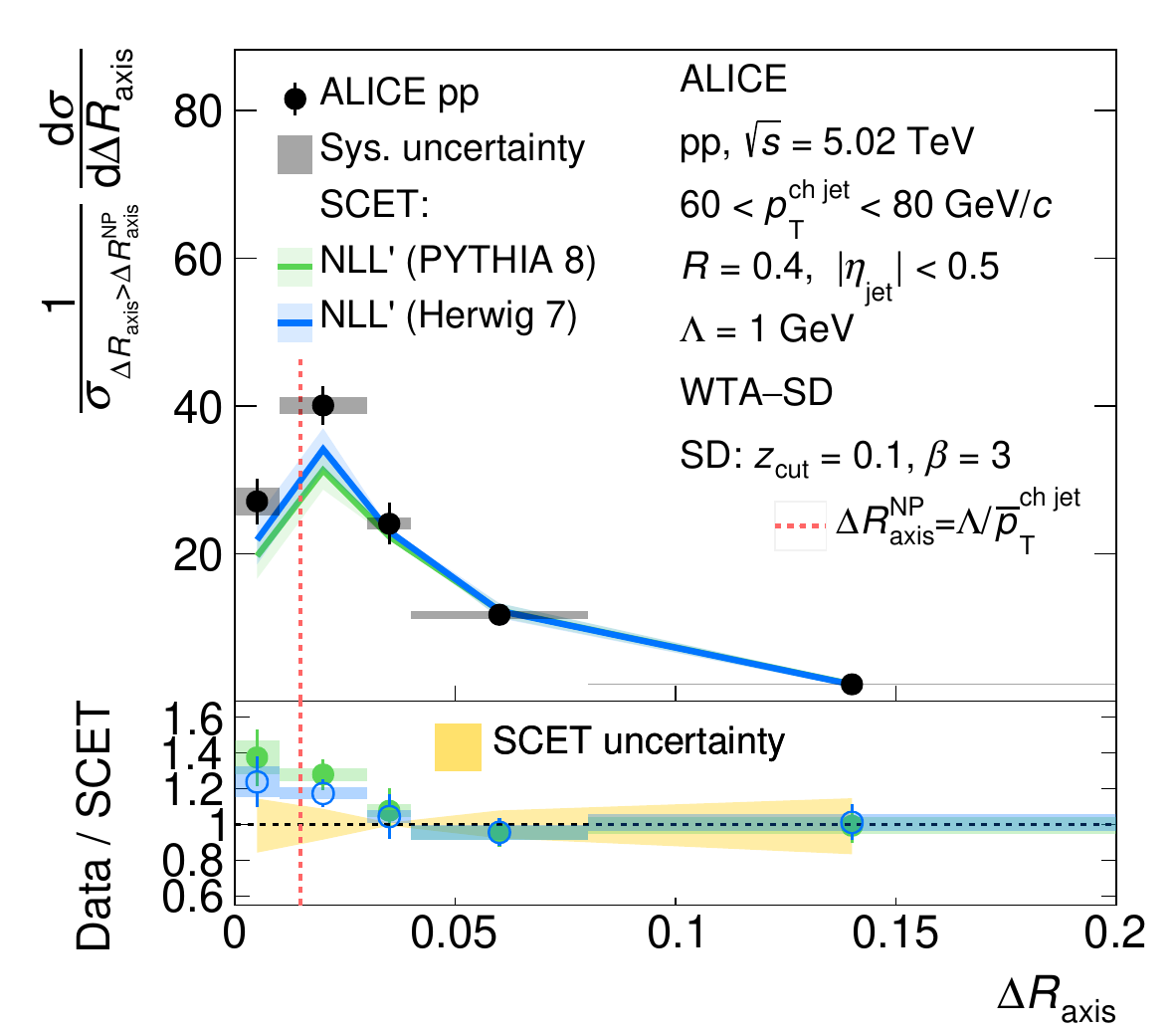}
\includegraphics[width=0.495\textwidth]{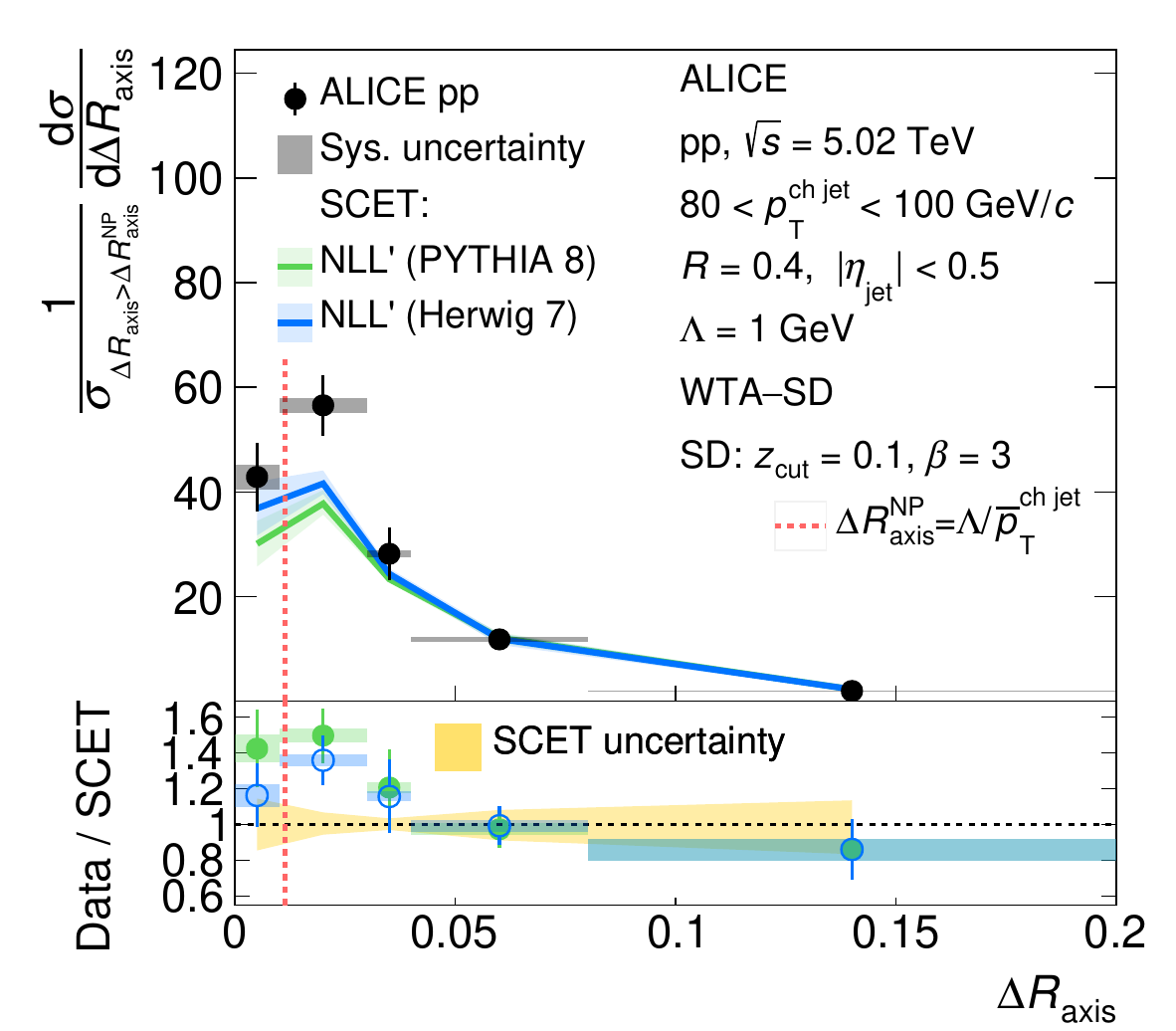}
\caption{Same as Figure~\ref{fig:SCET_Standard_WTA_R04}, for WTA$-$SD ($\zcut=0.1, \, \beta=3$) for $R=0.4$.}
\label{fig:SCET_WTA_SD_4_SD_zcut01_B3_R04}
\end{figure}

\begin{figure}[!htb]
\centering{}
\includegraphics[width=0.495\textwidth]{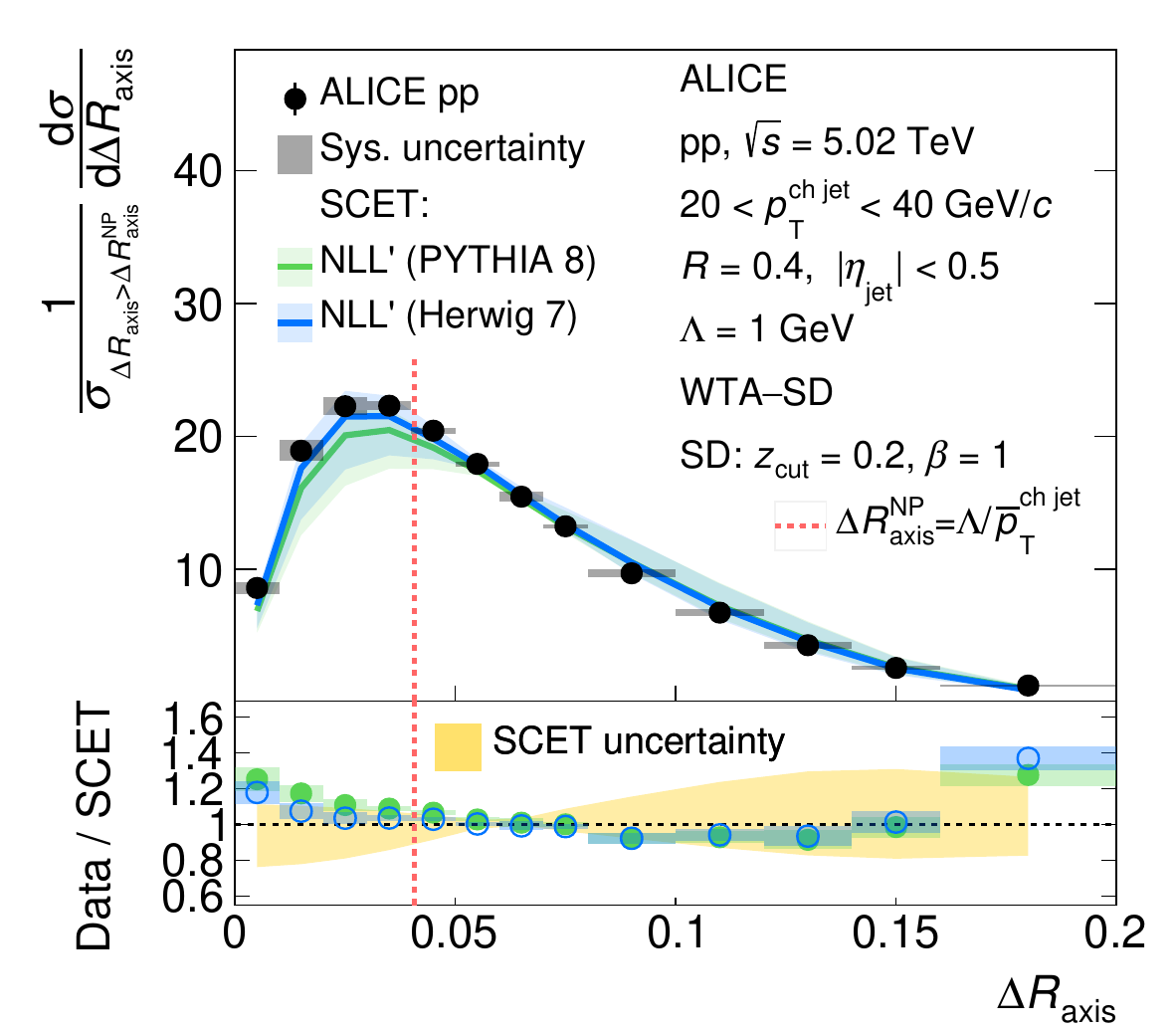}
\includegraphics[width=0.495\textwidth]{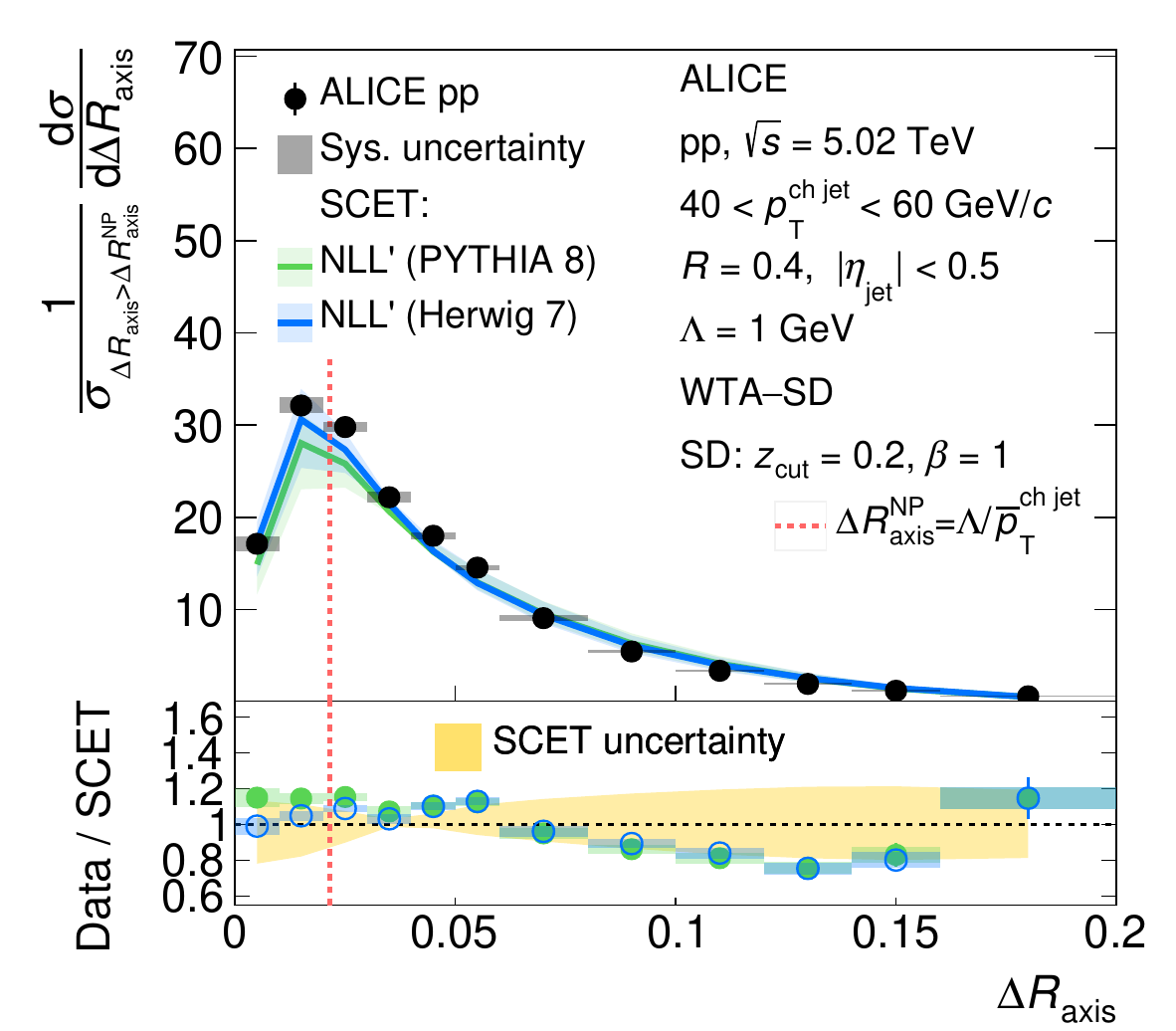}
\includegraphics[width=0.495\textwidth]{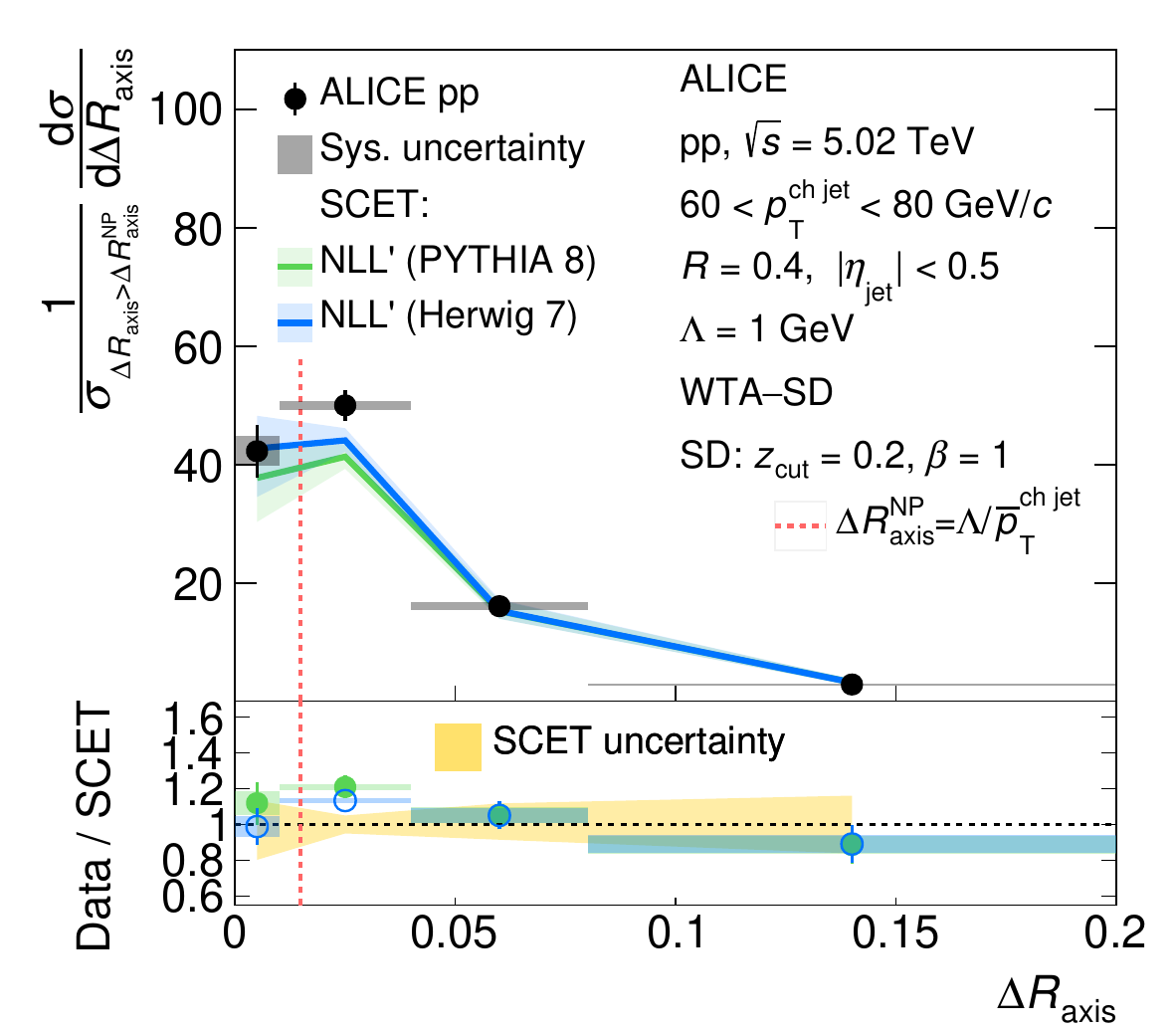}
\includegraphics[width=0.495\textwidth]{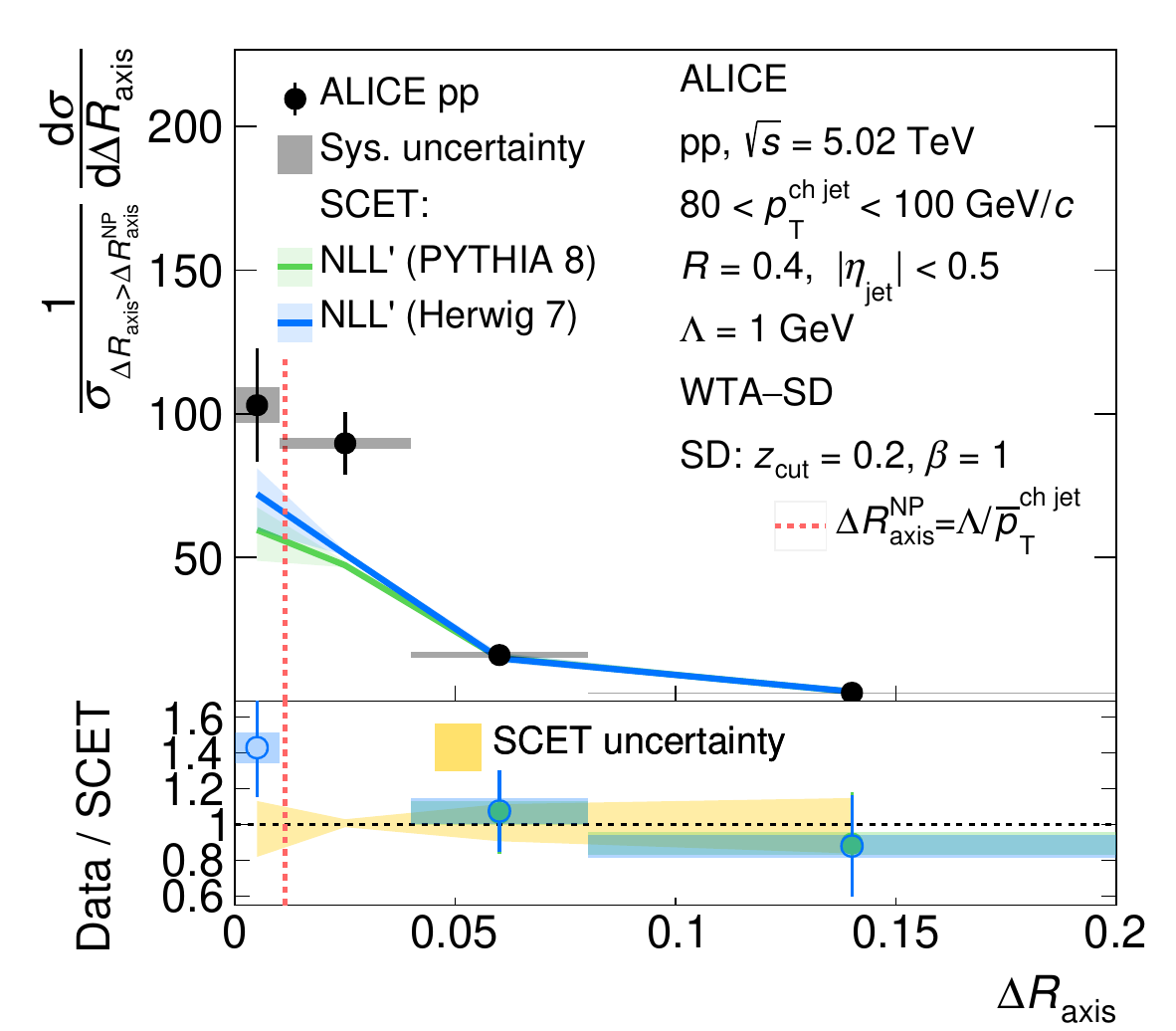}
\caption{Same as Figure~\ref{fig:SCET_Standard_WTA_R04}, for WTA$-$SD ($\zcut=0.2, \, \beta=1$) for $R=0.4$.}
\label{fig:SCET_WTA_SD_6_SD_zcut02_B1_R04}
\end{figure}

\begin{figure}[!htb]
\centering{}
\includegraphics[width=0.495\textwidth]{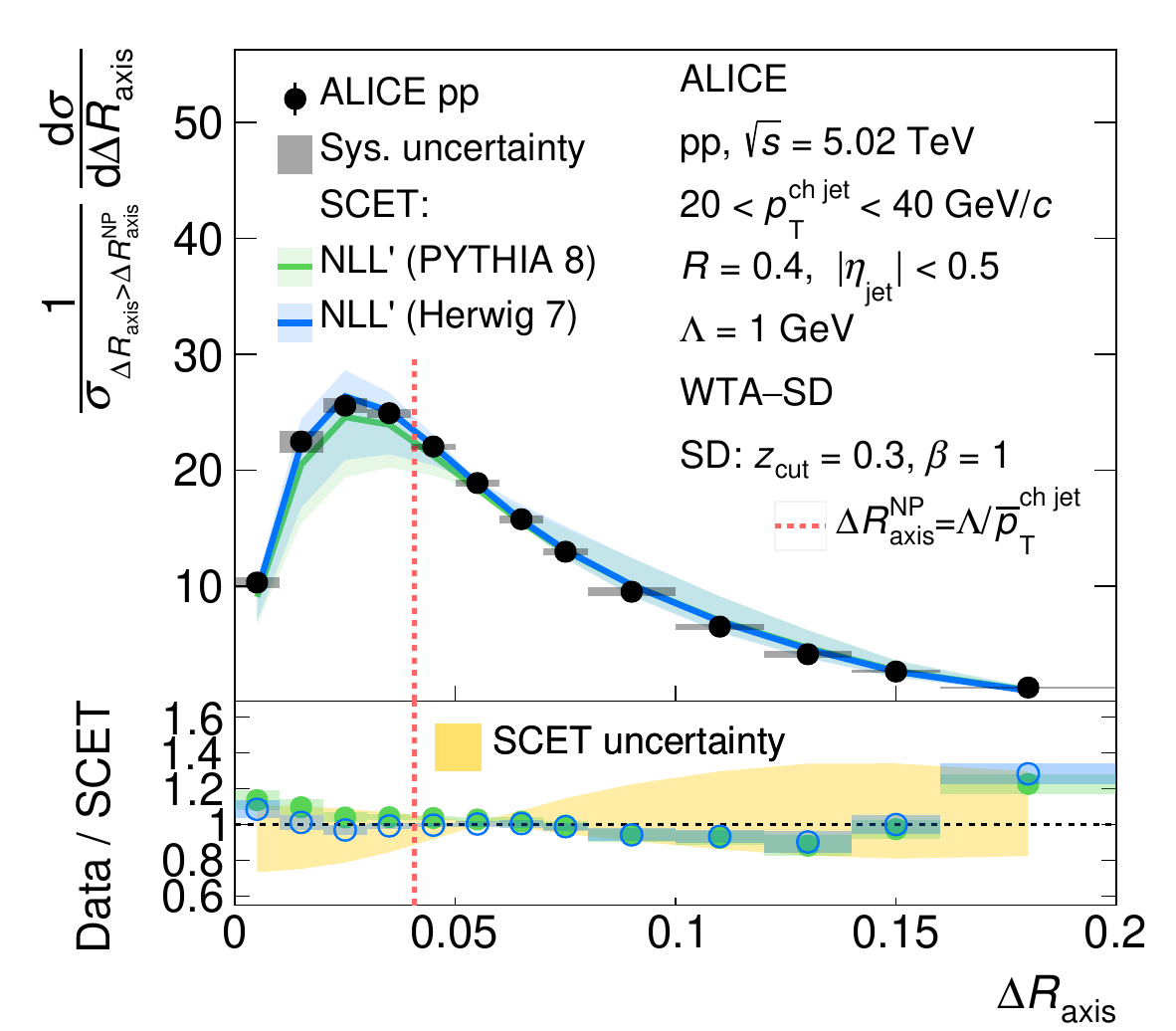}
\includegraphics[width=0.495\textwidth]{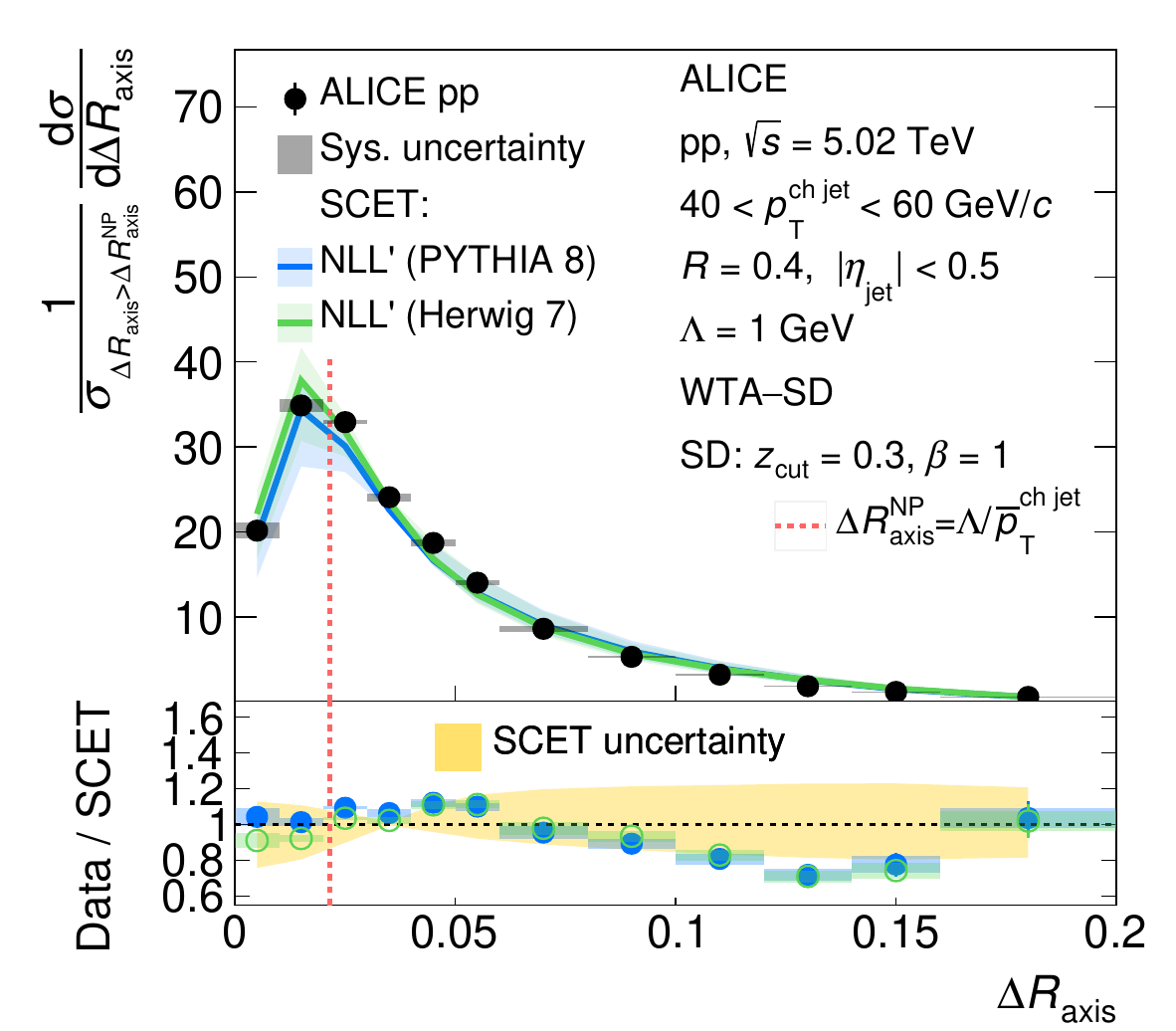}
\includegraphics[width=0.495\textwidth]{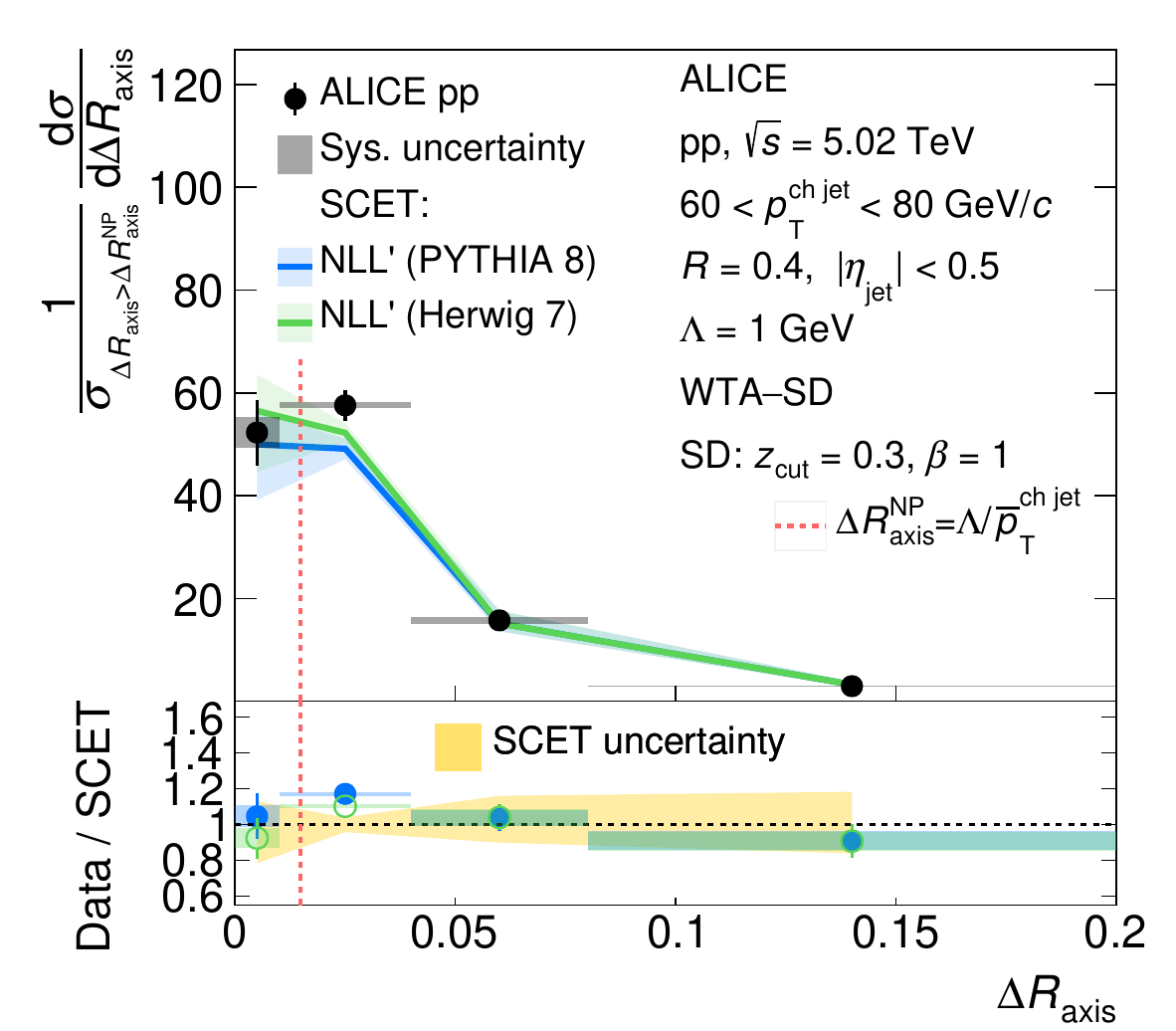}
\includegraphics[width=0.495\textwidth]{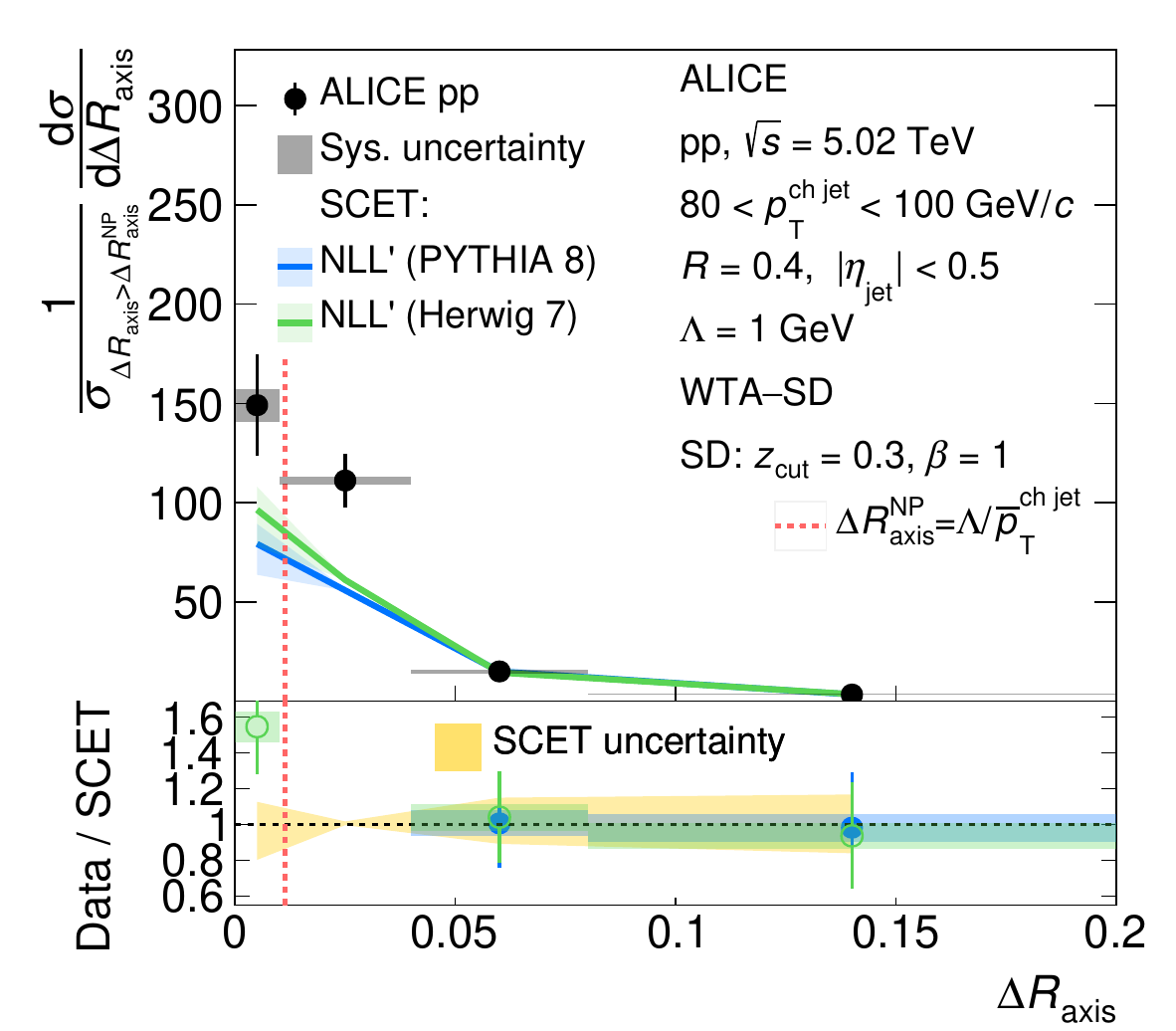}
\caption{Same as Figure~\ref{fig:SCET_Standard_WTA_R04}, for WTA$-$SD ($\zcut=0.3, \, \beta=1$) for $R=0.4$.}
\label{fig:SCET_WTA_SD_10_SD_zcut03_B1_R04}
\end{figure}


\begin{figure}[!htb]
\centering{}
\includegraphics[width=0.495\textwidth]{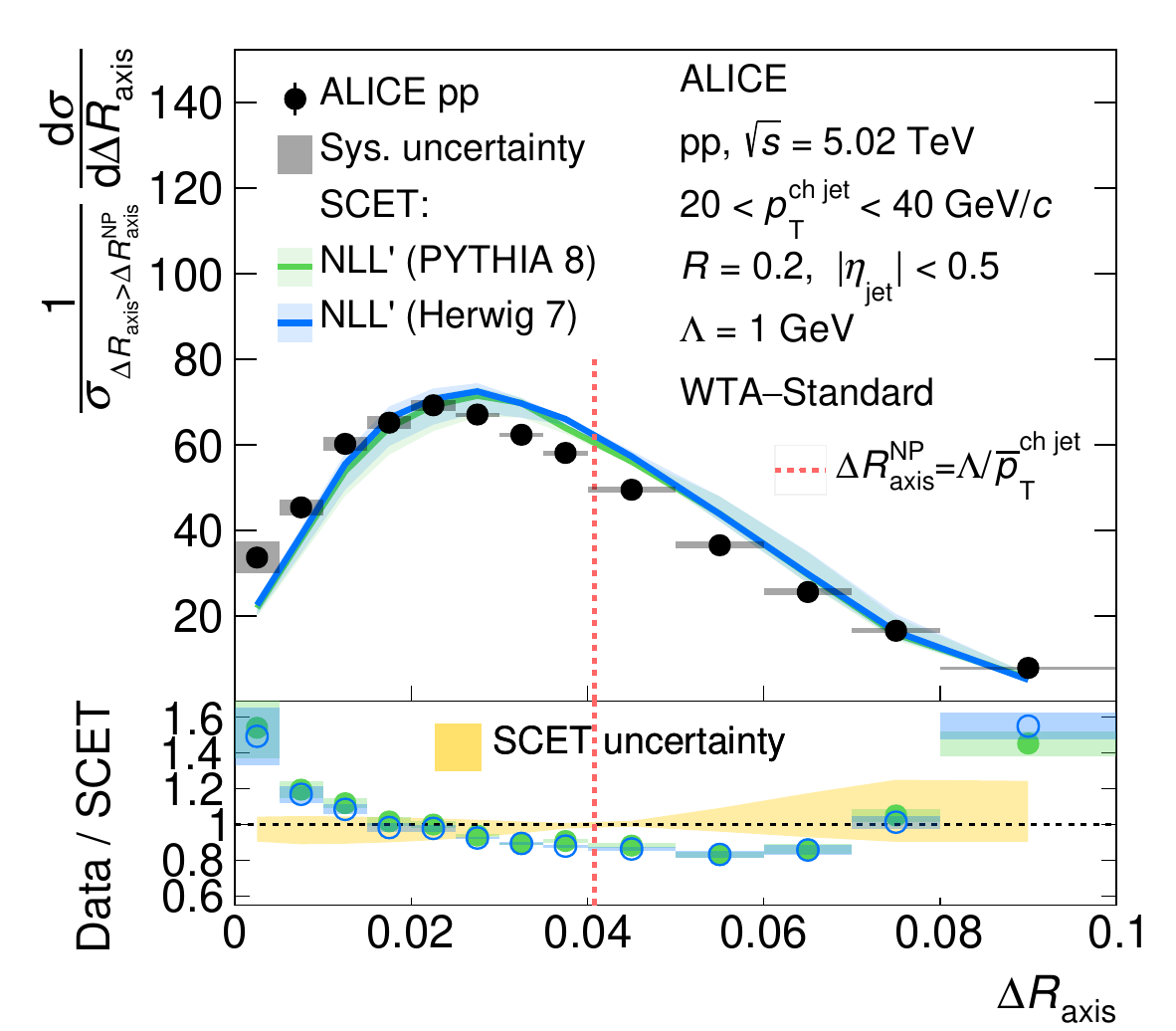}
\includegraphics[width=0.495\textwidth]{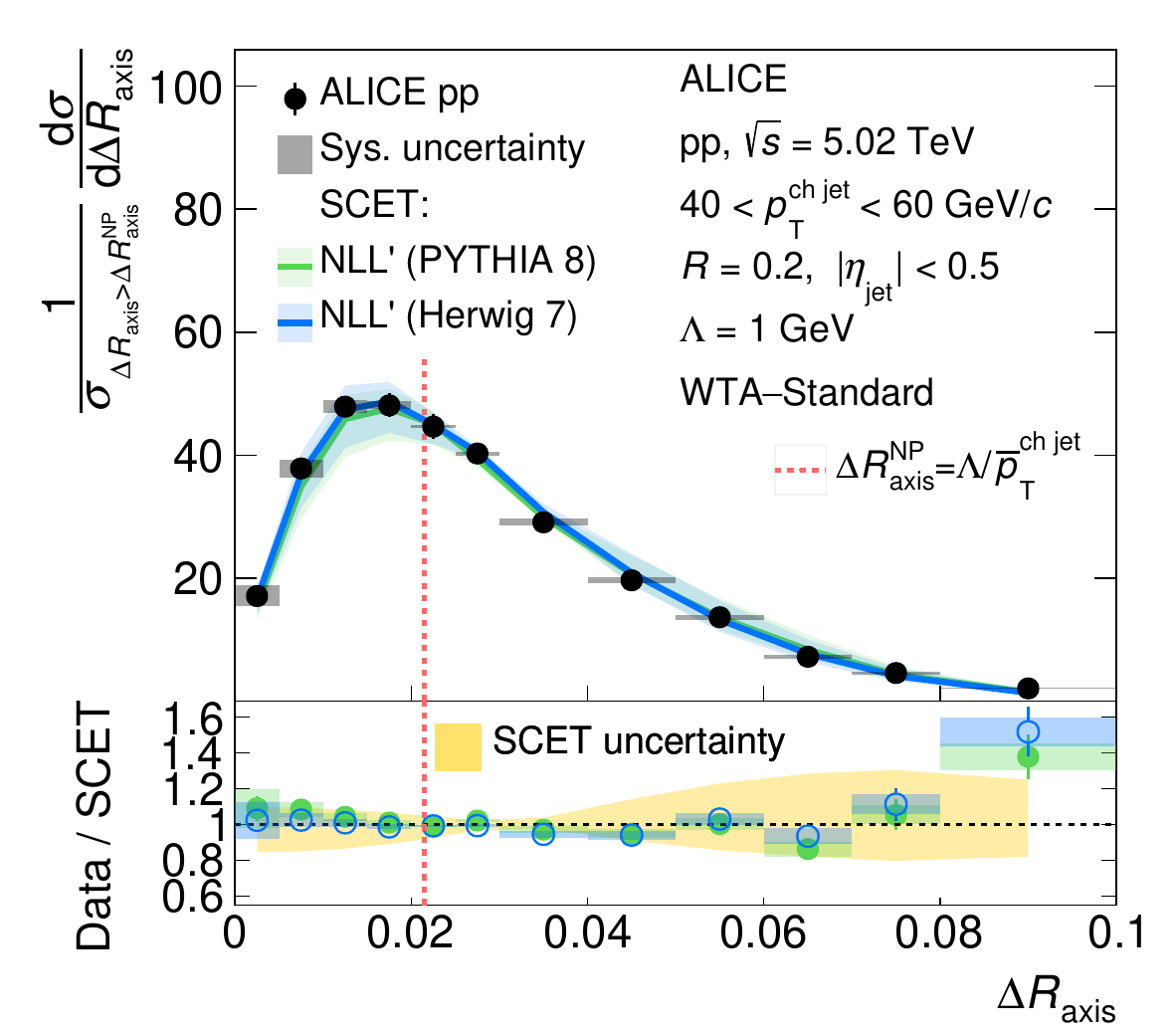}
\includegraphics[width=0.495\textwidth]{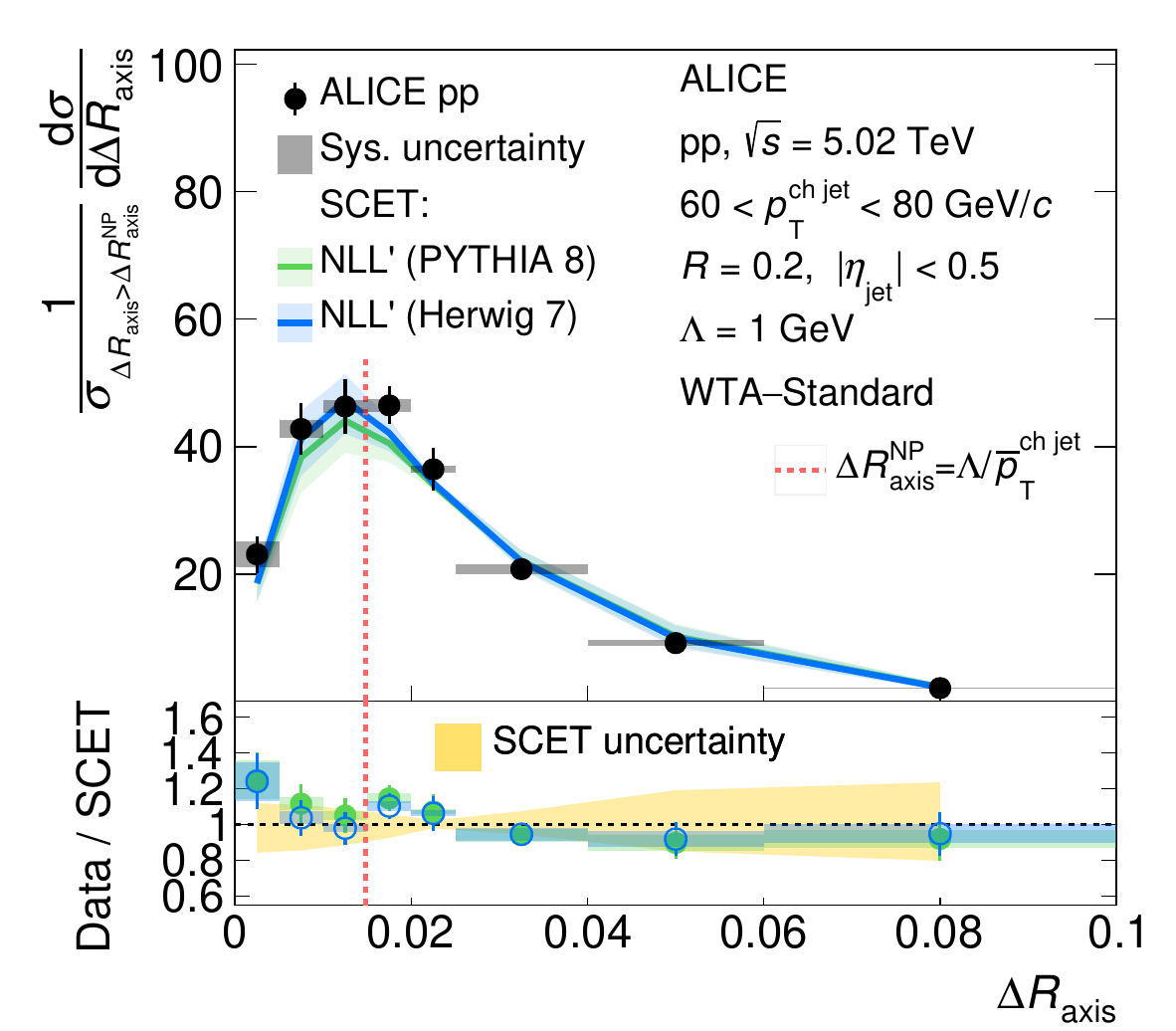}
\includegraphics[width=0.495\textwidth]{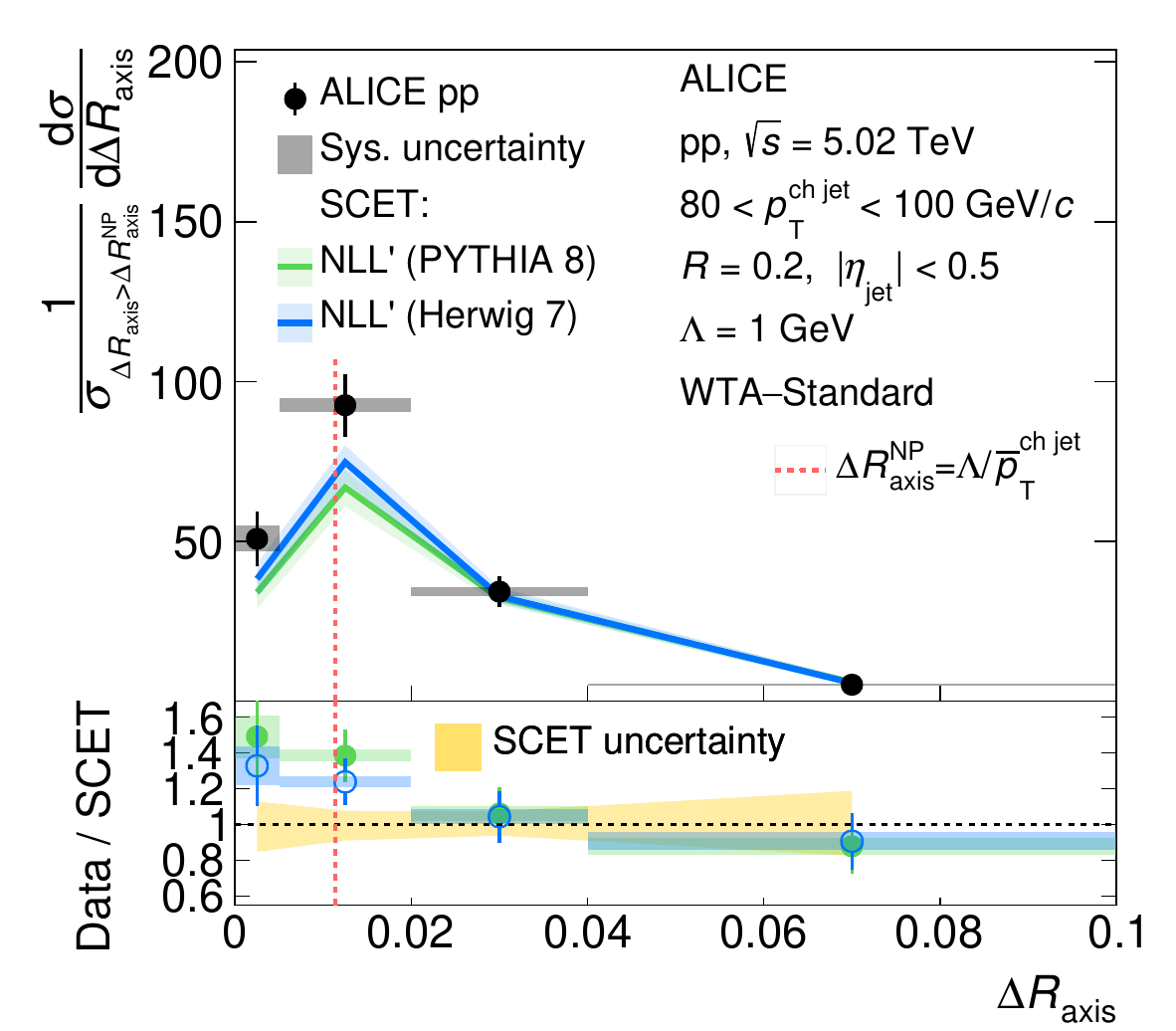}
\caption{Same as Figure~\ref{fig:SCET_Standard_WTA_R04}, for WTA\textendash{}Standard for $R=0.2$.}
\label{fig:SCET_Standard_WTA_R02}
\end{figure}

\begin{figure}[!htb]
\centering{}
\includegraphics[width=0.495\textwidth]{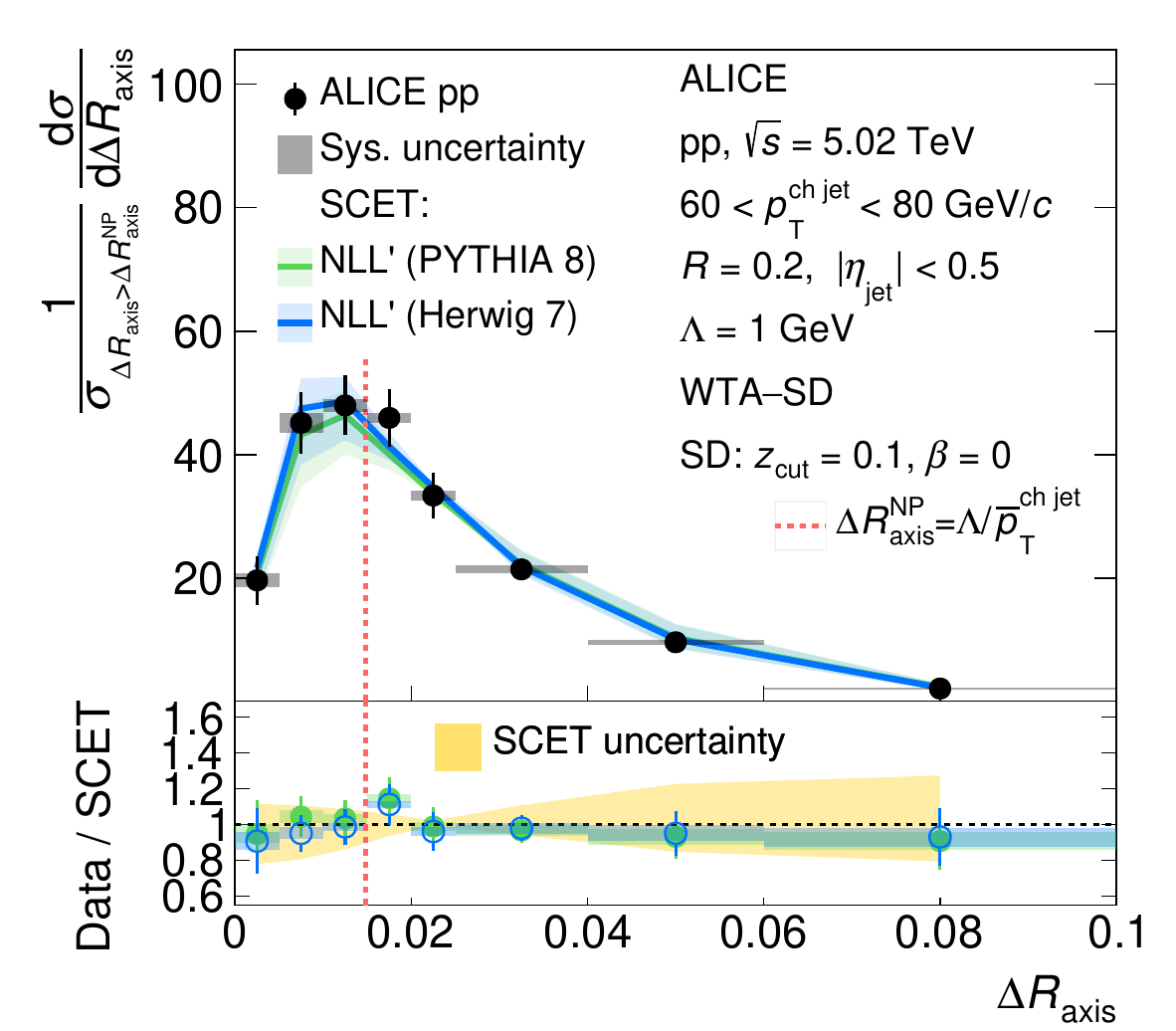}
\includegraphics[width=0.495\textwidth]{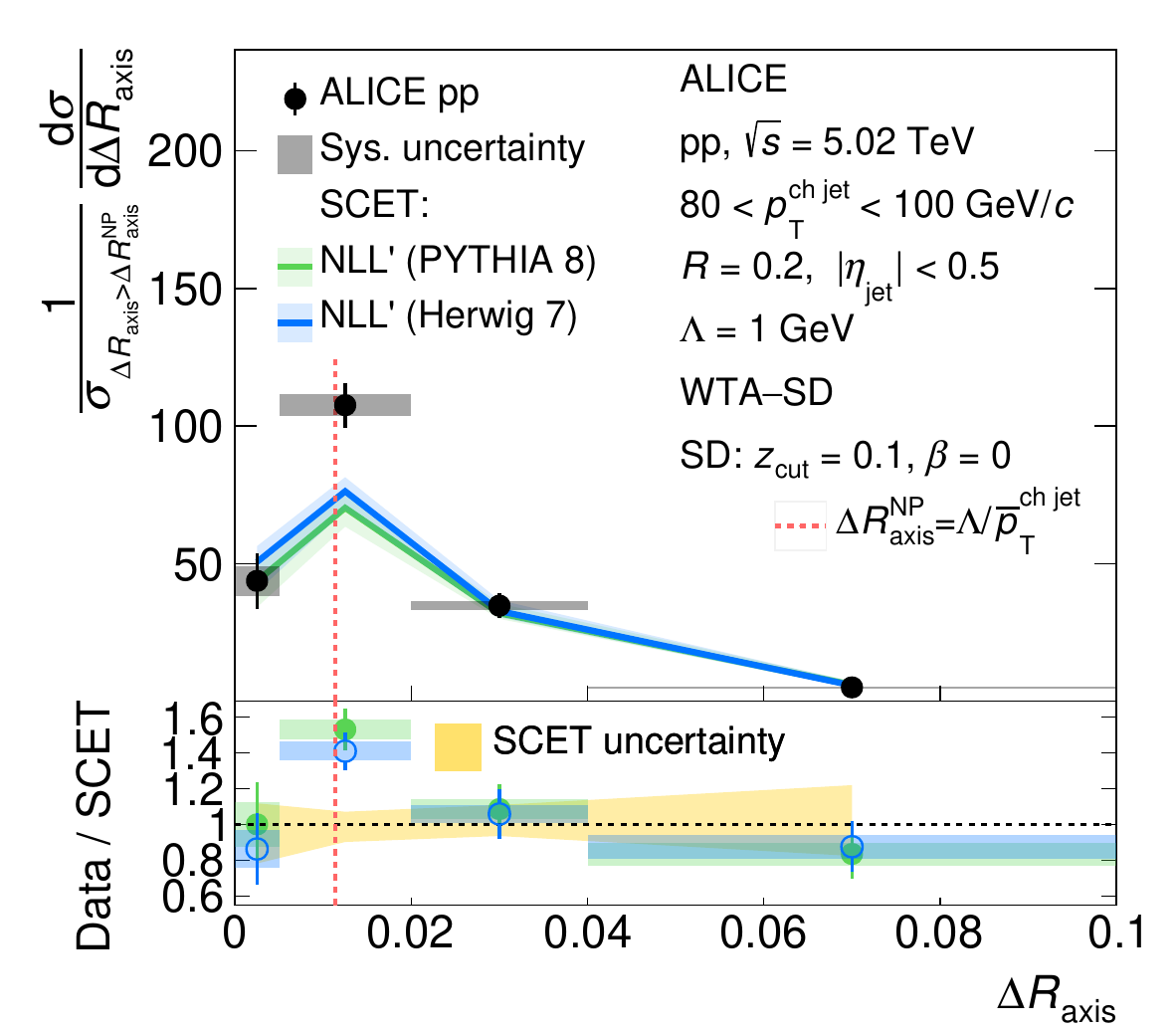}
\caption{Same as Figure~\ref{fig:SCET_Standard_WTA_R04}, for WTA$-$SD ($\zcut=0.1, \, \beta=0$) for $R=0.2$.}
\label{fig:SCET_WTA_SD_1_SD_zcut01_B0_R02}
\end{figure}

\begin{figure}[!htb]
\centering{}
\includegraphics[width=0.495\textwidth]{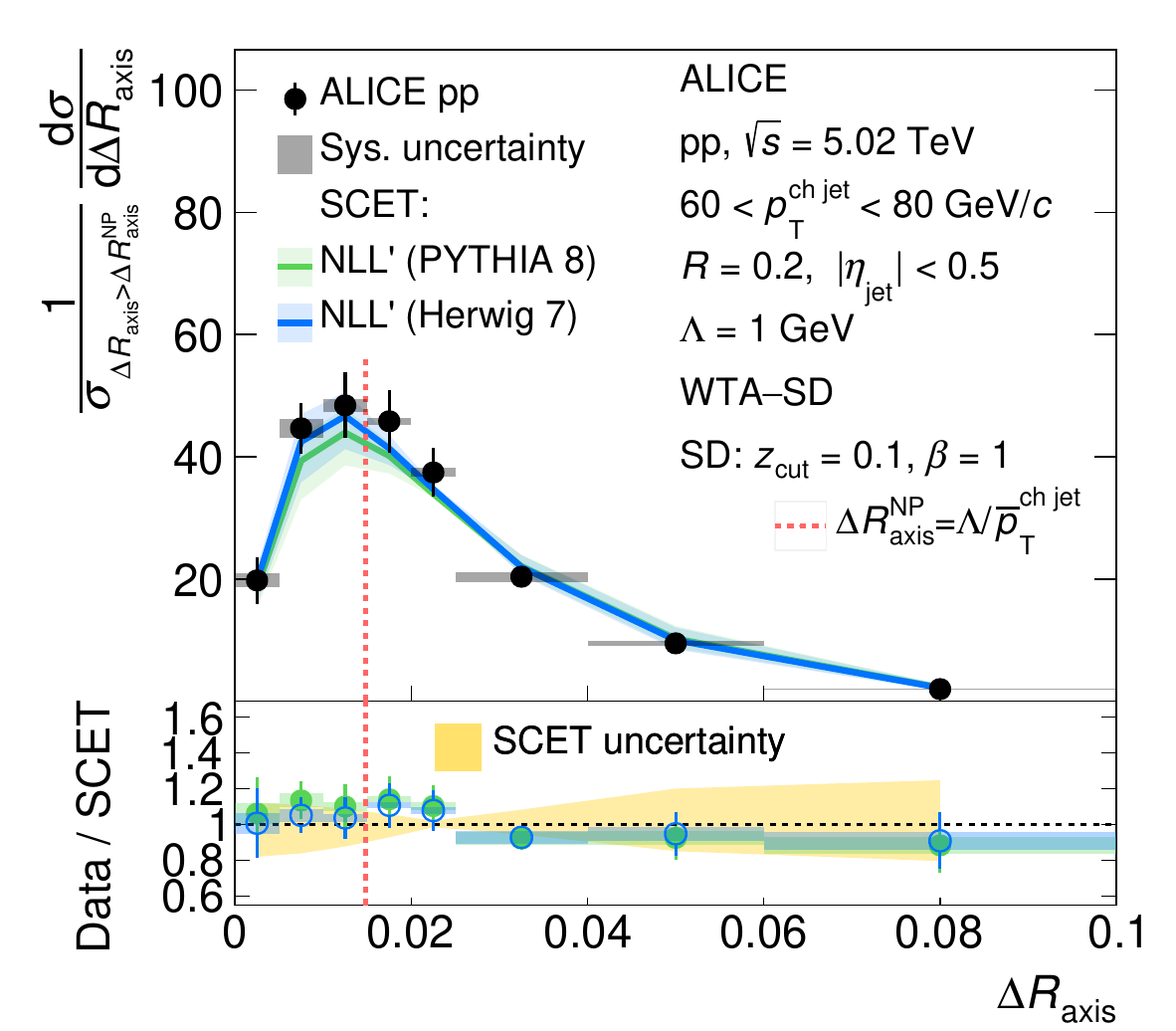}
\includegraphics[width=0.495\textwidth]{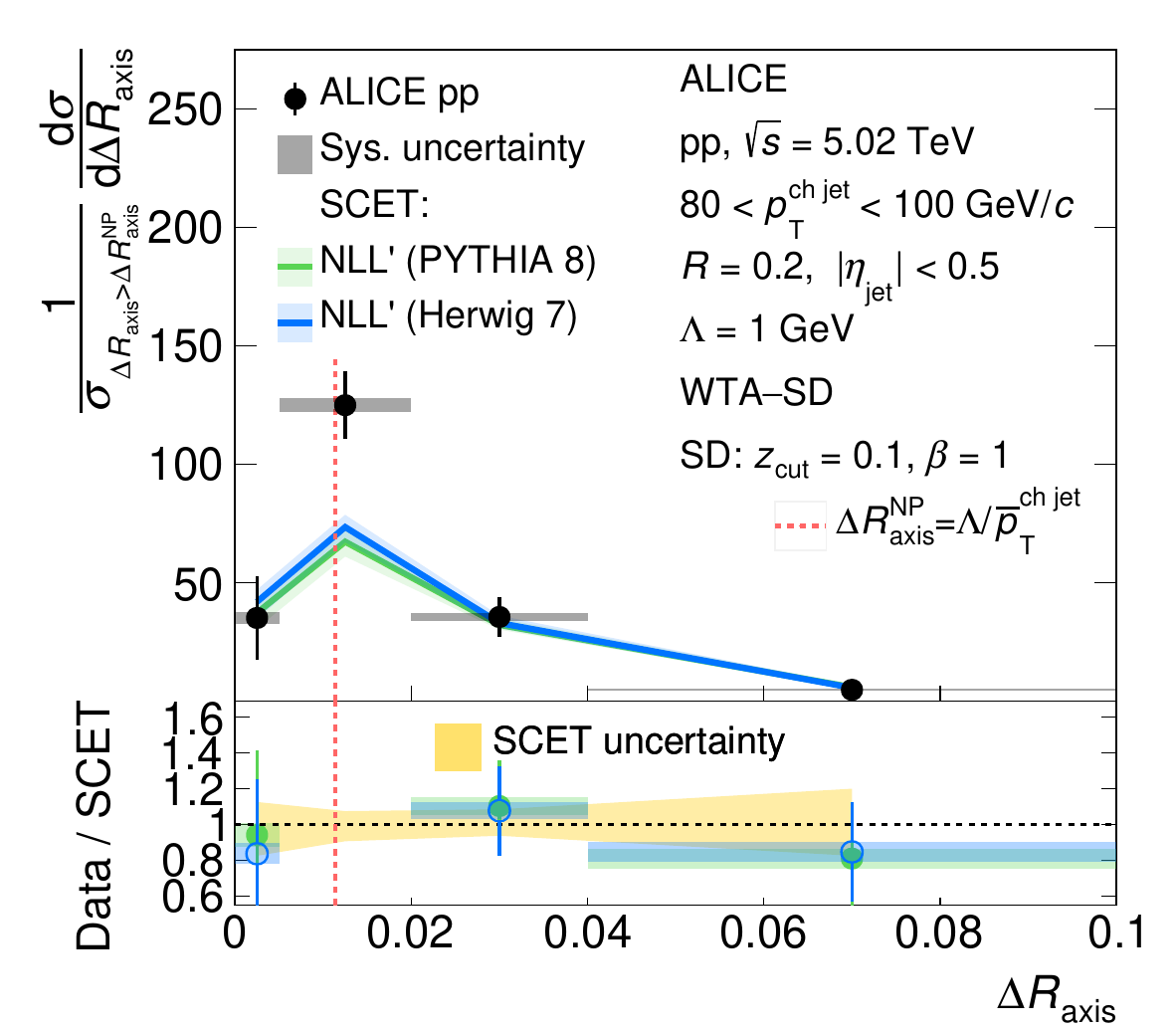}
\caption{Same as Figure~\ref{fig:SCET_Standard_WTA_R04}, for WTA$-$SD ($\zcut=0.1, \, \beta=1$) for $R=0.2$.}
\label{fig:SCET_WTA_SD_2_SD_zcut01_B1_R02}
\end{figure}

\begin{figure}[!htb]
\centering{}
\includegraphics[width=0.495\textwidth]{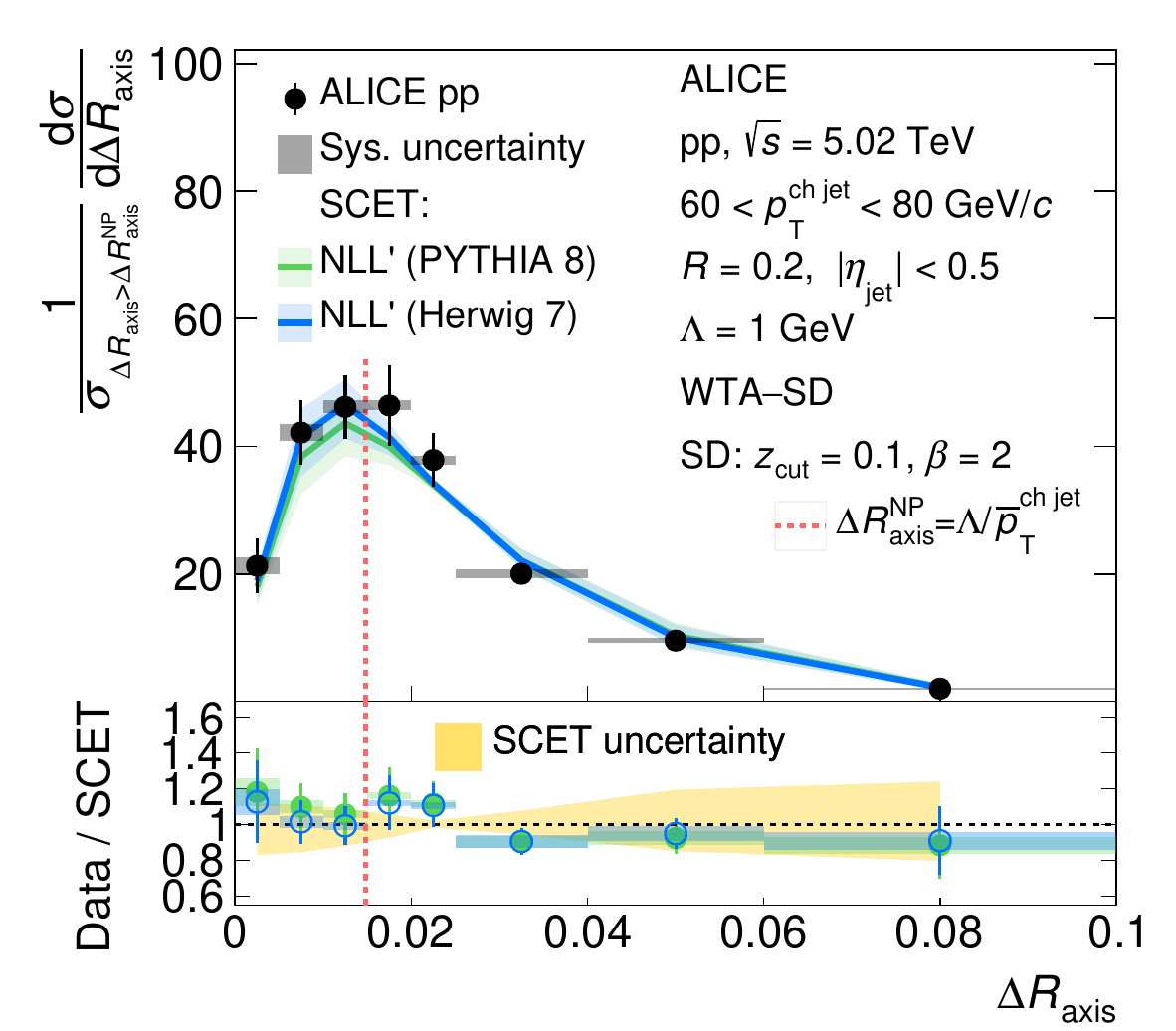}
\includegraphics[width=0.495\textwidth]{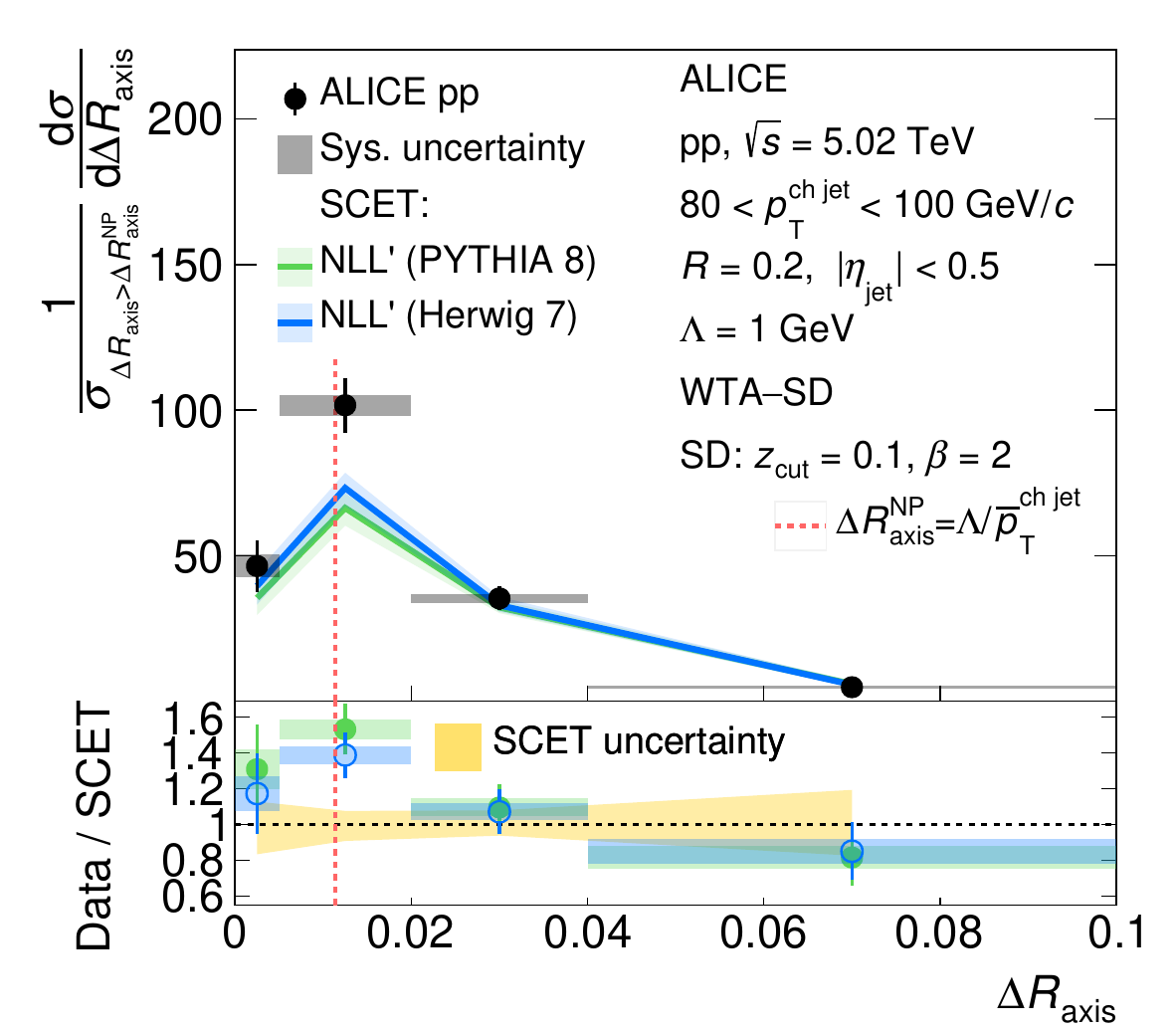}
\caption{Same as Figure~\ref{fig:SCET_Standard_WTA_R04}, for WTA$-$SD ($\zcut=0.1, \, \beta=2$) for $R=0.2$.}
\label{fig:SCET_WTA_SD_3_SD_zcut01_B2_R02}
\end{figure}

\begin{figure}[!htb]
\centering{}
\includegraphics[width=0.495\textwidth]{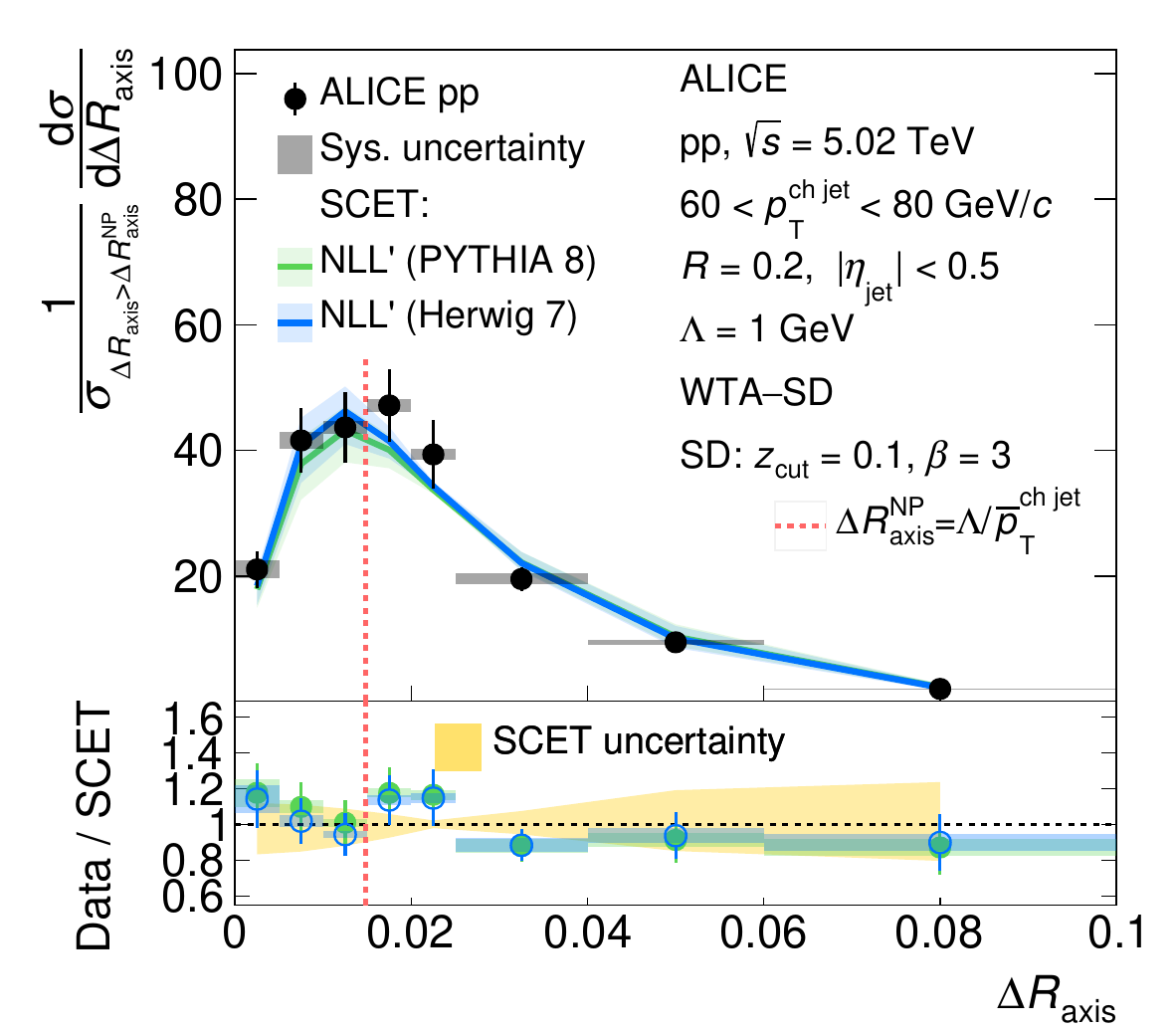}
\includegraphics[width=0.495\textwidth]{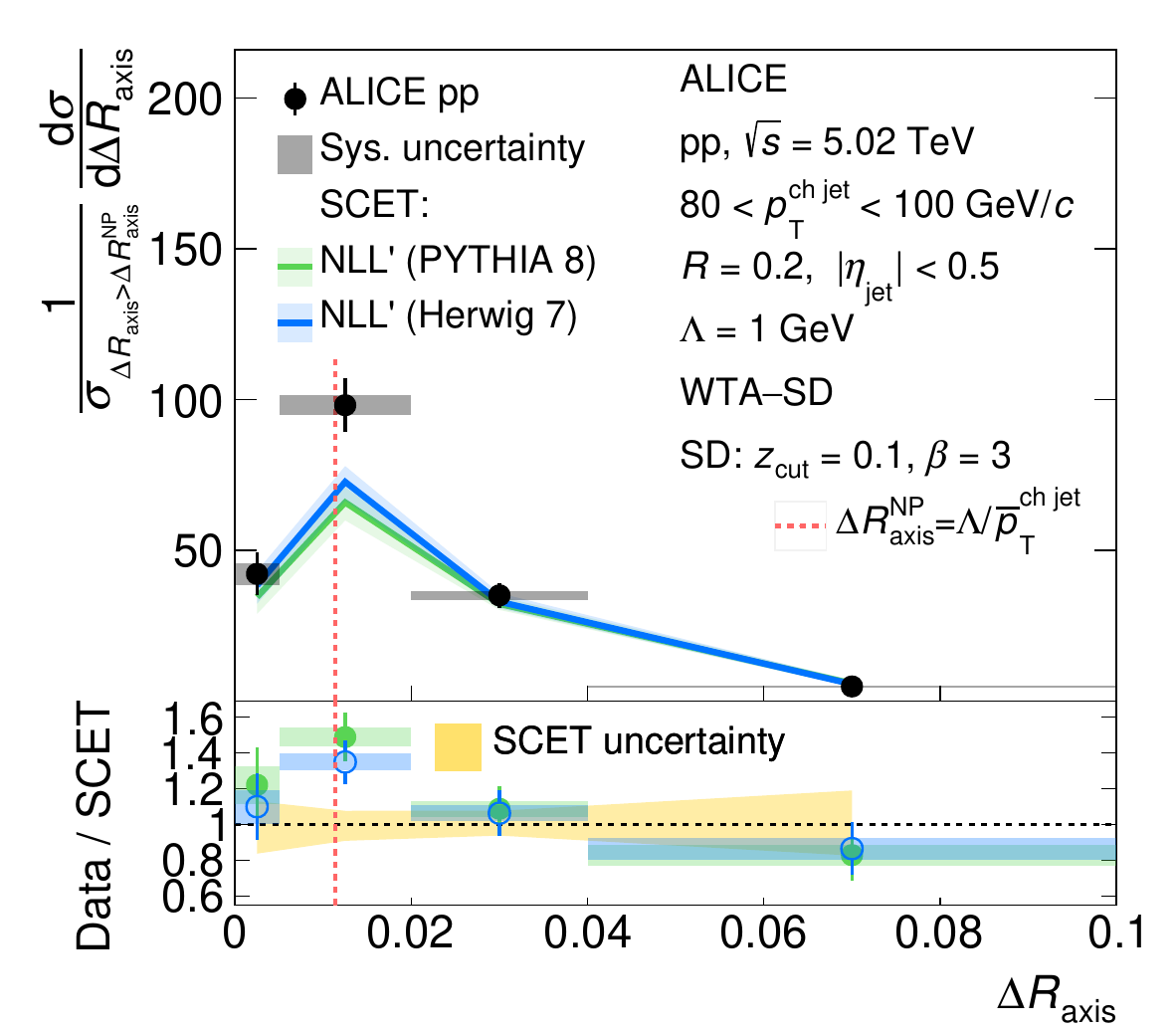}
\caption{Same as Figure~\ref{fig:SCET_Standard_WTA_R04}, for WTA$-$SD ($\zcut=0.1, \, \beta=3$) for $R=0.2$.}
\label{fig:SCET_WTA_SD_4_SD_zcut01_B3_R02}
\end{figure}

\begin{figure}[!htb]
\centering{}
\includegraphics[width=0.495\textwidth]{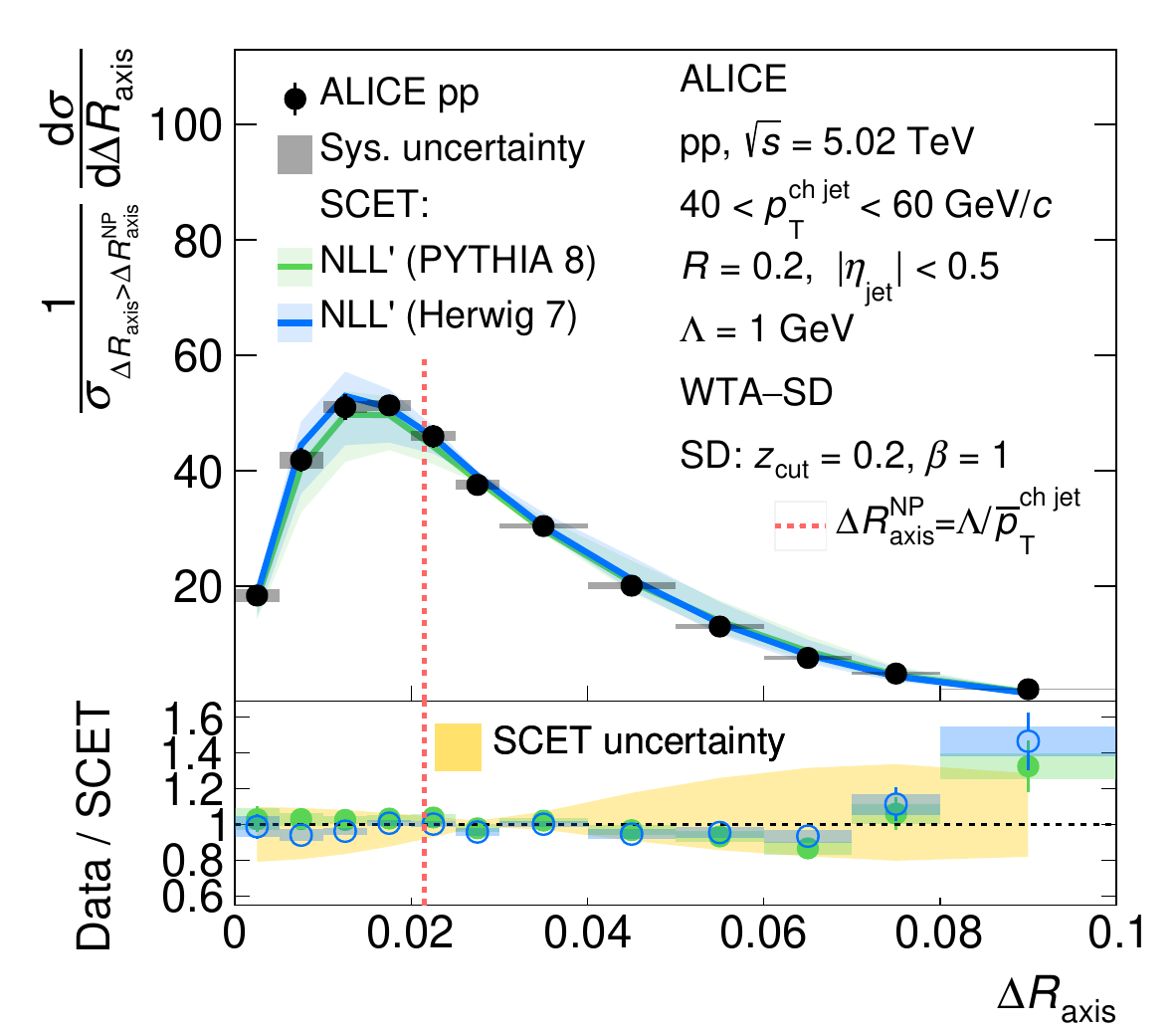}
\includegraphics[width=0.495\textwidth]{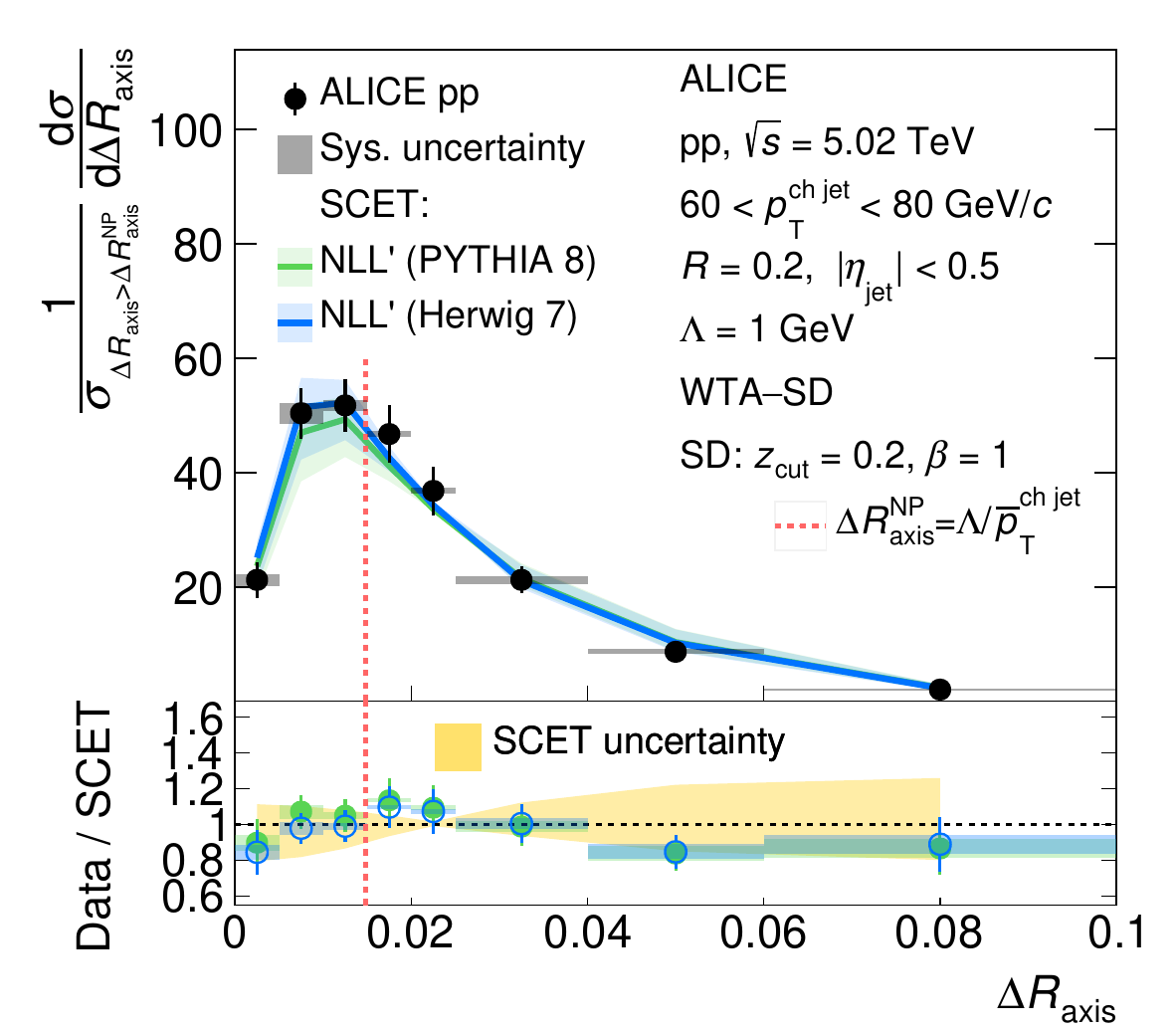}
\includegraphics[width=0.495\textwidth]{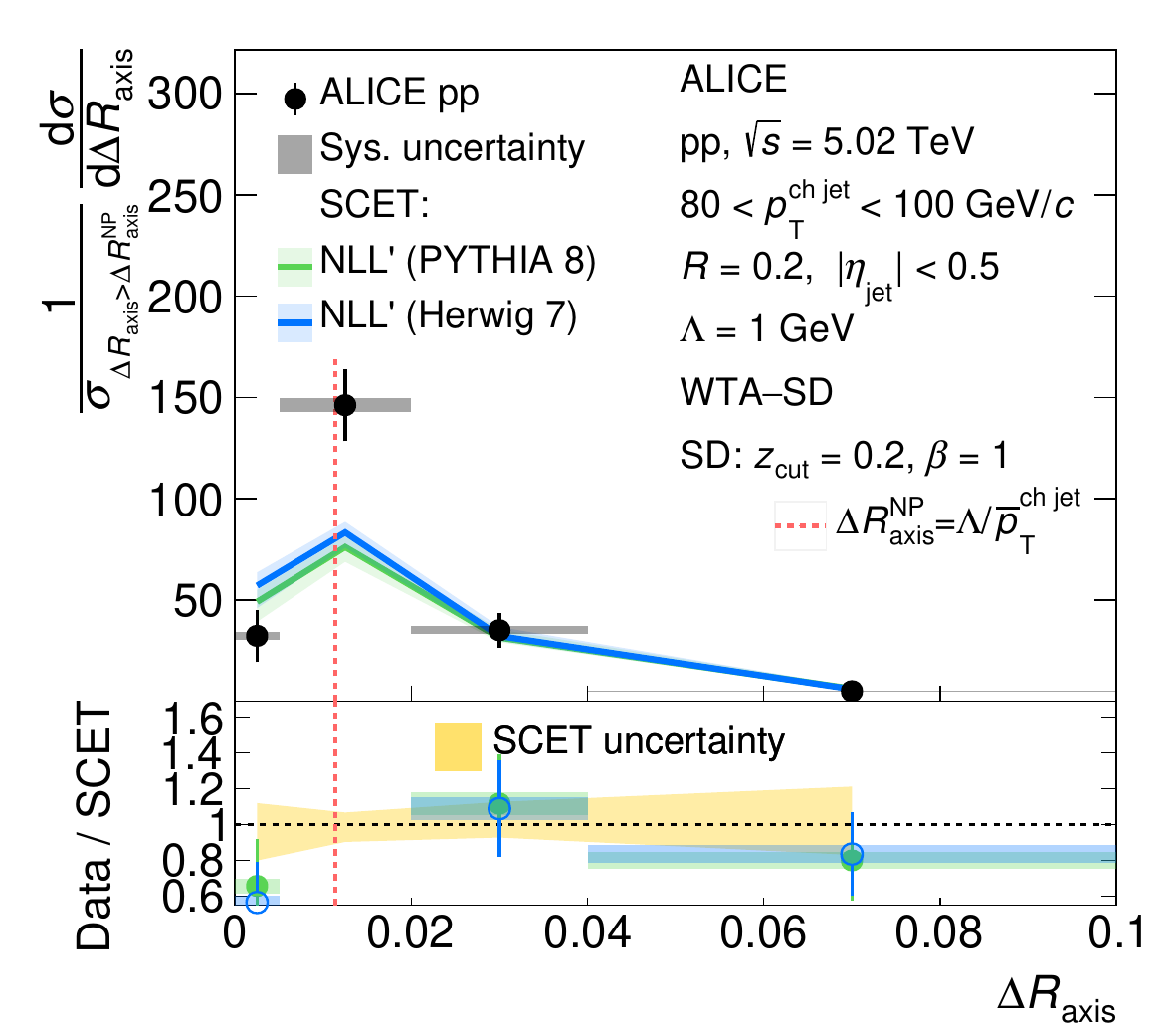}
\caption{Same as Figure~\ref{fig:SCET_Standard_WTA_R04}, for WTA$-$SD ($\zcut=0.2, \, \beta=1$) for $R=0.2$.}
\label{fig:SCET_WTA_SD_6_SD_zcut02_B1_R02}
\end{figure}

\begin{figure}[!htb]
\centering{}
\includegraphics[width=0.495\textwidth]{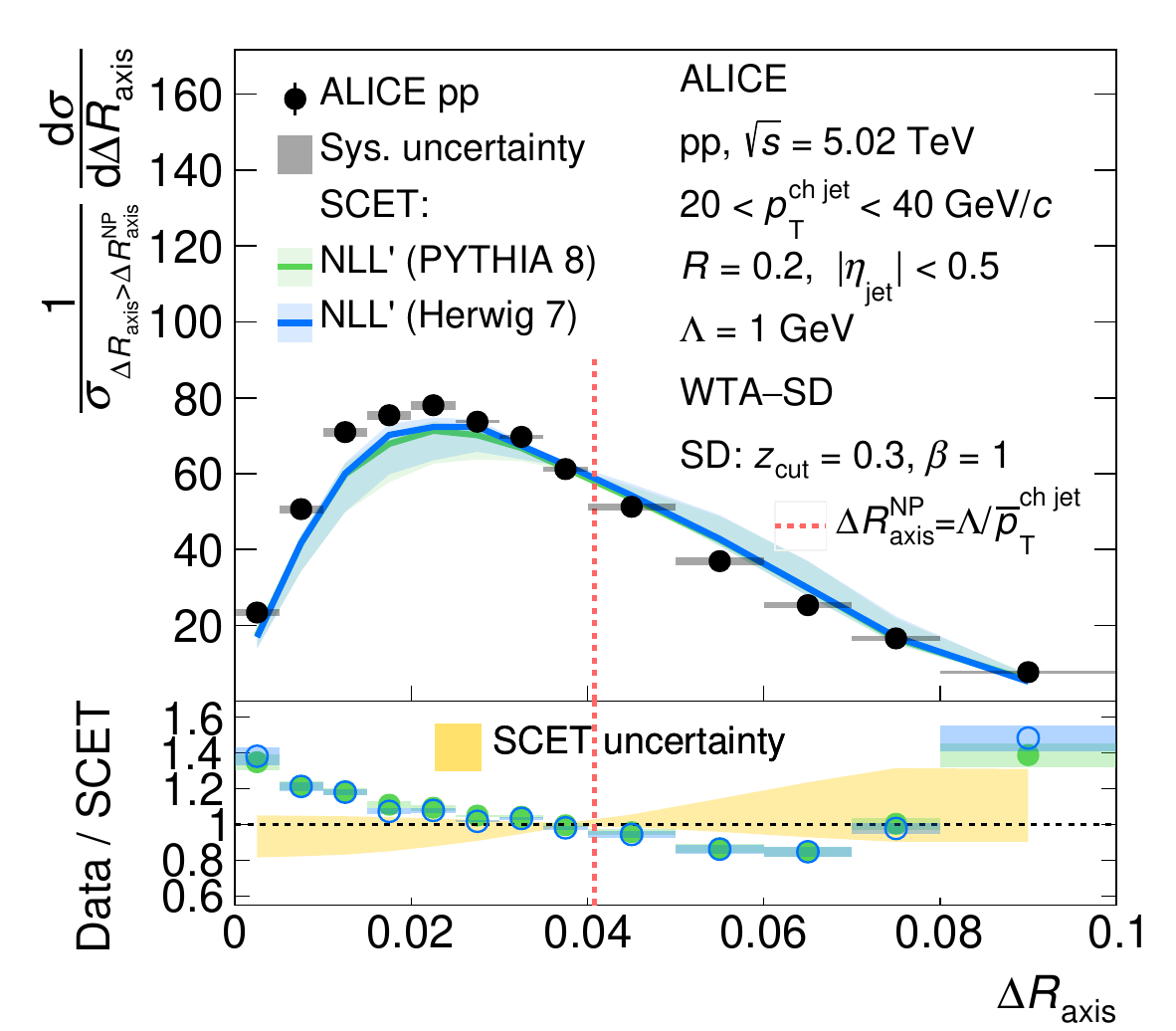}
\includegraphics[width=0.495\textwidth]{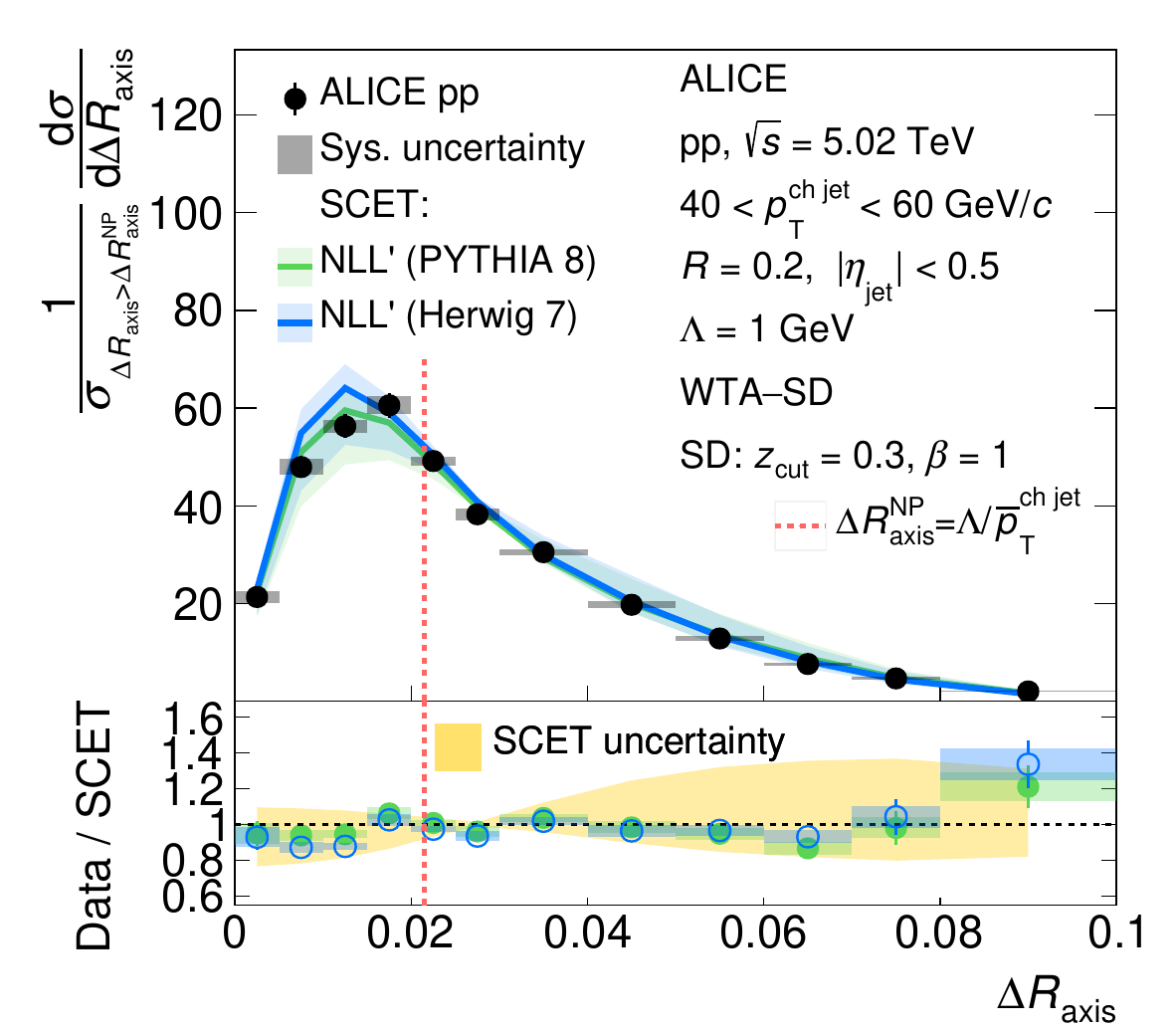}
\includegraphics[width=0.495\textwidth]{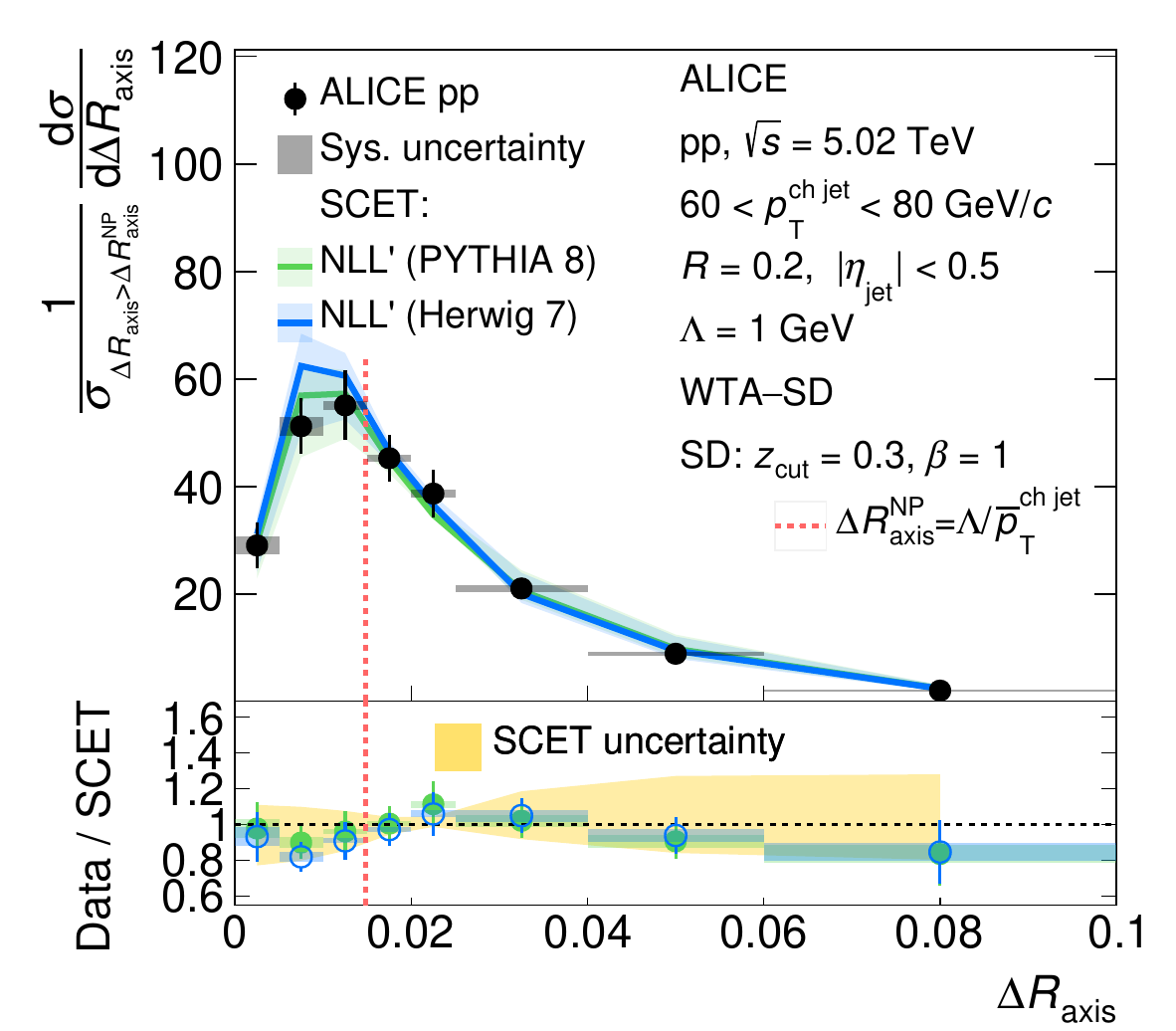}
\includegraphics[width=0.495\textwidth]{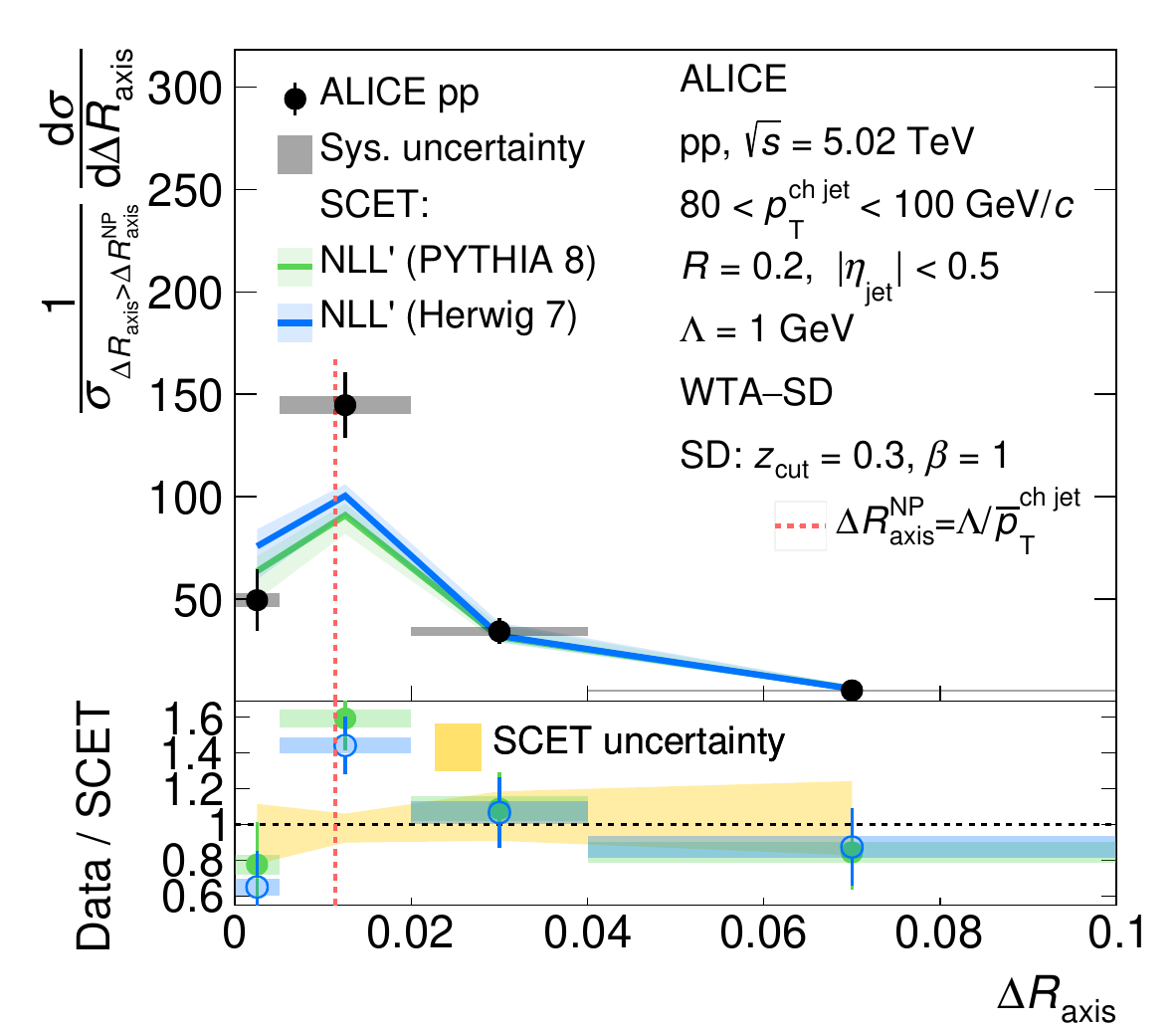}
\caption{Same as Figure~\ref{fig:SCET_Standard_WTA_R04}, for WTA$-$SD ($\zcut=0.3, \, \beta=1$) for $R=0.2$.}
\label{fig:SCET_WTA_SD_10_SD_zcut03_B1_R02}
\end{figure}

%
\clearpage
\section{The ALICE Collaboration}
\label{app:collab}
\begin{flushleft} 
\small

S.~Acharya\,\orcidlink{0000-0002-9213-5329}\,$^{\rm 125}$, 
D.~Adamov\'{a}\,\orcidlink{0000-0002-0504-7428}\,$^{\rm 86}$, 
A.~Adler$^{\rm 69}$, 
G.~Aglieri Rinella\,\orcidlink{0000-0002-9611-3696}\,$^{\rm 32}$, 
M.~Agnello\,\orcidlink{0000-0002-0760-5075}\,$^{\rm 29}$, 
N.~Agrawal\,\orcidlink{0000-0003-0348-9836}\,$^{\rm 50}$, 
Z.~Ahammed\,\orcidlink{0000-0001-5241-7412}\,$^{\rm 132}$, 
S.~Ahmad\,\orcidlink{0000-0003-0497-5705}\,$^{\rm 15}$, 
S.U.~Ahn\,\orcidlink{0000-0001-8847-489X}\,$^{\rm 70}$, 
I.~Ahuja\,\orcidlink{0000-0002-4417-1392}\,$^{\rm 37}$
% ...

\section*{Affiliation Notes}

$^{\rm I}$ Deceased\\
$^{\rm II}$ Also at: Max-Planck-Institut f\"{u}r Physik, Munich, Germany\\
$^{\rm III}$ Also at: Italian National Agency for New Technologies, Energy and Sustainable Economic Development (ENEA), Bologna, Italy\\
$^{\rm IV}$ Also at: Dipartimento DET del Politecnico di Torino, Turin, Italy\\
$^{\rm V}$ Also at: Department of Applied Physics, Aligarh Muslim University, Aligarh, India\\
$^{\rm VI}$ Also at: Institute of Theoretical Physics, University of Wroclaw, Poland\\
$^{\rm VII}$ Also at: An institution covered by a cooperation agreement with CERN\\

\section*{Collaboration Institutes}

$^{1}$ A.I. Alikhanyan National Science Laboratory (Yerevan Physics Institute) Foundation, Yerevan, Armenia\\
$^{2}$ AGH University of Science and Technology, Cracow, Poland\\
$^{3}$ Bogolyubov Institute for Theoretical Physics, National Academy of Sciences of Ukraine, Kiev, Ukraine\\
$^{4}$ Bose Institute, Department of Physics  and Centre for Astroparticle Physics and Space Science (CAPSS), Kolkata, India\\
$^{5}$ California Polytechnic State University, San Luis Obispo, California, United States\\
$^{6}$ Central China Normal University, Wuhan, China\\
$^{7}$ Centro de Aplicaciones Tecnol\'{o}gicas y Desarrollo Nuclear (CEADEN), Havana, Cuba\\
$^{8}$ Centro de Investigaci\'{o}n y de Estudios Avanzados (CINVESTAV), Mexico City and M\'{e}rida, Mexico\\
$^{9}$ Chicago State University, Chicago, Illinois, United States\\
$^{10}$ China Institute of Atomic Energy, Beijing, China\\
$^{11}$ Chungbuk National University, Cheongju, Republic of Korea\\
$^{12}$ Comenius University Bratislava, Faculty of Mathematics, Physics and Informatics, Bratislava, Slovak Republic\\
$^{13}$ COMSATS University Islamabad, Islamabad, Pakistan\\
$^{14}$ Creighton University, Omaha, Nebraska, United States\\
$^{15}$ Department of Physics, Aligarh Muslim University, Aligarh, India\\
$^{16}$ Department of Physics, Pusan National University, Pusan, Republic of Korea\\
$^{17}$ Department of Physics, Sejong University, Seoul, Republic of Korea\\
$^{18}$ Department of Physics, University of California, Berkeley, California, United States\\
$^{19}$ Department of Physics, University of Oslo, Oslo, Norway\\
$^{20}$ Department of Physics and Technology, University of Bergen, Bergen, Norway\\
$^{21}$ Dipartimento di Fisica, Universit\`{a} di Pavia, Pavia, Italy\\
$^{22}$ Dipartimento di Fisica dell'Universit\`{a} and Sezione INFN, Cagliari, Italy\\
$^{23}$ Dipartimento di Fisica dell'Universit\`{a} and Sezione INFN, Trieste, Italy\\
$^{24}$ Dipartimento di Fisica dell'Universit\`{a} and Sezione INFN, Turin, Italy\\
$^{25}$ Dipartimento di Fisica e Astronomia dell'Universit\`{a} and Sezione INFN, Bologna, Italy\\
$^{26}$ Dipartimento di Fisica e Astronomia dell'Universit\`{a} and Sezione INFN, Catania, Italy\\
$^{27}$ Dipartimento di Fisica e Astronomia dell'Universit\`{a} and Sezione INFN, Padova, Italy\\
$^{28}$ Dipartimento di Fisica `E.R.~Caianiello' dell'Universit\`{a} and Gruppo Collegato INFN, Salerno, Italy\\
$^{29}$ Dipartimento DISAT del Politecnico and Sezione INFN, Turin, Italy\\
$^{30}$ Dipartimento di Scienze MIFT, Universit\`{a} di Messina, Messina, Italy\\
$^{31}$ Dipartimento Interateneo di Fisica `M.~Merlin' and Sezione INFN, Bari, Italy\\
$^{32}$ European Organization for Nuclear Research (CERN), Geneva, Switzerland\\
$^{33}$ Faculty of Electrical Engineering, Mechanical Engineering and Naval Architecture, University of Split, Split, Croatia\\
$^{34}$ Faculty of Engineering and Science, Western Norway University of Applied Sciences, Bergen, Norway\\
$^{35}$ Faculty of Nuclear Sciences and Physical Engineering, Czech Technical University in Prague, Prague, Czech Republic\\
$^{36}$ Faculty of Physics, Sofia University, Sofia, Bulgaria\\
$^{37}$ Faculty of Science, P.J.~\v{S}af\'{a}rik University, Ko\v{s}ice, Slovak Republic\\
$^{38}$ Frankfurt Institute for Advanced Studies, Johann Wolfgang Goethe-Universit\"{a}t Frankfurt, Frankfurt, Germany\\
$^{39}$ Fudan University, Shanghai, China\\
$^{40}$ Gangneung-Wonju National University, Gangneung, Republic of Korea\\
$^{41}$ Gauhati University, Department of Physics, Guwahati, India\\
$^{42}$ Helmholtz-Institut f\"{u}r Strahlen- und Kernphysik, Rheinische Friedrich-Wilhelms-Universit\"{a}t Bonn, Bonn, Germany\\
$^{43}$ Helsinki Institute of Physics (HIP), Helsinki, Finland\\
$^{44}$ High Energy Physics Group,  Universidad Aut\'{o}noma de Puebla, Puebla, Mexico\\
$^{45}$ Horia Hulubei National Institute of Physics and Nuclear Engineering, Bucharest, Romania\\
$^{46}$ Indian Institute of Technology Bombay (IIT), Mumbai, India\\
$^{47}$ Indian Institute of Technology Indore, Indore, India\\
$^{48}$ INFN, Laboratori Nazionali di Frascati, Frascati, Italy\\
$^{49}$ INFN, Sezione di Bari, Bari, Italy\\
$^{50}$ INFN, Sezione di Bologna, Bologna, Italy\\
$^{51}$ INFN, Sezione di Cagliari, Cagliari, Italy\\
$^{52}$ INFN, Sezione di Catania, Catania, Italy\\
$^{53}$ INFN, Sezione di Padova, Padova, Italy\\
$^{54}$ INFN, Sezione di Pavia, Pavia, Italy\\
$^{55}$ INFN, Sezione di Torino, Turin, Italy\\
$^{56}$ INFN, Sezione di Trieste, Trieste, Italy\\
$^{57}$ Inha University, Incheon, Republic of Korea\\
$^{58}$ Institute for Gravitational and Subatomic Physics (GRASP), Utrecht University/Nikhef, Utrecht, Netherlands\\
$^{59}$ Institute of Experimental Physics, Slovak Academy of Sciences, Ko\v{s}ice, Slovak Republic\\
$^{60}$ Institute of Physics, Homi Bhabha National Institute, Bhubaneswar, India\\
$^{61}$ Institute of Physics of the Czech Academy of Sciences, Prague, Czech Republic\\
$^{62}$ Institute of Space Science (ISS), Bucharest, Romania\\
$^{63}$ Institut f\"{u}r Kernphysik, Johann Wolfgang Goethe-Universit\"{a}t Frankfurt, Frankfurt, Germany\\
$^{64}$ Instituto de Ciencias Nucleares, Universidad Nacional Aut\'{o}noma de M\'{e}xico, Mexico City, Mexico\\
$^{65}$ Instituto de F\'{i}sica, Universidade Federal do Rio Grande do Sul (UFRGS), Porto Alegre, Brazil\\
$^{66}$ Instituto de F\'{\i}sica, Universidad Nacional Aut\'{o}noma de M\'{e}xico, Mexico City, Mexico\\
$^{67}$ iThemba LABS, National Research Foundation, Somerset West, South Africa\\
$^{68}$ Jeonbuk National University, Jeonju, Republic of Korea\\
$^{69}$ Johann-Wolfgang-Goethe Universit\"{a}t Frankfurt Institut f\"{u}r Informatik, Fachbereich Informatik und Mathematik, Frankfurt, Germany\\
$^{70}$ Korea Institute of Science and Technology Information, Daejeon, Republic of Korea\\
$^{71}$ KTO Karatay University, Konya, Turkey\\
$^{72}$ Laboratoire de Physique des 2 Infinis, Ir\`{e}ne Joliot-Curie, Orsay, France\\
$^{73}$ Laboratoire de Physique Subatomique et de Cosmologie, Universit\'{e} Grenoble-Alpes, CNRS-IN2P3, Grenoble, France\\
$^{74}$ Lawrence Berkeley National Laboratory, Berkeley, California, United States\\
$^{75}$ Lund University Department of Physics, Division of Particle Physics, Lund, Sweden\\
$^{76}$ Nagasaki Institute of Applied Science, Nagasaki, Japan\\
$^{77}$ Nara Women{'}s University (NWU), Nara, Japan\\
$^{78}$ National and Kapodistrian University of Athens, School of Science, Department of Physics , Athens, Greece\\
$^{79}$ National Centre for Nuclear Research, Warsaw, Poland\\
$^{80}$ National Institute of Science Education and Research, Homi Bhabha National Institute, Jatni, India\\
$^{81}$ National Nuclear Research Center, Baku, Azerbaijan\\
$^{82}$ National Research and Innovation Agency - BRIN, Jakarta, Indonesia\\
$^{83}$ Niels Bohr Institute, University of Copenhagen, Copenhagen, Denmark\\
$^{84}$ Nikhef, National institute for subatomic physics, Amsterdam, Netherlands\\
$^{85}$ Nuclear Physics Group, STFC Daresbury Laboratory, Daresbury, United Kingdom\\
$^{86}$ Nuclear Physics Institute of the Czech Academy of Sciences, Husinec-\v{R}e\v{z}, Czech Republic\\
$^{87}$ Oak Ridge National Laboratory, Oak Ridge, Tennessee, United States\\
$^{88}$ Ohio State University, Columbus, Ohio, United States\\
$^{89}$ Physics department, Faculty of science, University of Zagreb, Zagreb, Croatia\\
$^{90}$ Physics Department, Panjab University, Chandigarh, India\\
$^{91}$ Physics Department, University of Jammu, Jammu, India\\
$^{92}$ Physics Program and International Institute for Sustainability with Knotted Chiral Meta Matter (SKCM2), Hiroshima University, Hiroshima, Japan\\
$^{93}$ Physikalisches Institut, Eberhard-Karls-Universit\"{a}t T\"{u}bingen, T\"{u}bingen, Germany\\
$^{94}$ Physikalisches Institut, Ruprecht-Karls-Universit\"{a}t Heidelberg, Heidelberg, Germany\\
$^{95}$ Physik Department, Technische Universit\"{a}t M\"{u}nchen, Munich, Germany\\
$^{96}$ Politecnico di Bari and Sezione INFN, Bari, Italy\\
$^{97}$ Research Division and ExtreMe Matter Institute EMMI, GSI Helmholtzzentrum f\"ur Schwerionenforschung GmbH, Darmstadt, Germany\\
$^{98}$ Saga University, Saga, Japan\\
$^{99}$ Saha Institute of Nuclear Physics, Homi Bhabha National Institute, Kolkata, India\\
$^{100}$ School of Physics and Astronomy, University of Birmingham, Birmingham, United Kingdom\\
$^{101}$ Secci\'{o}n F\'{\i}sica, Departamento de Ciencias, Pontificia Universidad Cat\'{o}lica del Per\'{u}, Lima, Peru\\
$^{102}$ Stefan Meyer Institut f\"{u}r Subatomare Physik (SMI), Vienna, Austria\\
$^{103}$ SUBATECH, IMT Atlantique, Nantes Universit\'{e}, CNRS-IN2P3, Nantes, France\\
$^{104}$ Sungkyunkwan University, Suwon City, Republic of Korea\\
$^{105}$ Suranaree University of Technology, Nakhon Ratchasima, Thailand\\
$^{106}$ Technical University of Ko\v{s}ice, Ko\v{s}ice, Slovak Republic\\
$^{107}$ The Henryk Niewodniczanski Institute of Nuclear Physics, Polish Academy of Sciences, Cracow, Poland\\
$^{108}$ The University of Texas at Austin, Austin, Texas, United States\\
$^{109}$ Universidad Aut\'{o}noma de Sinaloa, Culiac\'{a}n, Mexico\\
$^{110}$ Universidade de S\~{a}o Paulo (USP), S\~{a}o Paulo, Brazil\\
$^{111}$ Universidade Estadual de Campinas (UNICAMP), Campinas, Brazil\\
$^{112}$ Universidade Federal do ABC, Santo Andre, Brazil\\
$^{113}$ University of Cape Town, Cape Town, South Africa\\
$^{114}$ University of Houston, Houston, Texas, United States\\
$^{115}$ University of Jyv\"{a}skyl\"{a}, Jyv\"{a}skyl\"{a}, Finland\\
$^{116}$ University of Kansas, Lawrence, Kansas, United States\\
$^{117}$ University of Liverpool, Liverpool, United Kingdom\\
$^{118}$ University of Science and Technology of China, Hefei, China\\
$^{119}$ University of South-Eastern Norway, Kongsberg, Norway\\
$^{120}$ University of Tennessee, Knoxville, Tennessee, United States\\
$^{121}$ University of the Witwatersrand, Johannesburg, South Africa\\
$^{122}$ University of Tokyo, Tokyo, Japan\\
$^{123}$ University of Tsukuba, Tsukuba, Japan\\
$^{124}$ University Politehnica of Bucharest, Bucharest, Romania\\
$^{125}$ Universit\'{e} Clermont Auvergne, CNRS/IN2P3, LPC, Clermont-Ferrand, France\\
$^{126}$ Universit\'{e} de Lyon, CNRS/IN2P3, Institut de Physique des 2 Infinis de Lyon, Lyon, France\\
$^{127}$ Universit\'{e} de Strasbourg, CNRS, IPHC UMR 7178, F-67000 Strasbourg, France, Strasbourg, France\\
$^{128}$ Universit\'{e} Paris-Saclay Centre d'Etudes de Saclay (CEA), IRFU, D\'{e}partment de Physique Nucl\'{e}aire (DPhN), Saclay, France\\
$^{129}$ Universit\`{a} degli Studi di Foggia, Foggia, Italy\\
$^{130}$ Universit\`{a} del Piemonte Orientale, Vercelli, Italy\\
$^{131}$ Universit\`{a} di Brescia, Brescia, Italy\\
$^{132}$ Variable Energy Cyclotron Centre, Homi Bhabha National Institute, Kolkata, India\\
$^{133}$ Warsaw University of Technology, Warsaw, Poland\\
$^{134}$ Wayne State University, Detroit, Michigan, United States\\
$^{135}$ Westf\"{a}lische Wilhelms-Universit\"{a}t M\"{u}nster, Institut f\"{u}r Kernphysik, M\"{u}nster, Germany\\
$^{136}$ Wigner Research Centre for Physics, Budapest, Hungary\\
$^{137}$ Yale University, New Haven, Connecticut, United States\\
$^{138}$ Yonsei University, Seoul, Republic of Korea\\
$^{139}$  Zentrum  f\"{u}r Technologie und Transfer (ZTT), Worms, Germany\\
$^{140}$ Affiliated with an institute covered by a cooperation agreement with CERN\\
$^{141}$ Affiliated with an international laboratory covered by a cooperation agreement with CERN.\\

\end{flushleft} 
\end{document}